\documentclass[12pt,preprint]{aastex}
\usepackage{emulateapj5,apjfonts}
\usepackage{apjfonts}
\usepackage{graphicx}
\usepackage{amsmath}
\usepackage{color}
\usepackage{onecolfloat}

\newcommand{\lsim}{\lower0.6ex\vbox{\hbox{$ \buildrel{\textstyle <}\over{\sim}\ $}}}
\newcommand{\gsim}{\lower0.6ex\vbox{\hbox{$ \buildrel{\textstyle >}\over{\sim}\ $}}}
\newcommand{\beq}{\begin{equation}}
\newcommand{\eeq}{\end{equation}}

\renewcommand\tabcolsep{0.12cm}

\begin{document}
\def\head{
  \vbox to 0pt{\vss
                    \hbox to 0pt{\hskip 440pt\rm LA-UR-09-04193\hss}
                   \vskip 25pt}

\submitted{The Astrophysical Journal, submitted}

\lefthead{Bhattacharya et. al.}
\righthead{Mass Function Predictions Beyond $\Lambda$CDM}

\title{Mass Function Predictions Beyond $\Lambda$CDM}

\author{Suman Bhattacharya\altaffilmark{1}, Katrin Heitmann\altaffilmark{2}, 
Martin White\altaffilmark{3}, Zarija Luki\'c\altaffilmark{1}, Christian~Wagner\altaffilmark{4}, 
and Salman Habib\altaffilmark{1}}

\affil{$^1$ T-2, Theoretical Division, Los
Alamos National Laboratory, Los Alamos, NM 87545}
\affil{$^2$ ISR-1, ISR Division,  Los
Alamos National Laboratory, Los Alamos, NM 87545}
\affil{$^3$ Departments of Physics and Astronomy, University of
California, Berkeley, CA 94720}
\affil{$^4$ Institut de Ci\'encies del Cosmos (ICC), Universitat de
  Barcelona, Marti i Franqu\'es 1, E-08028, Barcelona}

\date{today}


\begin{abstract} 

  The statistics of dark matter halos is an essential
  component of precision cosmology. The mass distribution of halos, as
  specified by the halo mass function, is a key input for several
  cosmological probes. The sizes of $N$-body simulations are now such
  that, for the most part, results need no longer be
  statistics-limited, but are still subject to various systematic
  uncertainties. Discrepancies in the results of simulation campaigns
  for the halo mass function remain in excess of statistical
  uncertainties and of roughly the same size as the error limits set
  by near-future observations; we investigate and discuss some of the
  reasons for these differences. Quantifying error sources and
  compensating for them as appropriate, we carry out a high-statistics
  study of dark matter halos from 67 $N$-body simulations to
  investigate the mass function and its evolution for a reference
  $\Lambda$CDM cosmology and for a set of $w$CDM cosmologies. For the
  reference $\Lambda$CDM cosmology (close to WMAP5), we quantify the
  breaking of universality in the form of the mass function as a function of
  redshift, finding an evolution of as much as $10\%$ away from the
  universal form between redshifts $z=0$ and $z=2$. For cosmologies
  very close to this reference we provide a fitting formula to our
  results for the (evolving) $\Lambda$CDM mass function over a mass
  range of $6\cdot 10^{11}-3\cdot 10^{15}$~M$_{\odot}$ to an estimated
  accuracy of about 2\%. The set of $w$CDM cosmologies is taken from
  the Coyote Universe simulation suite. The mass functions from this
  suite (which includes a $\Lambda$CDM cosmology and others with
  $w\simeq-1$) are described by the fitting formula for the reference
  $\Lambda$CDM case at an accuracy level of 10\%, but with clear
  systematic deviations. We argue  that, as a consequence, fitting
  formulae based on a universal form  for the mass function may have
  limited utility in high precision cosmological applications. 

\end{abstract}


\keywords{Cosmology: large-scale structure of universe --- methods:
$N$-body simulations} } \twocolumn[\head]
\section{Introduction}
\label{section:introduction}

The current paradigm for the formation of cosmological structure is
based on the gravitational amplification of primordial density
fluctuations in an expanding Universe. The nonlinear transformation of
dark matter overdensities -- via a hierarchical dynamical process --
into clumpy distributions called halos, and the subsequent infall of
baryons leading to the formation of stars and galaxies within these
halos, rounds out the present picture of the formation of observed
structure. Although there is no precise mathematical definition of a
`halo', several operational definitions -- depending on the particular
applications of interest -- have been employed in practice.

The spatio-temporal statistics of halos and sub-halos, as well as of
their mass distribution (and its evolution), together provide most of
the descriptive framework within which fit all of the structure
formation-based probes of cosmology. The mass function alone is a very
useful probe in determining cosmological parameters. Because large and
massive halos form very late, the high-mass tail of the mass function
-- the regime of cluster-scale masses -- is exponentially sensitive to
dark energy-related parameters~\citep{haiman01}. Additionally, the
redshift evolution of the cluster mass function depends strongly on
the cosmological parameters in a way that is complementary to other
probes. The cluster mass function can also be used to measure the
normalization of primordial density fluctuations, $\sigma_8$, and
search for hints of primordial non-Gaussianity (see,
e.g.,~\citealt{dalal07, oguri09}). On cluster mass scales and smaller,
the mass function, both directly and indirectly, plays an important
role in halo models of galaxy formation and statistics, as applied to
a wide range of redshifts and objects (predictions of bias, early
galaxies, groups, quasars, spatial statistics of luminous galaxies,
etc.).

An important motivation for the precision determination of the mass
function is the existence of several ongoing and upcoming surveys that
aim to detect clusters via their optical, X-ray and Sunyaev-Zel'dovich
(SZ) effect
signatures~\citep{maxbcg,abbott05,vikhlinin08,act,spt,bartlett08,fang07}.
The number of detected clusters from the individual surveys will range
from thousands to tens of thousands. To maximally extract cosmological
information from these cluster surveys, the mass function must be
specified to better than a few percent accuracy for a range of
cosmologies. As discussed by~\cite{cunha09a} and by~\cite{wu}, the
current theoretical uncertainty in the determination of the mass
function can lead to a considerable degradation in the constraints on
cosmological parameters. The investigations have also pointed to the
usefulness of determining the mass function over a wide range of
masses, extending down into the group scale.

Because massive halos are very nonlinear and dynamically nontrivial
objects, a fully satisfactory first principles approach to determine
the structure and statistics of halos does not yet exist. It follows,
therefore, that our current theoretical understanding of the halo mass
function is somewhat limited. (This situation may be contrasted to
that of nonlinear perturbation theory for the matter power spectrum,
where independent of whether individual approaches fail or succeed,
the actual problem is well defined conceptually and mathematically.)
From the analytical standpoint, the only viable approach to the mass
function is still that based on the (heuristic) Press-Schechter (PS)
excursion set model \citep{ps80,bond91} and its extensions (see
\citealt{zentner07} for a review). Although this work has been
valuable in suggesting functional forms and representations for the
mass function and in analyzing such effects as the scaling of
finite-volume corrections, it has not independently yielded
predictions for the mass function that are anywhere close to the
accuracies that are now required. (For a recent critical assessment,
see~\citealt{robertson08}). Moreover, it is hard to imagine how
additional dynamics, gas physics, and feedback mechanisms can be
modeled within such a framework. Therefore, it appears that a
sufficiently accurate prediction for the mass function of halos can
only be achieved using high resolution simulations, and modeling a
range of physics tailored to specific applications.

Numerical simulations have become a standard tool to determine the
halo mass function over the last decade. Several groups have used
suites of simulations to calibrate the halo mass function over an
increasingly wider range of masses and redshifts~\citep{J01,evrard02,
  white02,reed03,warren05,heitmann06,reed06,
  lukic07,tinker08,millenium2,crocce09}; see~\cite{J01} for references
to previous work. A key aspect of the calibration of the mass function
is the use of $\ln\sigma^{-1}(M,z)$ as the central variable, instead
of the halo mass, $M$. Here $\sigma^2(M,z)$ is the variance of the
linear density field, extrapolated by linear theory to the redshift of
interest, $z$, and smoothed by a spherical top-hat filter of radius
$R$, which on average encloses a mass $M$
$(R=[3M/4\pi\rho_b(z)]^{1/3})$. The associated scaled differential
mass function $f(\sigma,z;X)$~\citep{J01} is
\begin{equation}
f(\sigma,z;X)=\frac{M}{\rho_b}\frac{dn_X(M,z)}{d\ln[\sigma^{-1}(M,z)]},
\label{univ}
\end{equation}
where $X$ labels the cosmological model and particular halo
definition. The variable $\ln\sigma^{-1}(M,z)$ appears naturally in
the PS approach and extensions thereof, presenting a relatively simple
form for $f(\sigma,z;X)$, in fact one with no dependence on
cosmological epoch and parameters. \cite{J01} found that for a certain
fixed definition of halo, independent of cosmology, their simulation
results covering redshifts from $z=0-4$, and across different
cosmologies, could be well fitted by this ``universal'' form of the
mass function at accuracies of order 20\%. Recent work has shown that
mass function universality is apparently not valid beyond the $5-10\%$
level~\citep{reed06, lukic07,tinker08, cohn07,crocce09,courtin10}.

Efforts to study this issue further quickly encounter a host of
complications. Even independent of such significant physics issues as
mass-observable mapping and baryonic effects, it turns out that the
choice of halo definition and systematic errors in simulations can
easily have as large an effect as that being investigated. Thus,
despite the major effort expended in numerical determination of the
halo mass function, the present situation cannot be considered to be
fully satisfactory, as we discuss in Section~\ref{section:halos}.
Among other sources of error, the effects of finite force resolution
and finite sampling error must be carefully dealt with in order to
obtain a converged result. 

Beyond this point, there is a further cautionary note to keep in mind
in terms of precision determination of the mass function: most
simulation campaigns have focused on a single cosmology at a time.
Therefore, even though results are often quoted in the universal form
of Equation~(\ref{univ}) with small statistical errors, in the absence
of rigorous testing of the universal ansatz they cannot be directly
applied to cosmologies other than those considered specifically (and
even in this case, the actual systematic errors have often turned out
to be larger than originally estimated).

Motivated by these considerations, it is important to first establish
just how accurately various mass functions can be computed and what
the systematic errors are in the most fundamental situation -- the
gravity-only $N$-body case. Once an accurate mass function for a
particular $\Lambda$CDM case has been established, it is important to
consider a range of observationally relevant redshifts and of
cosmologies around that reference point (see, e.g., the discussion in
\citealt{heitmann09}), to understand and explore the range of
applicability -- and limitations -- of the (almost) universal
description described above. Therefore, the major aims of this paper
are: (i) to carefully consider the systematic effects due to numerical
errors on the mass function and either avoid or correct for them; (ii)
based on these results, establish an accurate prediction for the mass
function of a reference $\Lambda$CDM model at $z=0$; and (iii) extend
the investigation to a larger redshift range and provide an accurate
prediction for more general $w$CDM models (where the dark energy
equation of state parameter, $w$, is constant in time, but $w\neq
-1$).

As a first step, the choice of halo definition has to be considered.
For the most part, numerical simulations use two different techniques
to identify halos: friends-of-friends (FOF) or spherical overdensity
(SO). In the FOF method, halos are found by a percolation technique
where particles belong to the same halo if they are within a certain
distance (the linking length $b$) of each other. The linking length is
typically chosen between $b=0.15$ and $b=0.2$, where $b$ is defined
with respect to the mean interparticle spacing. The FOF definition of
halos approximately traces isodensity contours and connects more
directly to the simulated mass distribution; it is often used in
cluster SZ studies. However, the choice of linking length is an issue:
too large a linking length can connect neighboring overdensities in a
possibly unrealistic manner. The SO method measures the mass in
spherical shells around the center of the halo (which is usually
determined from the potential minimum of the halo or from the most
bound particle) until the density in the shells falls below a certain
threshold which is given with respect to either critical or background
density. Typically, values for $m_{200}$ to $m_{500}$ (or higher in
the case of clusters) are measured (with respect to $\rho_c$). The
spherical overdensity method is particularly convenient for providing
predictions for certain kinds of observations, e.g. X-ray cluster
masses where one is concerned primarily with studying the inner,
virialized region of a halo. The major disadvantage of the SO method
is the crudeness of the spherical approximation and that neighboring
halos can overlap.

Because isolated, relaxed halos are well-fit by the
Navarro-Frenk-White (NFW) profile~\citep{nfw}, SO and FOF masses are
strongly correlated~\citep{white01, tinker08}. In fact, if the halo
concentrations are known, it has been shown that -- in cosmological
simulations -- a one-to-one mapping for the two halo definitions
exists at the 5\% level of accuracy~\citep{lukic08}. However, a fair
fraction of halos in simulations are irregular. For currently favored
cosmologies, 15-20\% of $b=0.2$ FOF halos have irregular substructure
or have two or more major halo components linked
together~\citep{lukic08}. For such irregular halos, not only does the
simple mapping between SO and FOF halos fail, it is not obvious just
how to define an appropriate halo mass (lower $b$ to what value, or
correspondingly, what choice of overdensity criterion to use?).

In the absence of a compelling theoretical motivation, most numerical
studies of the mass function have used FOF masses with linking length
$b=0.2$ following the convention set by~\cite{J01} who noted that this
definition led to a universal form for the mass function (for a
systematic investigation, see \citealt{white02, tinker08}). While noting its
possible deficiencies, we retain this convention here in order to
better compare our results with other work.

Our study of the FOF mass function uses a large suite of runs for a
single reference $\Lambda$CDM cosmology (very close to the WMAP5
parameters from \citealt{wmap5}) and a set of $w$CDM cosmological
simulations -- the ``Coyote Universe'' suite named after the
supercomputer it was run on -- that include a different $\Lambda$CDM
model and a few others close to $\Lambda$CDM. The latter set of
simulations represents a simple step beyond $\Lambda$CDM, where the
dark energy equation of state parameter is treated in a purely
phenomenological context. Allowing for dark energy evolution (as
required by quintessence models, for example) opens up a large
parameter space that near-future observations are unlikely to be
sensitive to. We, therefore, defer this extension to future work.

In order to carefully control errors, we have followed the criteria
for starting redshift, and mass and force resolution as presented
in~\cite{lukic07}. These criteria ensure that halos of a certain size
and at certain redshifts can be resolved reliably.
In~\cite{heitmann09} similar criteria were laid out to obtain the
matter power spectrum at 1\% accuracy out to scales $k\sim
1~h$Mpc$^{-1}$. These criteria are also obeyed by the simulations used
here. As discussed further in Section~\ref{section:halos}, the
high-mass tail of the mass function is particularly susceptible to
systematic errors in the determination of individual halo masses.
These systematic errors can arise from the effects of finite force
resolution and we study and characterize these effects. Overall, the
contribution of various errors in typical cosmological simulations
breaks down basically as follows: (i) too low starting redshift, $\sim
10\%$ \citep{lukic07}, (ii) halo sampling errors, $\sim 5\%$, (iii)
transfer function approximations, $\sim 5\%$, (iv) finite volume
effects, $\sim 1\%$, (v) force resolution effects, $\sim 1\%$. Of
these, (i) and (iii) are trivially avoidable, and the others can be
controlled at least to the percent level.

In this paper, after compensating for finite sampling and force
resolution limitations, we present quantitative results for the mass
function from the reference $\Lambda$CDM simulations and for a suite
of $w$CDM cosmologies designed explicitly to bracket the currently
observationally relevant range of cosmological parameters
\citep{heitmann09}. We provide a fitting formula describing our
reference $\Lambda$CDM simulation data at the 2\% accuracy level at
the current epoch (we also compute the halo model prediction for the
large scale halo bias from the mass function fit). We trace the
evolution of the mass function between redshifts $z=0-2$, which
represents up to a 10\% breaking of the universal description of the
mass function across a representative range of masses.

We then turn to investigating the variation of the mass function as a
function of cosmological parameters from the suite of $w$CDM
simulations. This simulation suite, while it lacks some of the
statistical power of the reference $\Lambda$CDM runs, provides a good
test of the validity of the universal description of the mass
function. At $z=0$, we find that universality for different
cosmologies holds to no better than at the 10\% level, with clear
systematic deviations (in both directions) from the quasi-universal
form fitted to the reference $\Lambda$CDM runs. Thus, while not
adequate for future precision studies (as is the case for all current
mass function fits), our analytic form provides a good estimate for
the mass function for $w$CDM cosmologies at the accuracy of currently
available data.

\begin{table}[t]
\begin{center} 
\caption{\label{tab:basic} Parameters for the 38 cosmological models}    
\hspace{0cm}
\begin{tabular}{cccccccc}
\tableline\tableline
\#& $\omega_m$ & $\omega_b$ & $n_s$ & $-w$ & $\sigma_8$ & $h$ & M\\
 & & & & & & & $10^{14} M_\odot$\\
\hline 
0 & 0.1296 &  0.0224 & 0.9700 & 1.000 &  0.8000 & 0.7200 & 7.00\\
1 & 0.1539 & 0.0231 & 0.9468 & 0.816 & 0.8161  & 0.5977 & 13.3\\
2 & 0.1460 & 0.0227 & 0.8952 & 0.758 & 0.8548 & 0.5970 & 15.9\\
3 & 0.1324 & 0.0235 & 0.9984 & 0.874 & 0.8484 & 0.6763 & 9.96\\
4 & 0.1381 & 0.0227 & 0.9339 & 1.087 & 0.7000 & 0.7204 & 4.42\\
5 & 0.1358 & 0.0216 & 0.9726 & 1.242 & 0.8226 & 0.7669 & 7.20\\
6 & 0.1516 & 0.0229 & 0.9145 & 1.223 & 0.6705 & 0.7040 & 4.27\\
7 & 0.1268 & 0.0223 & 0.9210 & 0.700 & 0.7474 & 0.6189 & 7.30\\
8 & 0.1448 & 0.0223 & 0.9855 & 1.203 & 0.8090 & 0.7218 & 8.04\\
9 & 0.1392 & 0.0234 & 0.9790 & 0.739 & 0.6692 & 0.6127 & 4.98\\
10 & 0.1403 & 0.0218 & 0.8565 & 0.990 & 0.7556 & 0.6695 & 7.58\\
11 & 0.1437 & 0.0234 & 0.8823 & 1.126 & 0.7276 & 0.7177 & 5.64\\
12 & 0.1223 & 0.0225 & 1.0048 & 0.971 & 0.6271 & 0.7396 & 2.26\\
13 & 0.1482 & 0.0221 & 0.9597 & 0.855 & 0.6508 & 0.6107 & 4.78\\
14 & 0.1471 & 0.0233 & 1.0306 & 1.010 & 0.7075 & 0.6688 & 5.42\\
15 & 0.1415 & 0.0230 & 1.0177 & 1.281 & 0.7692 & 0.7737 & 5.47\\
16 & 0.1245 & 0.0218 & 0.9403 & 1.145 & 0.7437 & 0.7929 & 4.22\\
17 & 0.1426 & 0.0215 & 0.9274 & 0.893 & 0.6865 & 0.6305 & 5.50\\
18 & 0.1313 & 0.0216 & 0.8887 & 1.029 & 0.6440 & 0.7136 & 3.05\\
19 & 0.1279 & 0.0232 & 0.8629 & 1.184 & 0.6159 & 0.8120 & 1.88\\ 
20 & 0.1290 & 0.0220 & 1.0242 & 0.797 & 0.7972 & 0.6442 & 8.24\\ 
21 & 0.1335 & 0.0221 & 1.0371 & 1.165 & 0.6563 & 0.7601 & 2.80\\ 
22 & 0.1505 & 0.0225 & 1.0500 & 1.107 & 0.7678 & 0.6736 & 7.46\\ 
23 & 0.1211 & 0.0220 & 0.9016 & 1.261 & 0.6664 & 0.8694 & 2.19\\ 
24 & 0.1302 & 0.0226 & 0.9532 & 1.300 & 0.6644 & 0.8380 & 2.44\\ 
25 & 0.1494 & 0.0217 & 1.0113 & 0.719 & 0.7398 & 0.5724 & 9.09\\ 
26 & 0.1347 & 0.0232 & 0.9081 & 0.952 & 0.7995 & 0.6931 & 8.24\\ 
27 & 0.1369 & 0.0224 & 0.8500 & 0.836 & 0.7111 & 0.6387 & 6.28\\ 
28 & 0.1527 & 0.0222 & 0.8694 & 0.932 & 0.8068 & 0.6189 & 12.6\\ 
29 & 0.1256 & 0.0228 & 1.0435 & 0.913 & 0.7087 & 0.7067 & 4.14\\ 
30 & 0.1234 & 0.0230 & 0.8758 & 0.777 & 0.6739 & 0.6626 & 4.09\\ 
31 & 0.1550 & 0.0219 & 0.9919 & 1.068 & 0.7041 & 0.6394 & 6.25\\ 
32 & 0.1200 & 0.0229 & 0.9661 & 1.048 & 0.7556 & 0.7901 & 4.30\\
33 & 0.1399 & 0.0225 & 1.0407 & 1.147 & 0.8645  & 0.7286 & 9.37\\ 
34 & 0.1497 & 0.0227 & 0.9239 & 1.000 & 0.8734 & 0.6510 & 14.5\\  
35 & 0.1485 & 0.0221 & 0.9604 & 0.853 & 0.8822 & 0.6100 & 16.4\\ 
36 & 0.1216 & 0.0233 & 0.9387 & 0.706 & 0.8911 & 0.6421 & 12.9\\
37 & 0.1495 & 0.0228 & 1.0233 & 1.294 & 0.9000 & 0.7313 & 11.7\\
\tableline\tableline
\vspace{-1.5cm}

\tablecomments{See text for more details and \cite{heitmann09} for the
  model\\ selection procedure. To obtain good statistics over a wide range of halo
  masses,\\ the reference $\Lambda$CDM case, model~0, was augmented by a
  set of additional runs\\ (see Table~\ref{tab:sim_specs}). The
  rightmost column shows the mass corresponding to $1/\sigma=1.8$\\ for
  each cosmology at $z=0$.}

\end{tabular}
\end{center}
\end{table}

Our results demonstrate that the goal of determining the mass function
at the percent level of accuracy will require a much more intensive
program of simulations in the future, sampling both cosmological and
physical modeling parameters (baryonic physics, feedback), along with
well-controlled statistical errors. As has been emphasized earlier
\citep{lukic07}, a possible solution is mass function emulation from a
large, but finite set of simulations, using techniques that have been
shown to be successful in high-dimensional regression problems
\citep{habib07}.

The paper is organized as follows. In Section~\ref{section:simulation}
we describe the simulation suite used in this paper, encompassing 67
high-resolution simulations for 38 different cosmologies. Several
overlapping-volume $\Lambda$CDM simulations are used to understand and
control systematic errors. These errors and their ramifications for
the accuracy of the mass determination of halos, and how these
translate to limiting the accuracy of the mass function itself, are
discussed in Section~\ref{section:halos}. In
Section~\ref{section:massfunction} we present our results for the mass
function for the reference $\Lambda$CDM model at different redshifts
and provide a new fitting form for the mass function matching our
simulations at the 2\% level as well as the associated mass
function-derived halo bias. We extend our discussion in
Section~\ref{section:mf_wcdm} to the wider set of $w$CDM cosmologies
and investigate how well the mass function fit derived for the
reference cosmology holds for this broader class of models. We
conclude in Section~\ref{section:disc}. We discuss error control
issues and provide relevant details in Appendix~\ref{app}.

\section{Simulation Suite}
\label{section:simulation}

\begin{table*}[t]
\begin{center} 
\caption{\label{tab:sim_specs}  Specifications of  the Simulation runs}
\begin{tabular}{ccccccccccc}
\tableline\tableline
 Box size & Name & $n_p$ &  $m_p$ & $n_h^{\rm min}$ &$\epsilon$ & Code & $z_{\rm in}$ & $z_{\rm out}$ &$N_{\rm runs}$ & ICs\\
\hline
\hline 
&&&&$\Lambda$CDM &&&&\\
\hline
1000  Mpc &  C   & 1500$^3$  &   $1.1\cdot 10^{10}$~M$_\odot$ & 400 & 24 kpc  &  TreePM                  & 100/75       & 0     & 2 &ZA/2LPT\\
1736 Mpc & B    & 1200$^3$  &   $1.1\cdot 10^{11}$~M$_\odot$& 400 & 51 kpc   &  TreePM                  & 100  & 0, 1 & 6 & 2LPT\\
2778 Mpc & A    & 1024$^3$  &   $7.2\cdot 10^{11}$~M$_\odot$& 400 & 97 kpc   &  TreePM                  & 100  & 0, 1 & 10 & 2LPT\\
178   Mpc & GS &   512$^3$   &   $1.5\cdot 10^{9}$~M$_\odot$& 400 &14 kpc   & {\small GADGET-2}  &  211 & 0, 1, 2 & 10 & ZA\\
1300  Mpc & G    & 1024$^3$  &  $7.4\cdot 10^{10}$~M$_\odot$ & 400 &50 kpc  & {\small GADGET-2}  &  211 & 0, 1, 2 & 2 & ZA\\
\hline
&&&& $w$CDM &&&&\\
\hline
1300 Mpc & Coyote & 1024$^3$ & varies & 400 &50 kpc & {\small GADGET-2} & 211 & 0, 1, 2 &  37 & ZA\\

\tableline\tableline
\vspace{-1.5cm}

\tablecomments{Box size, mass and force resolution for the different
  runs; the upper section of the table describes the reference
  $\Lambda$CDM simulation suite (model 0 in Table~\ref{tab:basic})
  while the lower section specifies the Coyote Universe runs (models 1
  - 37 in Table~\ref{tab:basic}). The total number of particles is
  denoted by $n_p$, the particle mass by $m_p$, $n_h^{\rm min}$ the
  number of particles in the smallest halo kept, $\epsilon$ the force
  resolution, and $N_{\rm runs}$ the number of realizations. For some
  simulations we used the Zel'dovich approximation~\citep{zel} (see
  also discussions in \citealt{lukic07} and \citealt{heitmann08b}) to
  generate the initial condtions and 2LPT~(\citealt{bouchet95},
  \citealt{2lpt}) for others.} 

\label{tab:sim}
\end{tabular}
\end{center}
\end{table*}

Our simulation suite spans a wide range of observationally relevant
$w$CDM cosmologies, as specified in Table~\ref{tab:basic}. For each
model we have results from a 1.3~Gpc box simulation, run with $1024^3$
particles, with  masses of $\approx 10^{10}$~M$_\odot$, exact values
depending on the specific cosmology.  We vary five cosmological
parameters within the following boundaries: 
\begin{eqnarray}
\label{eq:range}
0.120 \leq & \omega_m & \leq 0.155,\nonumber\\
0.0215 \leq & \omega_b & \leq 0.0235,\nonumber\\
0.85 \leq & n_s & \leq 1.05,\\
-0.130 \leq & w & \leq -0.70,\nonumber\\
0.61 \leq & \sigma_8 & \leq 0.90.\nonumber
\end{eqnarray}
The Hubble parameter $h$ is fixed for models 1-37 by imposing the
cosmic microwave background constraint, $\ell_A=\pi d_{\rm ls}/r_s =
302.4$, where $d_{\rm ls}$ is the distance to the last scattering
surface and $r_s$ is the sound horizon. For a detailed description of
the model selection process, see~\cite{heitmann09}. The simulations
are carried out with {\small GADGET-2}~\citep{Spr05}, a tree-particle
mesh (tree-PM) code. For a detailed discussion and comparison of
different $N$-body methods used for cosmological simulations,
including {\small GADGET-2}, see, e.g.~\cite{heitmann08a}. We use a
2048$^3$ PM grid and a (Gaussian) smoothing of 1.5 grid cells. The
force matching is set to six times the smoothing scale, the tree
opening criterion being set to 0.5\%. The softening length is 50~kpc.

For model 0, the reference $\Lambda$CDM cosmology, we have carried out
additional simulations for different box sizes and multiple
realizations in order to cover a wide halo mass range between
$6\cdot10^{11}$~M$_\odot$ to $3\cdot$10$^{15}$~M$_\odot$ with good
statistics. Results from the overlapping volume boxes are also useful
in understanding systematic errors. Details about these $\Lambda$CDM
simulations are given in Table~\ref{tab:sim_specs}. In addition to
{\small GADGET-2} we use a second tree-PM code for a subset of these
simulations, described in \cite{white02}. The algorithmic structure of
this code is very similar to {\small GADGET-2} and the code was also
part of the code comparison carried out in~\cite{heitmann08a}. Aside
from the main simulation runs, we also use a PM simulation with
identical cosmological parameter settings as for the ``G'' run solely
to study the impact of force resolution on individual halo masses.

\section{Halos in Numerical Simulations}
\label{section:halos}

\begin{figure}[b]
\centerline{
\includegraphics[width=8cm]{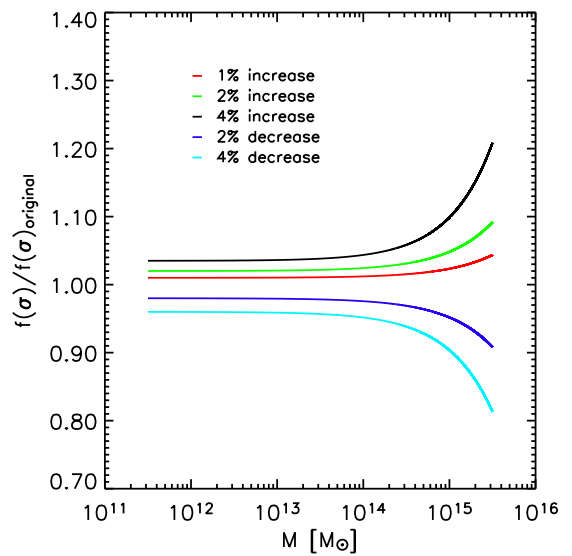}}
\caption{Sensitivity of the mass function to systematic shifts in
  individual halo masses. Changes are shown relative to a baseline
  mass function, taken to be the fitting form of Table~\ref{tab:eqn}.
  A small shift of 2\% in the halo masses can lead to changes of up to
  5-10\% in the high mass tail of the mass function.}
\label{fig:corr2}
\end{figure}

An important question to understand before proceeding further is how
uncertainties in individual halo masses -- defined any way one chooses
-- translate into errors in the mass function itself.  To investigate
this sensitivity we carry out some simple experiments. Introducing a
Gaussian (or other symmetric) noise with $\sigma=4\%$ in individual
halo mass measurements can make a small difference in the mass
function at high masses (at the percent level) and should not be a
concern. The picture changes substantially, however, if the halo
masses are given small systematic shifts. The error induced in the
mass function can become much larger than the level of systematic
error introduced in individual halo masses. We demonstrate this effect
in Figure~\ref{fig:corr2}, where systematic shifts in halo masses of
1\%, 2\%, and 4\% are studied. An increase of the masses by only 4\%
results in a difference in the mass function of 15\% at high masses,
which is quite significant.

These results, arising from the exponential sensitivity of the mass
function at high masses, demonstrate an important point: Given the
level of uncertainties in halo masses due to numerical errors and the
definition of the halo mass itself, it is not possible to derive a
mass function prediction from simulations at sub-percent or percent
accuracy without a consistent study of how individual halo masses vary
as a function of simulation parameters like force resolution, time
step size, starting redshift, etc. Fortunately, it is already known
that ($b$=0.2) FOF halo masses, over the mass range of interest, are
relatively robust to changes in simulation parameters; as demonstrated
in \cite{heitmann05}, individual halo masses as computed by six
different codes with varying resolutions and time-stepping schemes
typically agree to better than 2\%. Additionally, \cite{lukic07} have
provided criteria for running simulations so as to minimize systematic
errors from a variety of possibilities. Our task is to ascertain
whether error control can be further improved systematically.

\begin{figure*}[t]
\includegraphics[width=7.5cm]{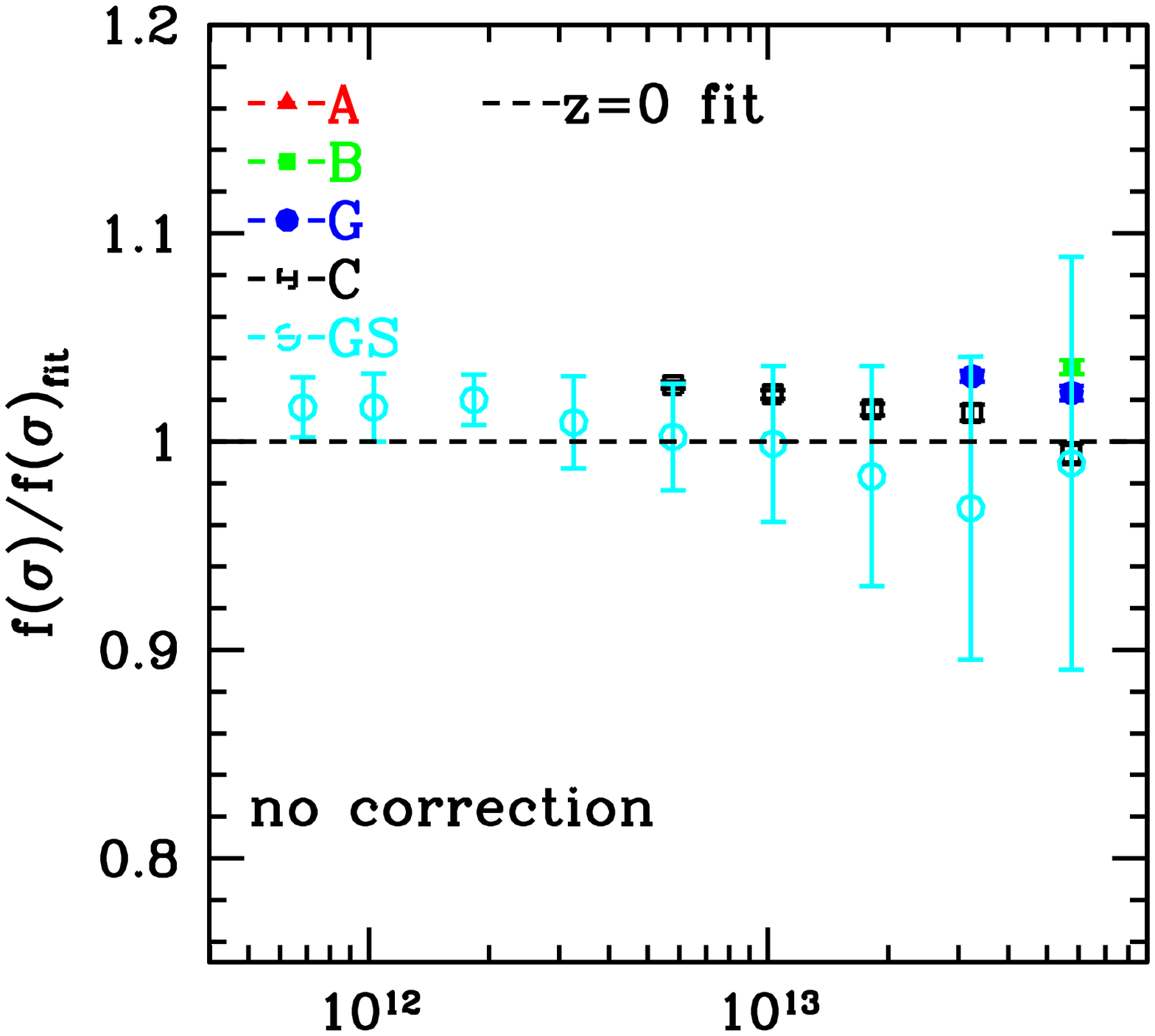}
\hspace{-2.2cm}
\includegraphics[width=7.5cm]{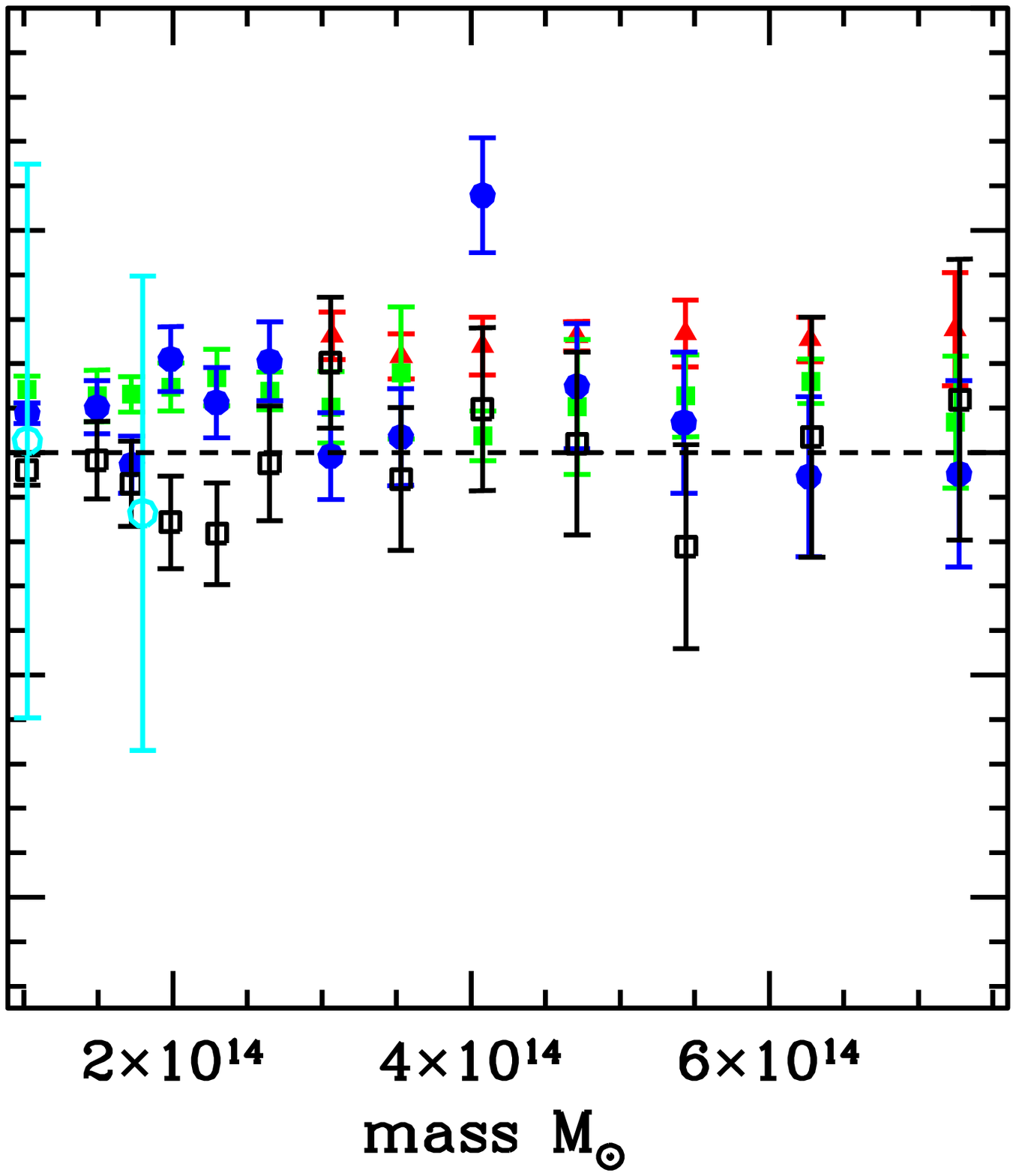}
\hspace{-2.2cm}
\includegraphics[width=7.5cm]{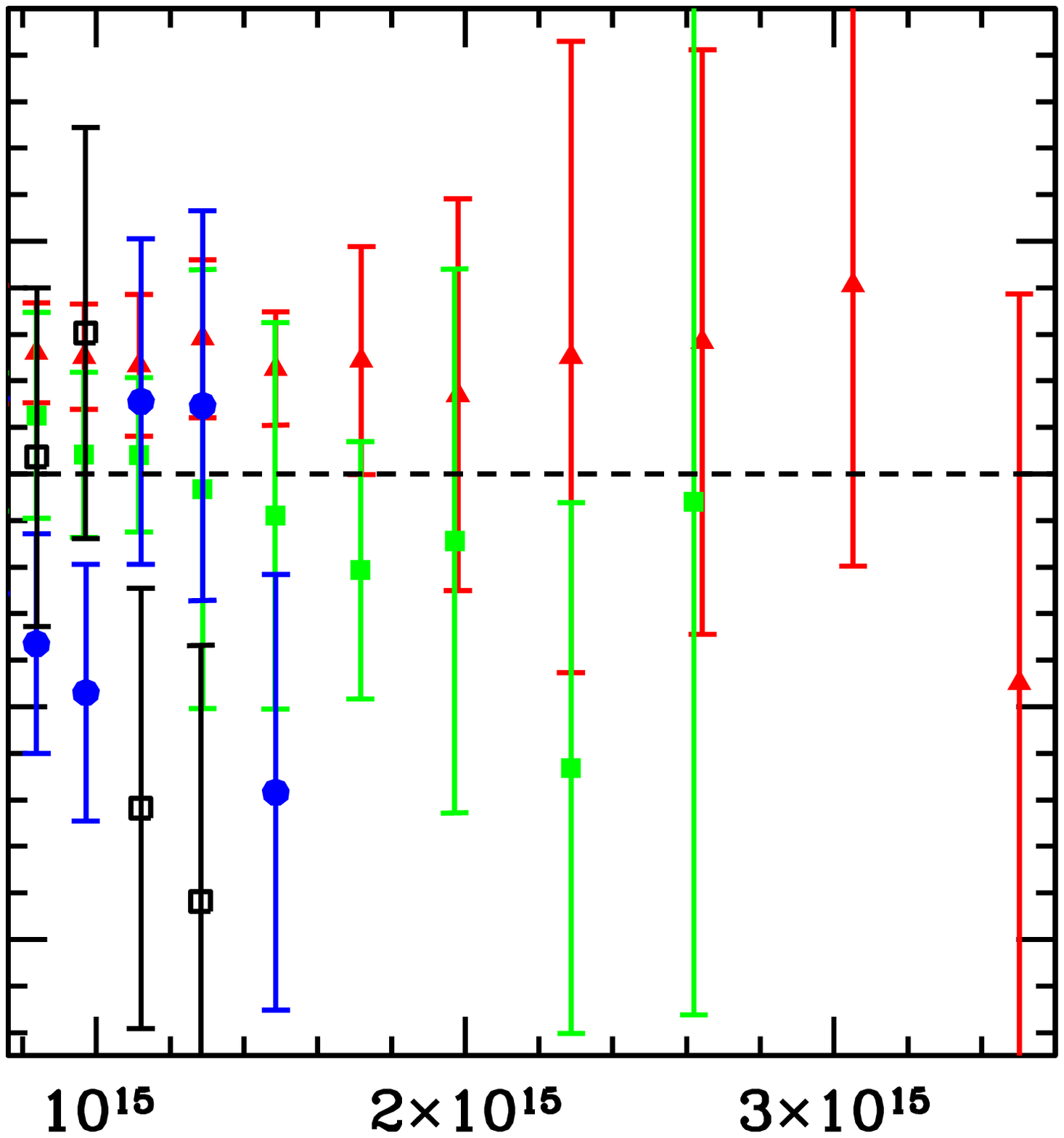}\\
\vspace{-1.2cm}

\includegraphics[width=7.5cm]{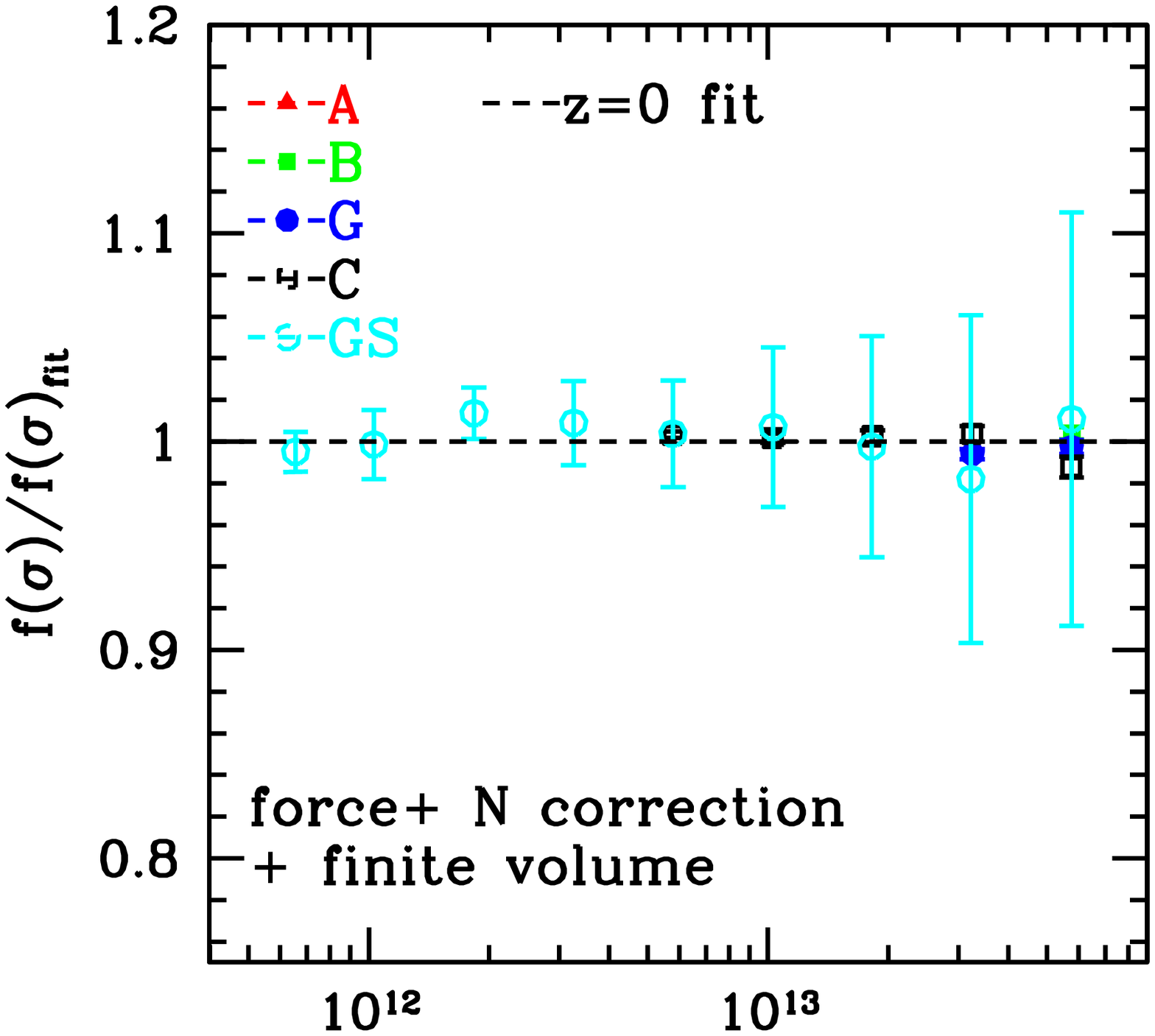}
\hspace{-2.2cm}
\includegraphics[width=7.5cm]{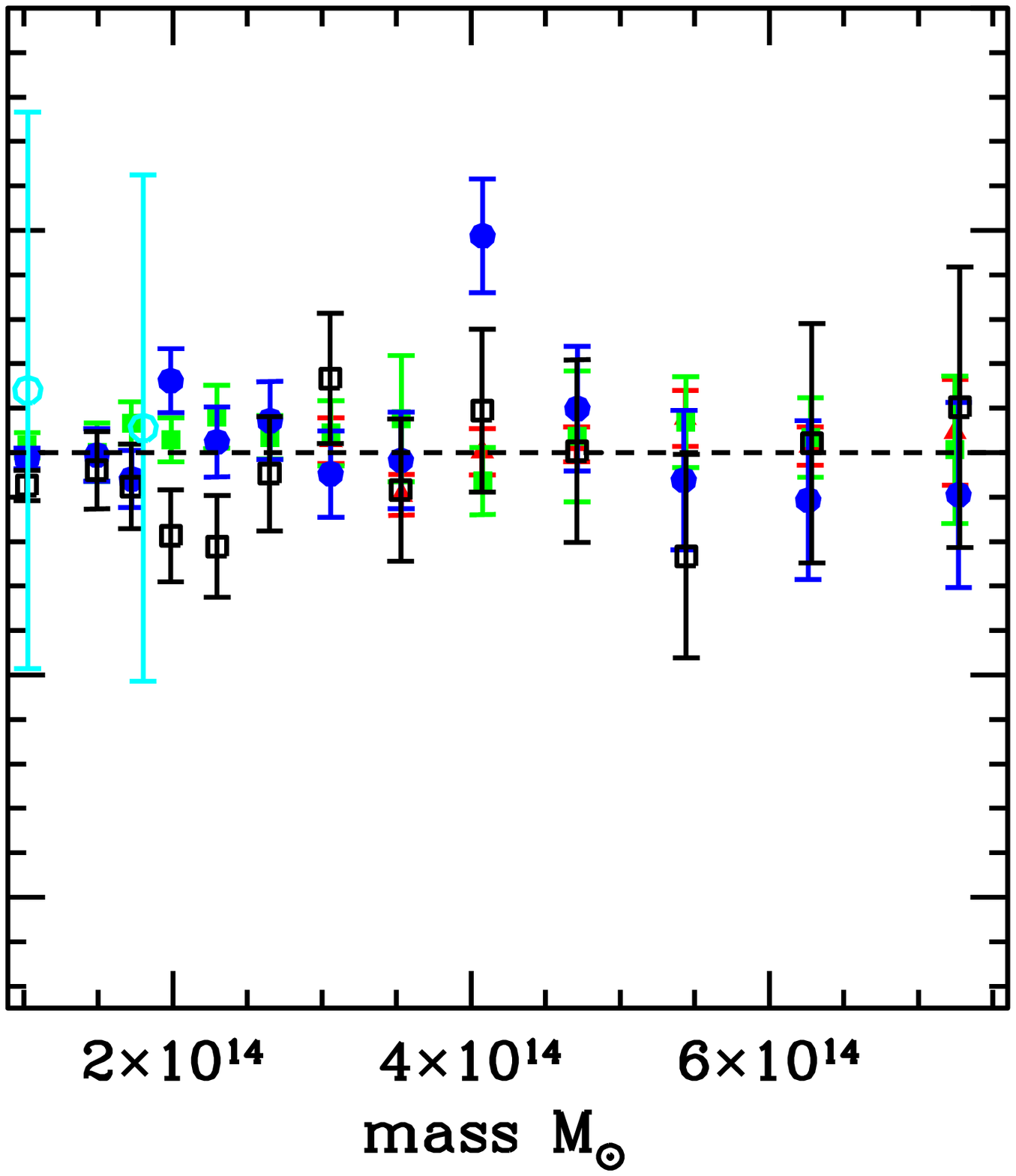}
\hspace{-2.2cm}
\includegraphics[width=7.5cm]{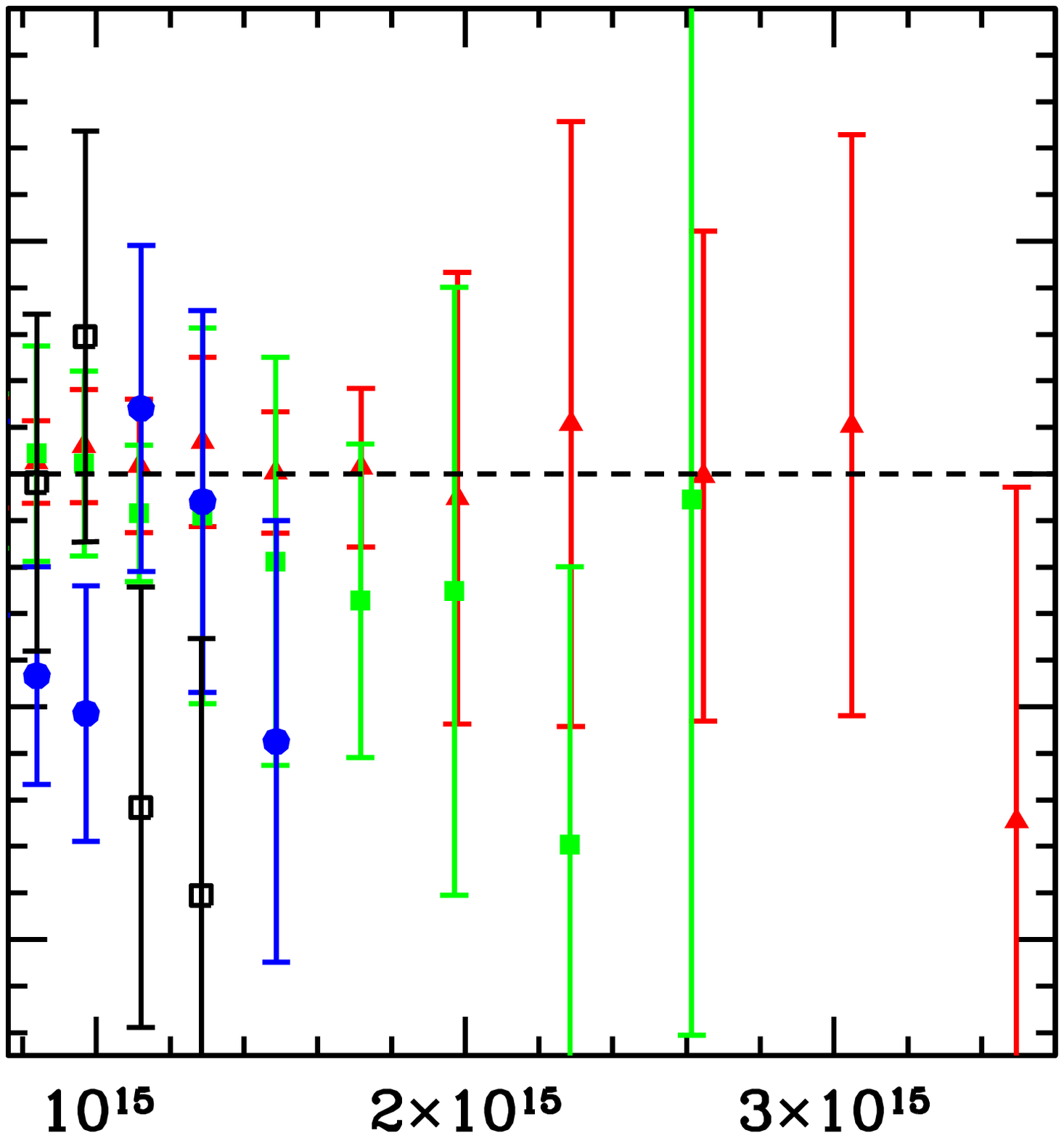}\\
\vspace{-1cm}

\caption{Uncorrected (upper panel) and corrected (lower panel) results
  for the ratio of the mass function to our best fit model at $z=0$.
  Most of the difference is due to the correction for the finite particle
  sampling of the halos. A detailed description of the error bars is
  given in the appendix.}
\label{fig:mass_corr}
\end{figure*}

We focus here on three main possible {\em systematic} errors in the
mass function (based on experience from \cite{heitmann05} and
\cite{lukic07}): (i) small systematic errors in individual halo masses
due to force resolution, focusing attention on the high mass tail,
(ii) the known systematic bias in individual FOF masses as a function
of the number of particles in the FOF halo~\citep{warren05}, and (iii)
systematic errors from missing long wavelength power due to the
necessarily finite box size. We also note the easily avoidable pitfall
of using approximate fits for the transfer function rather than the
numerical solution (see Appendix~\ref{app} for details). As discussed
below, and in more detail in Appendix~\ref{app}, all of these effects
can induce systematic errors in the mass function and need to be taken
into account. (Figure~\ref{fig:mass_corr} shows the effect of various
corrections.)

All numerical simulations are necessarily run with a finite force
resolution, balancing the need for high spatial resolution with
suppression of noise/collisional artifacts. Below the chosen force
softening length, the forces between the particles asymptote to zero.
As a result, the halos that are formed tend to be ``puffed out'' in a
simulation with coarse force resolution. The net effect is that -- for
heavy halos, with scale radii significantly greater than the softening
length -- the mass of a given halo in a simulation with coarse
resolution tends to be greater compared to a simulation with better
resolution. This implies that the halo mass and hence the mass
function determined from a simulation study would be higher compared
to the ``ideal'' case of infinite force (and mass) resolution.
Although this effect is known to be small \citep{lukic07}, it can
certainly be significant at the percent level of error in the mass
function. We also note that the effect of force resolution depends on
the details of how halos are found using the SO and FOF algorithms.
Here we focus on the case of FOF halos; a discussion of the impact of
force resolution on SO halos is given in \cite{tinker08}. In
Appendix~\ref{app} we provide a detailed analysis of  the errors due
to finite force resolution and how to correct for them.  In our
simulations, we find that finite force resolution effects can be
accounted for by introducing a corrected rescaled mass $M_c$ via 
\begin{equation}
M_c/M= 1.0- 0.04(\epsilon/650 ~{\rm kpc}).
\end{equation}
Here $M$ is the uncorrected halo mass and $\epsilon$ is the force
resolution measured in kpc of the different runs as specified in
Table~\ref{tab:sim}. In our case, the biggest correction applied is
$\approx 0.6\%$ for individual halo masses for the $A$ runs (with a
force resolution of 97~kpc). This results in a systematic lowering of
approximately 2\% in the high-mass tail of the mass function
(primarily run A) as shown in Figure~\ref{fig:force_corr} in the
Appendix.

As stated earlier, we identify halos with a standard FOF algorithm,
with a linking length, $b=0.2$. Although halos with only a small
number of particles ($\sim 20$) can be reliably found with the FOF
algorithm, accurate mass estimation requires keeping many more
particles within individual halos. Aside from simple considerations of
particle shot noise, there is an inherent systematic error and scatter
in the definition of an FOF halo mass with particle number, even in
the absence of all other limitations, as pointed out by
\cite{warren05}. For ideal NFW halos (and for isolated relaxed halos
in simulations), this effect was studied by~\cite{lukic08} and
represents the best possible scenario. To avoid problems with too few
particles in halos, we restrict attention to halos with at least 400
particles. With this cutoff, the agreement in the mass function for
the overlap regions across the various boxes is within a few percent.
After first applying the FOF sampling correction suggested by
\cite{warren05}, we find that a slightly modified correction of the
form $n_h^{\rm corr}=n_h(1-n_h^{-0.65})$ brings the results from the
nested boxes in good agreement, as shown in the Appendix in
Figure~\ref{fig:n_corr}. We stress that this adjustment is purely
empirical, targeted at matching halo masses in overlapping boxes. The
study in ~\cite{lukic08} shows that for NFW halos, the correction
depends on $n_h$ as well as on the halo concentration. Thus, in
principle, one may expect the FOF sampling correction to be more
complex than a simple compensation based on $n_h$; we leave a more
detailed analysis for future work. The use of the restriction $n_h\geq
400$ appears, however, to make the correction predominantly dependent 
on $n_h$ alone. The net correction in halo mass due to finite force
and mass resolution, for our simulation suite, can then be written as
\begin{equation} 
M_c/M= [1.0- 0.04(\epsilon/650 ~ {\rm kpc})](1-n_h^{-0.65}). 
\end{equation}

Last, we consider systematic errors due to the finite volume of
simulations. There are three sorts of effects of this type. The first
is simply that the number of halos at high masses will be poorly
sampled, and the mass function in this region will have large
statistical error bars due to shot noise (Poisson fluctuations). This
is purely a question of having sufficient total simulation volume. The
second effect is the fact that missing large-scale modes, with
$k<2\pi/L$ ($L$ is the box size in linear dimension), lead to a
suppression of structure formation, and hence of the mass function, as
reduced power is available for transfer from linear to nonlinear
scales as evolution proceeds. Mass functions measured from simulations
must therefore account for the infrared cutoff in the variance of
matter fluctuations $\sigma(M)$. The extended Press-Schechter approach
has been found to work well in compensating for this effect (see
\citealt{lukic07} and references therein), and we follow it here. For
our simulations, this volume correction is relevant only for the small
box (GS) set of simulations, where $L=178$~Mpc, and affects only the
low mass halos. Applying the EPS method to the smallest box shifts the
masses by 2\%. 

The third finite volume effect is related to the (effective) number of
independent realizations, i.e., the sample variance. Because halos are
biased tracers of the density field, the mass function in a given,
sufficiently large, target volume is sensitive to the local mean
density as set by the scale-independent bias. In a large-volume
simulation, local sub-volumes will have fluctuating average densities
and these will add another component to the mass function variance,
aside from shot noise. There are two ways to address this: (i) because
of the high covariance between the small number of low $k$ modes in a
simulation and its high $k$ modes, run either very large boxes where
the relevant low $k$ modes (in terms of power transfer) are sampled
sufficiently well, or (ii) run a sufficient number of statistically
independent large-volume boxes.

In either case, one can successfully estimate the variance using a
simple halo model prescription and linear theory for density
fluctuations \citep{hukravtsov03}; details are given in
Appendix~\ref{fnum}. An important point to note is that all simulation
boxes necessarily have a definite infrared cutoff set by the box size.
This can in fact be tuned to control the mass function variance:
smaller boxes have smaller variance because they have less
low-frequency power. Of course, one cannot make the box too small
because then the mass function will be biased low as in the EPS
discussion above.

If one runs only one box, then resampling techniques such as the
jackknife may have to be used to estimate the associated errors (see,
e.g., \citealt{tinker08, crocce09}). These techniques can be
susceptible to misestimation of errors for smaller boxes, due to the
assumed independence of subvolumes of the simulation box. To estimate
our errors, we prefer to use independent realizations for a given
simulation volume. This has the advantage that the individual modes of
fluctuations are truly independent across realizations and hence
averaging over them represents the true variance for each set of runs.
We note that for large boxes $\sim$2~Gpc, jacknife sampling works well
\citep{tinker08}. However, since we have 10 independent 2~Gpc box runs
(A), we can use the different realizations to estimate the errors.
(Additionally, the increased mass resolution in each box makes one
less susceptible to the finite sampling problem in determining FOF
halo masses.) The sample variance errors were computed by taking the
ratio of the variance for each of the runs divided by the number of
realizations of each run; details are discussed in
Appendix~\ref{fnum}. As stated already, with this method one also has
the extra flexibility of tuning the low frequency cut-off to optimize
the mass variance (if the only simulation goal is to produce an
accurate mass function).

Our results from this section are summarized in
Figure~\ref{fig:mass_corr}. The upper panel shows the raw simulation
results and the lower panel includes our three corrections for force
resolution, finite sampling, and finite volume. In order to show the
effects more clearly, we display the ratio of the mass function to our
best fit to the data, as derived in the next section. Once the
simulation parameters for initial redshift, force resolution, and box
size are chosen in a suitable way for studying the mass function, the
major correction required is the one due to finite mass resolution.

\section{Reference $\Lambda$CDM Model}
\label{section:massfunction}

\subsection{Mass Function at the Present Epoch}
\label{mf_0}

After having analyzed possible systematic errors and corrected for
them in all our simulations, we now investigate the mass function for
the reference $\Lambda$CDM cosmology specified in
Table~\ref{tab:basic} (model 0) at $z=0$. The effective simulation
volume, combining all of our runs, is approximately $250$~Gpc$^3$. The
simulations provide coverage of halo masses ranging from that relevant
for individual bright galaxies all the way to clusters, with good
statistics. Although the total volume is dominated by the set of runs
with box size of $2778$~Mpc, the other runs are very helpful in
checking for and investigating systematic errors as seen above, and in
extending the simulation reach to lower halo masses.

\begin{figure}[b]
\includegraphics[width=9.0cm]{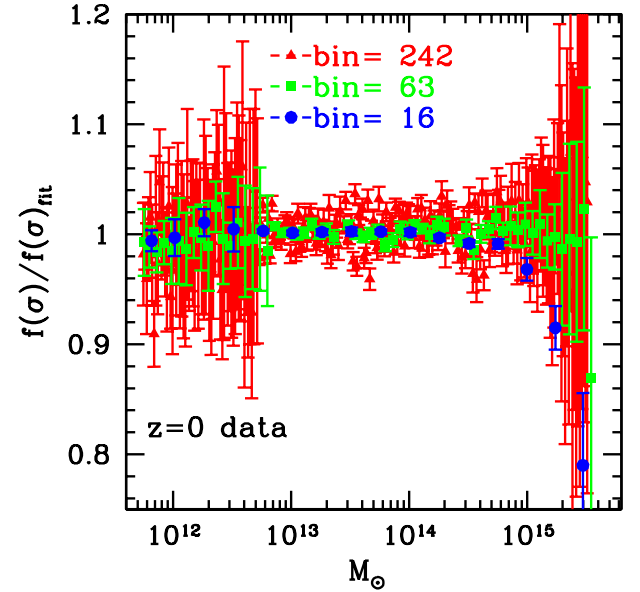}
\caption{Ratio of the mass function data at $z=0$ with respect to our
 $z=0$ fit for different choices of the number of bins. For halos of
 mass $\le 10^{14}$~M$_\odot$, four bins per decade in mass are
 sufficient for the mass function data to converge. For halos of mass
 $\ge 10^{14}$~M$_\odot$, the result converges for a choice of 15 bins
 per decade. }
\label{fig:bin}
\end{figure}

As previously discussed, a convenient form to express the scaled
differential mass function $f(\sigma,z)$ is \citep{J01}:
\begin{equation}
f(\sigma,z)=\frac{d\rho/\rho_b}{d\ln
\sigma^{-1}}=\frac{M}{\rho_b(z)}\frac{dn(M,z)}{d\ln
[\sigma^{-1}(M,z)]}.
\label{eq:diffmf}
\end{equation}
Here $n(M,z)$ is the number density of halos with mass $M$,
$\rho_b(z)$ is the background density at redshift $z$, and
$\sigma(M,z)$ is the variance of the linear matter power spectrum
$P(k)$ over a length $R$,
\begin{equation}
\sigma^2(M,z)\equiv
\frac{D^2(z)}{2\pi^2}\int_0^{\infty}{k^2P(k)W^2(kR(M))dk}, 
\label {eq:sigma}
\end{equation}
when smoothed on the scale $R(M)= (3M/4\pi\rho_b)^{1/3}$ with the
top-hat filter $W(x)=3[\sin(x)-x\cos(x)]/x^3$. We write $\sigma(M)
\equiv \sigma$ for brevity in the following. The redshift dependence
is encapsulated in the growth factor $D(z)$ which is normalized in
such a way that $D(0)=1$. As mentioned earlier, the advantage of this
definition of the mass function is that to a reasonable accuracy it
does not explicitly depend on redshift, power spectrum, or cosmology;
all of these are encapsulated in $\sigma(M,z)$.

A popular numerical fit for the differential mass function $f(\sigma)$
is given in \cite{st99} (ST hereafter).  The expression for the ST
mass function is 
\begin{equation}
f_{ST}(\sigma)= A\sqrt{\frac{2a}{\pi}} \exp \left [-
\frac{a\delta_c^2}{2\sigma^2}\right ]\left[ 1+ \left(
\frac{\sigma^2}{a\delta_c^2} \right)^p \right]\frac{\delta_c}{\sigma},
\label{eq:st}
\end{equation}
where $A,~a$, and $p$ are three parameters tuned to simulation results
with $a=0.707$ ($a=0.75$ is proposed as a better estimate in
\citealt{st02} ), $p=0.3$, and $A=0.3222$. The parameter $A$ is fixed
by the normalization condition that all dark matter particles reside
in halos, i.e.
\begin{equation}
\int_0^{\infty} {d \ln \sigma \, f (\sigma)} =1.
\label{eq:norm}
\end{equation}
We note that, as a practical matter, the lower halo mass cut-off in
numerical simulations is too large to test this particular assumption.
So, in principle, one could leave this constant as a free variable in
the fitting process. With the normalization condition fullfilled, the
ST mass function has two free parameters, $a$ and $p$. $\delta_c$ is
the density threshold for spherical collapse. In an Einstein-de Sitter
cosmology, $\delta_c=1.686$, independent of redshift. For $\Omega_m
\ne 1$, the value for $\delta_c$ shows insignificant dependence on
cosmology \citep{laceycole93}. We checked that including the cosmology
dependence in $\delta_c$ does not explain the redshift evolution of
$f(\sigma)$ seen in our simulations. In the following, we therefore
keep $\delta_c=1.686$ as a fixed value.

In order to obtain a fit for $f(\sigma,z)$, we need to compare
Equation~(\ref{eq:diffmf}) with the binned mass function obtained from
the simulations. We measure the number density of halos in a bin of
size $\Delta \ln M$ with mass limits $[M_1, M_2]$, as
\begin{equation}
n_{\rm bin}= \rho_b\int_{\Delta \ln M}\frac{1}{M} \frac{d\ln \sigma}{d\ln M}
f_{\rm data}(M, z)d\ln M, 
\label{mf_bin}
\end{equation} 
In the limit that $\Delta \ln M \rightarrow 0$, we can write
Equation~(\ref{mf_bin}) as 
\begin{equation}
n_{\rm bin}= \rho_b\frac{1}{M_{\rm bin}} \frac{d\ln \sigma(M_{\rm
bin},z)}{d\ln M_{\rm bin}} f_{\rm data}( M_{\rm bin}, z) \Delta \ln
M_{\rm bin},\nonumber
\end{equation}
and hence $f_{\rm data}( M_{\rm bin}, z)$ can be written as
\begin{equation}
f_{\rm data}( M_{\rm bin}, z)=\frac{n_{\rm bin}M_{\rm bin}}{\rho_b \Delta \ln M_{bin}}
\left(\frac{d\ln \sigma(M_{\rm bin},z)}{d\ln M_{\rm
      bin}}\right)^{-1}, 
\label{eq:fsigma_data}
\end{equation}
where $M_{\rm bin}= \sum M_i/N_{\rm bin}$, and the summation is over
all the halos in a bin. When combining all the boxes from the various
simulations, we vary $\Delta \ln M$ and make it small enough to ensure
that Equation~(\ref{eq:fsigma_data}) holds within the accuracy of the
data. The results are shown in Figure~\ref{fig:bin}. We find that four
bins per decade equally divided in $\ln$-space between $10^{11}- 10^{14}
$~M$_{\odot}$ and $15$ bins per decade between $10^{14}- 3\cdot
10^{15} $~M$_{\odot}$ are sufficient for the mass function to converge
within the measurement errors for $z=0$ and $z=1$. This corresponds to
a bin size of $\Delta \ln M= 0.25$ between $10^{11}- 10^{14}
$~M$_{\odot}$ and $\Delta \ln M = 0.06$ between $10^{14}- 10^{16}
$~M$_{\odot}$. For $z=2$, four bins per decade in mass ($\Delta \ln M=
0.25$) are sufficient for the measurements to converge within the
accuracy of the data.

\begin{figure}[t]
\includegraphics[width=9.0cm]{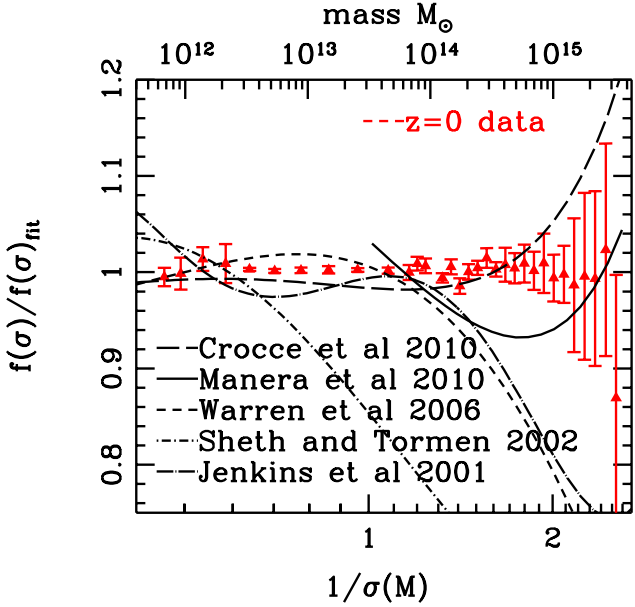}
\caption{Ratio of various mass function fits derived in previous
  studies with respect to the results of this paper at $z=0$. The
  binned numerical data are the points with error bars; the ratios are
  taken with respect to the analytic fit to the numerical data
  specified by Equation~(\ref{eq:st_mod}). Because the fits are based
  on runs with different cosmologies, exact correspondence cannot be
  expected (since the mass function is not universal).} 
\label{fig:ratio_otherfit}
\end{figure}

\begin{table*}[t]
\begin{center} 
\caption{\label{tab:fits}  Mass function fitting formulae derived in
  previous studies} 
\begin{tabular}{cccc}
\tableline\tableline
 Reference & Fitting function f($\sigma$) & Mass Range &  Redshift range \\
\hline\hline
\cite{st02} & $f_{ST}(\sigma)= 0.3222\sqrt{\frac{2(0.75)}{\pi}} \exp \left [-
\frac{0.75\delta_c^2}{2\sigma^2}\right ]\left[ 1+ \left(
\frac{\sigma^2}{0.75\delta_c^2} \right)^{0.3}
\right]\frac{\delta_c}{\sigma}$ & Unspecified&  Unspecified\\
\\
\cite{J01} & $0.315\exp\left[-| \ln \sigma^{-1} + 0.61 |^{3.8}\right]$
& $-1.2 \le \ln \sigma^{-1} \ge  1.05$ & z= 0-5\\
\\
\cite{warren05} & $ 0.7234\left( \sigma^{-1.625} + 0.2538\right)\exp
\left[-\frac{1.1982}{\sigma^2}\right]$ & $(10^{10}- 10^{15})$
h$^{-1}$M$_\odot$ & z=0\\ 
\\
\cite{reed06} & $0.3222\sqrt{\frac{2(0.707)}{\pi}}\left[1+
  \left(\frac{\sigma^2}{0.707\delta_c^2}  \right)^{0.3}
  +0.6G_1(\sigma) + 0.4G_2(\sigma)\right]$& $-0.5  \le \ln \sigma^{-1}
\ge 1.2$ &z=0-30\\ 
&$\times \frac{\delta_c}{\sigma}\exp\left[
  -\frac{0.764\delta_c^2}{2\sigma^2}- \frac{0.03}{(n_{eff}+3)^2\left(\delta_c/\sigma\right)^{0.6}}\right]$&\\ 
\\
\cite{manera09} & $f_{ST}(\sigma)= 0.3222\sqrt{\frac{2a}{\pi}} \exp \left [-
\frac{a\delta_c^2}{2\sigma^2}\right ]\left[ 1+ \left(
\frac{\sigma^2}{a\delta_c^2} \right)^{p}
\right]\frac{\delta_c}{\sigma}$ &$(3.3\times 10^{13}-3.3\times
10^{15})$ h$^{-1}$M$_\odot$ & z=0-0.5\\ 
\\
\cite{crocce09} & $A(z)\left[ \sigma^{-a(z)} +
  b(z)\right]\exp\left[-\frac{c(z)}{\sigma^2}\right]$ &
$(10^{10}-10^{15})$ h$^{-1}$M$_\odot$ &z=0-1\\ 
\\
\hline
\hline 
\tableline\tableline

\vspace{-1.5cm}

\tablecomments{Various fits from previous studies shown in
  Figure~\ref{fig:ratio_otherfit} and  \ref{fig:ratio_mice}  for
  friends-of-friends halos of linking length  $b=0.2$ are listed. For
  \cite{manera09}, the parameter  values are (a, p)= (0.709, 0.248) at
  z=0 and (0.724, 0.241) at z=0.5.  For \cite{crocce09}, the parameter
  values are $A(z)=  0.58(1+z)^{-0.13}, a(z)= 1.37(1+z)^{-0.15}, b(z)=
  0.3(1+z)^{-0.084}, c(z)=  1.036(1+z)^{-0.024}$. For \cite{reed06},
  $G_1(\sigma)= \exp\left[-\frac{(\ln
      \sigma^{-1}-0.4)^2}{2(0.6)^2}\right]$ and  $G_2(\sigma)=
  \exp\left[-\frac{(\ln  \sigma^{-1}-0.75)^2}{2(0.2)^2}\right]$} 
\label{tab:fits2}
\end{tabular}
\end{center}
\end{table*}

Figure~\ref{fig:ratio_otherfit} compares different mass function
expressions given previously as compared to our new simulation results
at $z=0$. The expressions for different fits are given in
Table~\ref{tab:fits}. Some of these expressions, derived from earlier
simulations, are significantly discrepant -- especially at high masses
-- at the 20\% level and higher. Results from more recent simulations
are in much better agreement, partly because the cosmologies run are
much closer. We note that exact correspondence cannot be expected
because the mass function is not universal; we defer a discussion of
this point to Section~\ref{section:mf_wcdm}.

In order to obtain a fitting form to our results, we begin with the ST
fit as the starting point, although, as shown in
Figure~\ref{fig:ratio_otherfit}, the ST mass function deviates from
the simulation results by as much as 40\% at the high mass end. As a
first step to improve the accuracy of the ST mass function, we drop
the normalization requirement and refit all three parameters $A$, $a$,
and $p$ to the numerical data. With this approach, the simulation data
can be fit to an accuracy of 10-15\%, significantly worse than the
statistical errors of our data set (see also \citealt{manera09}). The
remaining inadequacy of the ST fit can be addressed in different ways.
For example, \cite{warren05} introduced a fourth parameter into the ST
functional form and refitted the other three parameters to their
simulation data, finding a best fit mass function: 
\begin{equation}
  f_{W}(\sigma)= A_{W}\left( \frac{1}{\sigma^b } + c  \right)\exp\left[-
    \frac{d}{\sigma^2}\right], 
  \label{eq:warren}
\end{equation}
with $A_W= 0.7234,~b= 1.625,~c=0.2538$, and $d=1.1982$. As shown in
Figure~\ref{fig:ratio_otherfit}, this particular fit also severely
underestimates the mass function at high masses, by up to $\sim30\%$.
While adequate as a fitting form over a finite range,
Equation~(\ref{eq:warren}) diverges when the normalization condition
is imposed [Equation~(\ref{eq:norm})]. To avoid this, we present a new
fitting function for $f(\sigma)$. This is the simplest ST modification
that does not diverge but adds one extra parameter, ${\tilde q_0}$
(for ${\tilde q_0} = 1$ we recover the ST mass function): 
\begin{equation}
  f^{\rm mod}(\sigma,z=0)= \tilde A_0\sqrt{\frac{2}{\pi}} \exp
  \left [- \frac{\tilde a_0\delta_c^2}{2\sigma^2}\right ]\left[ 1+
    \left( \frac{\sigma^2}{\tilde a_0\delta_c^2} \right)^{\tilde
      p_0}\right]\left(\frac{\delta_c \sqrt {\tilde a}}{\sigma}\right)^{\tilde q_0}.
  \label{eq:st_mod}
\end{equation}
We use a $\chi^2$ technique to determine the best fit $f(\sigma)$ that
matches the mass function data obtained by combining all of the
$\Lambda$CDM runs. That is, we minimize
\begin{equation}
  \chi^2= \sum_{i=1}^N \frac{f(\sigma)^{\rm mod}- f(\sigma)_{\rm
    data}}{(\Delta f(\sigma)_{\rm data})^2}, 
\label{eq:chisq}
\end{equation}
where $f(\sigma)^{\rm mod}$, $f(\sigma)_{\rm data}$ and $\Delta
f(\sigma)_{\rm data}$ are given by Equations~(\ref{eq:st_mod}),
(\ref{eq:fsigma_data}), and (\ref{eq:err}) respectively.

Minimizing $\chi^2$ gives the best fit parameter values: $\tilde A_0=
0.333$, $ \tilde a_0= 0.788$, $\tilde p_0= 0.807$, and $\tilde
q_0=1.795$ with a $\chi^2$ per degree of freedom of 1.15. The
subscript ``0'' indicates that the best fit values are specified at
$z=0$. The results are summarized in Table~\ref{tab:eqn}. As mentioned
above, this expression does not diverge when the normalization
condition is imposed, however, the best fit does not lead to a
normalization of unity. As shown in Figure~\ref{fig:ratio}, this
modified expression agrees with the simulation data to better than 2\%
accuracy at $z=0$. As further discussed in Section~\ref{mf_z} a simple
redshift dependence has to be introduced into the fitting function to
obtain agreement at the same accuracy level at higher redshifts.

\subsection{Redshift Evolution and Universality}
\label{mf_z}

\begin{table}[t]
\begin{center} 
\caption{\label{tab:eqn}  Mass function fitting formula for the
  reference $\Lambda$CDM model (mass range:
  $6\times10^{11}$~M$_\odot$-- $3\times 10^{15}$~M$_\odot$; redshift
  range: $z=$0--2)}  
\begin{tabular}{c}
\tableline\tableline
$f^{\rm mod}(\sigma,z)= \tilde A\sqrt{\frac{2}{\pi}} \exp
\left [- \frac{\tilde a\delta_c^2}{2\sigma^2}\right ]\left[ 1+ \left(
\frac{\sigma^2}{\tilde a\delta_c^2} \right)^{\tilde
p}\right]\left(\frac{\delta_c \sqrt {\tilde a}}{\sigma}\right)^{\tilde q}$\\ 

\tableline\\
Redshift Evolution \\
\tableline\\
$\tilde A= \frac{0.333}{(1+z)^{0.11}},~
\tilde a= \frac{0.788}{(1+z)^{0.01}},~
\tilde p= \frac{0.807}{(1+z)^{0.0}},~
\tilde q= \frac{1.795}{(1+z)^{0.0}} $\\
\tableline\tableline
\end{tabular}
\end{center}
\end{table}

\begin{figure}[t]
\includegraphics[width=9.0cm]{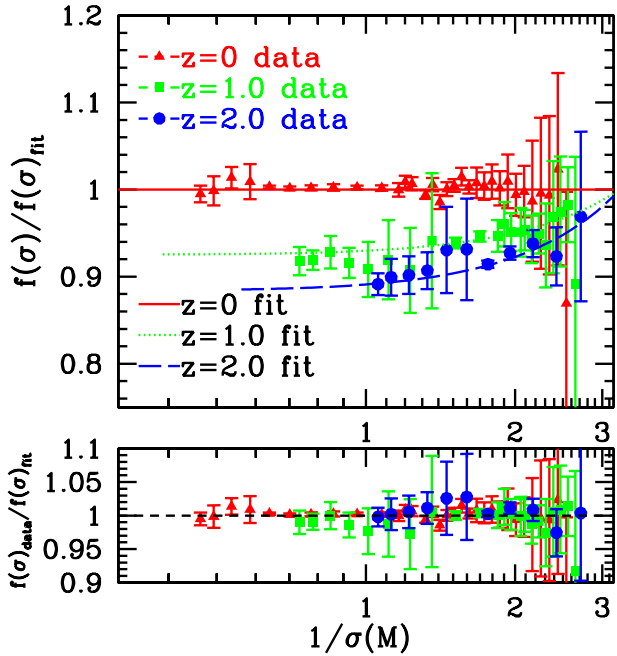}
\caption{Ratio of the mass function data to the $z=0$ fit of
  Equation~(\ref{eq:st_mod}) (reference flat red line). The $z=1$ and
  $z=2$ datasets demonstrate that redshift evolution is important and
  must be taken into account; the curves show the corresponding fits
  following the time-dependence as parameterized in
  Equations~(\ref{eq:z}). The lower panel shows the ratio of the
  measured mass function at the three different redshifts to the
  corresponding analytic fits.}
\label{fig:ratio}
\end{figure}

The $z=0$ mass function fit of Section~\ref{mf_0} has a default
universal form. However, as remarked previously, it is known that the
mass function deviates from universality -- as a function of redshift
-- for $\Lambda$CDM cosmologies. Recent results include those of
\cite{tinker08} who found the SO halo mass function to evolve by
$20-30$\% from redshift z=0 to 2.5. As shown in
Figure~\ref{fig:ratio}, this deviation can be as much as 10\% between
redshifts $z=0-2$ for FOF halos, in agreement with the results of
\cite{crocce09}. In this section we extend our fitting function to
include the redshift evolution of the mass function. We parameterize
the possible redshift evolution of each parameter via a simple
power-law form
\begin{eqnarray}
\tilde A= \tilde A_{0}/(1+z)^{\alpha_1},\nonumber\\
\tilde a= \tilde a_{0}/(1+z)^{\alpha_2} ,\nonumber \\
\tilde p= \tilde p_{0}/(1+z)^{\alpha_3} , \nonumber \\
\tilde q= \tilde q_{0}/(1+z)^{\alpha_4}. 
\label{eq:z}
\end{eqnarray}
In order to ensure that the expression for the redshift evolution
reproduces the mass function at any intermediate redshift when
interpolated or even extrapolated, we fit two redshift outputs at a
time. Thus we have three values for each parameter. The final set of
parameters is the average of the three values obtained using redshift
outputs in pairs. Figure~\ref{fig:ratio} shows that the power law
model of Equations~(\ref{eq:z}) is able to capture the redshift
evolution with an accuracy of better than 3\% within the range of $0.6
\le 1/\sigma \le 2.4$. We find that only two of the four parameters of
Equations~(\ref{eq:z}) show any redshift evolution. The best fit
values for the parameters $\alpha_i$ describing the redshift evolution
are $\alpha_1= 0.11$, $\alpha_2=0.01$, $\alpha_3= 0.0$, and $\alpha_4=
0.0$. We note that the non-universal redshift evolution is suppressed
at higher redshifts as the effect of the cosmological constant is
reduced and matter-domination takes over. To recap, our analytic
best-fit to the mass function data uses one extra shape parameter
compared to ST to match the $z=0$ data, and then introduces a simple
$z$-dependence (two more parameters) to capture non-universal
behavior.

We also explore the option of allowing for a redshift dependence,
$\delta_c(z)$, as determined by the spherical collapse model. For the
$\Lambda$CDM cosmology adopted here, the spherical collapse
calculation yields $\delta_c= 1.674$, 1.684, and 1.686 respectively
for $z=0$, 1, and 2. As expected, at $z=2$, $\delta_c$ approaches the
value expected for the $\Omega_m=1$ cosmology. Allowing for this
redshift dependence does not mitigate the redshift dependence of the
other fitting parameters; including $\delta_c(z)$ changes the value of
$\tilde a$ to $0.799/(1+z)^{0.024}$ (in fact increasing the redshift
dependence) while the other parameters remain unchanged. Therefore,
adding $\delta_c(z)$ does not help explain the $z$-dependence seen in
our simulations, and we do not include it in our fit.

\begin{figure}[t]
\includegraphics[width=9.0cm]{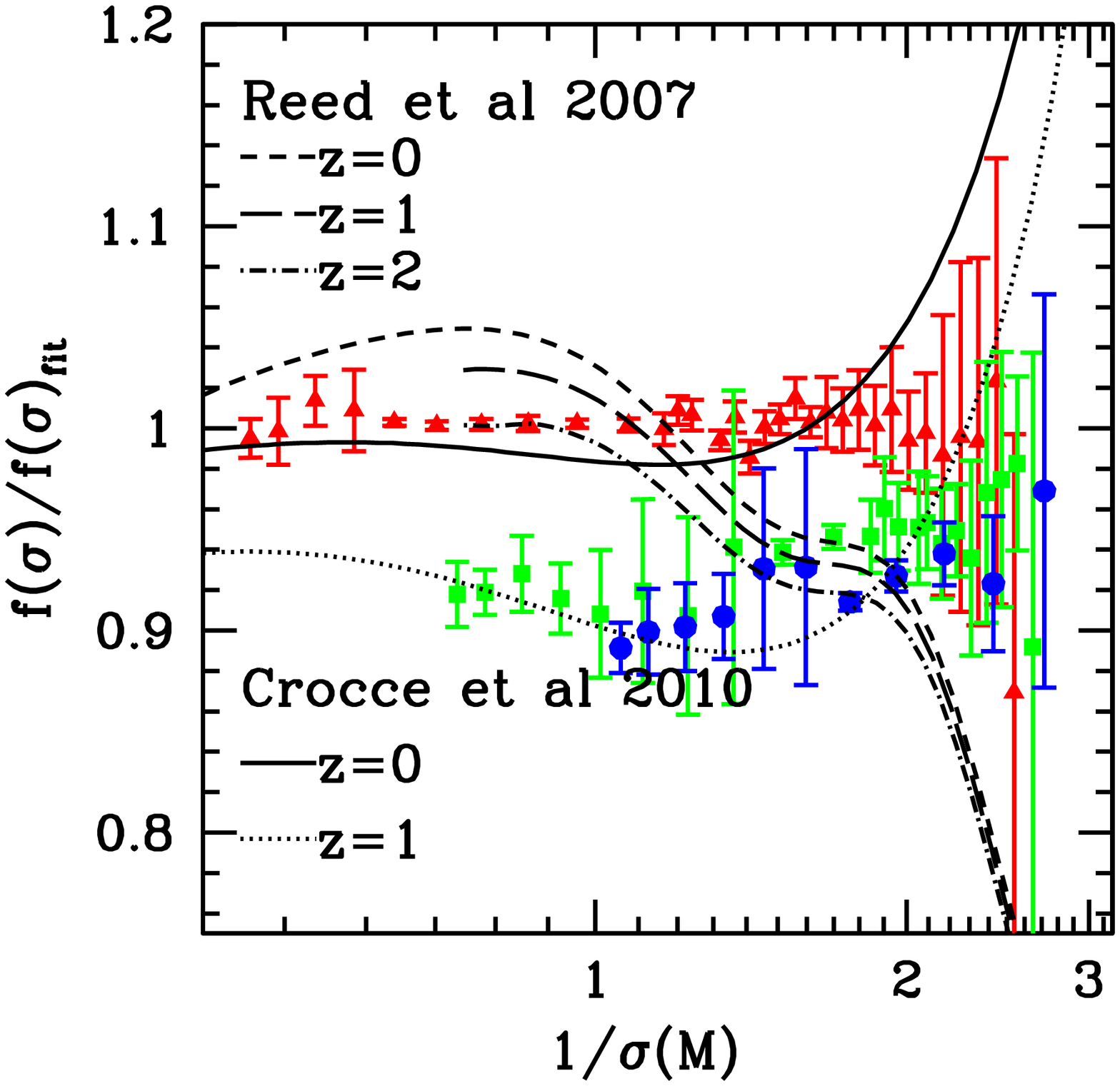}
\caption{Redshift dependent mass function fits as introduced by
  \cite{reed06} and \cite{crocce09} compared with the numerical data
  of this work. Aside from disagreement in the overall shape, the
  results of \cite{reed06} underestimate the amount of evolution
  indicating that high redshift evolution of the mass function is
  slower compared to that at lower redshift. The agreement with
  \cite{crocce09} is better (at the 4-5\% level), except for the
  runaway at high masses (see discussion in Section~\ref{mf_z}). }
\label{fig:ratio_mice}
\end{figure}

\begin{figure}[t]
\includegraphics[width=9.0cm]{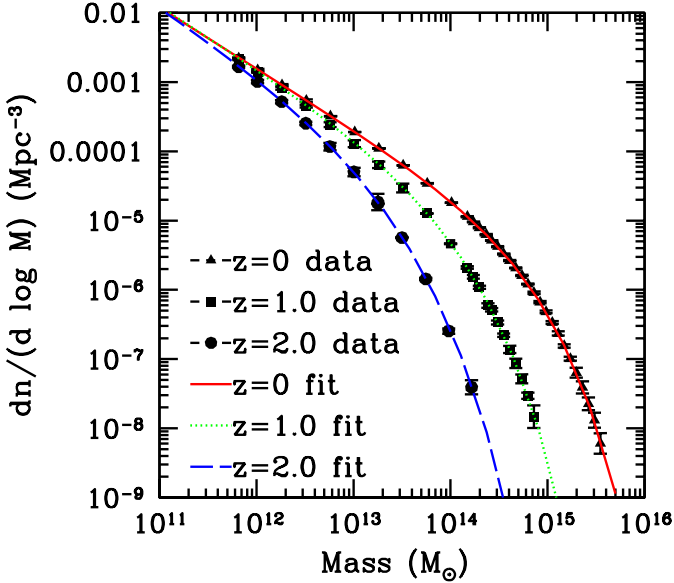}
\caption{Halo mass function as measured in our simulations at three
  different redshifts, $z=0$, 1, and 2 along with the analytic fit at
  each redshift.} 
\label{fig:mf}
\end{figure}

\begin{figure*}[t]
\includegraphics[width=5.8cm]{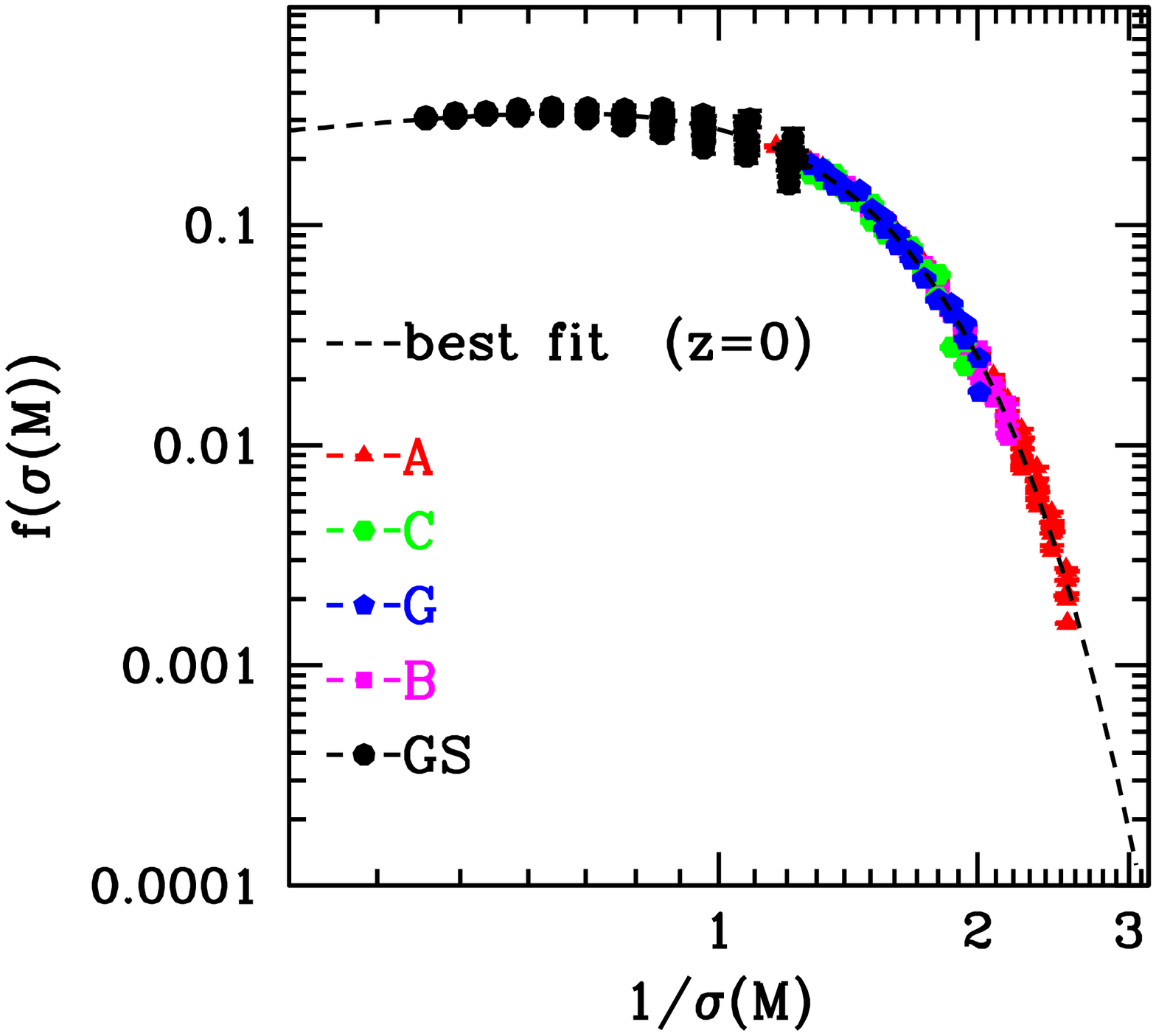}
\hspace{-1.7cm}
\includegraphics[width=5.8cm]{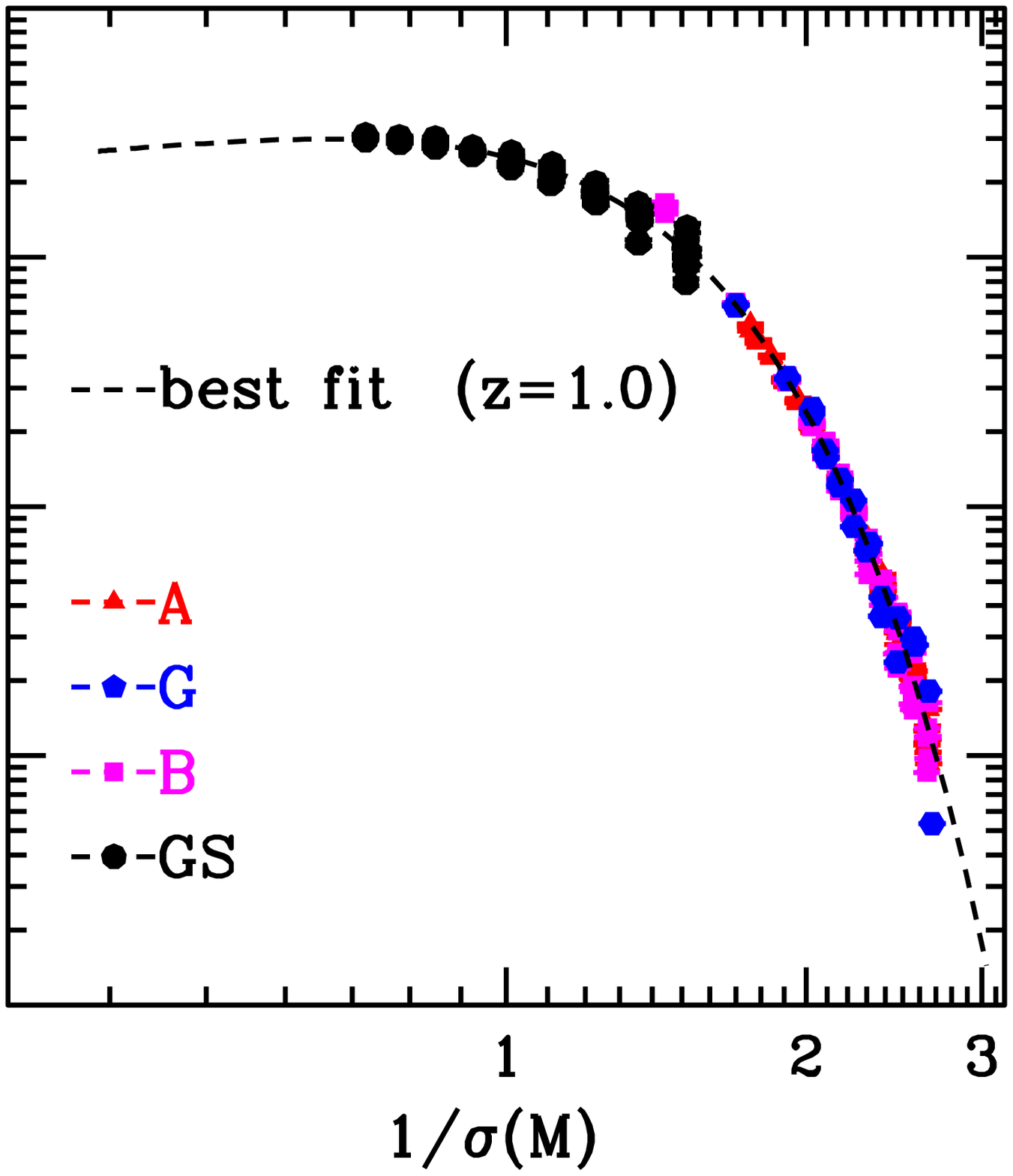}
\hspace{-1.7cm}
\includegraphics[width=5.8cm]{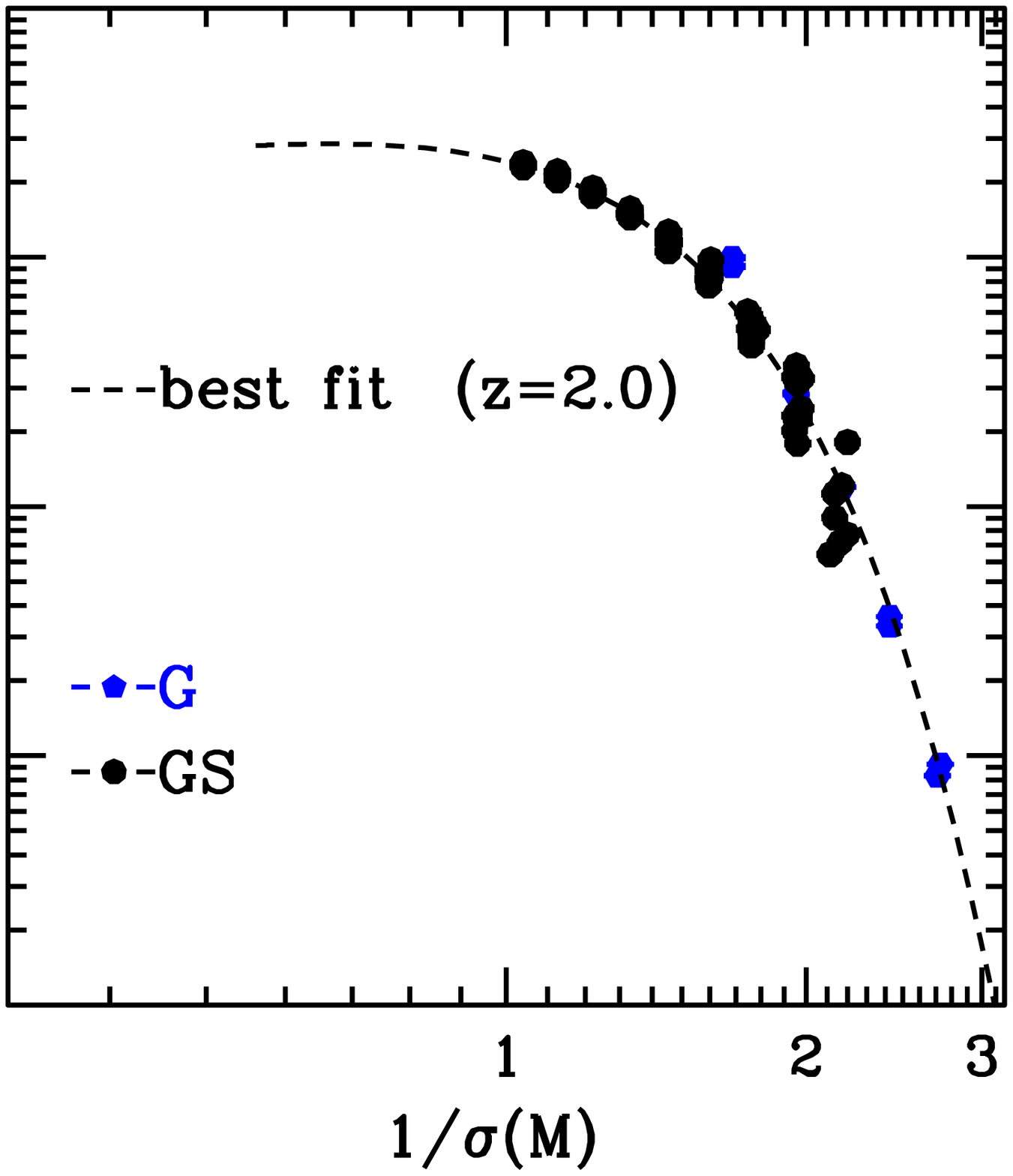}
\hspace{-1.7cm}
\includegraphics[width=5.8cm]{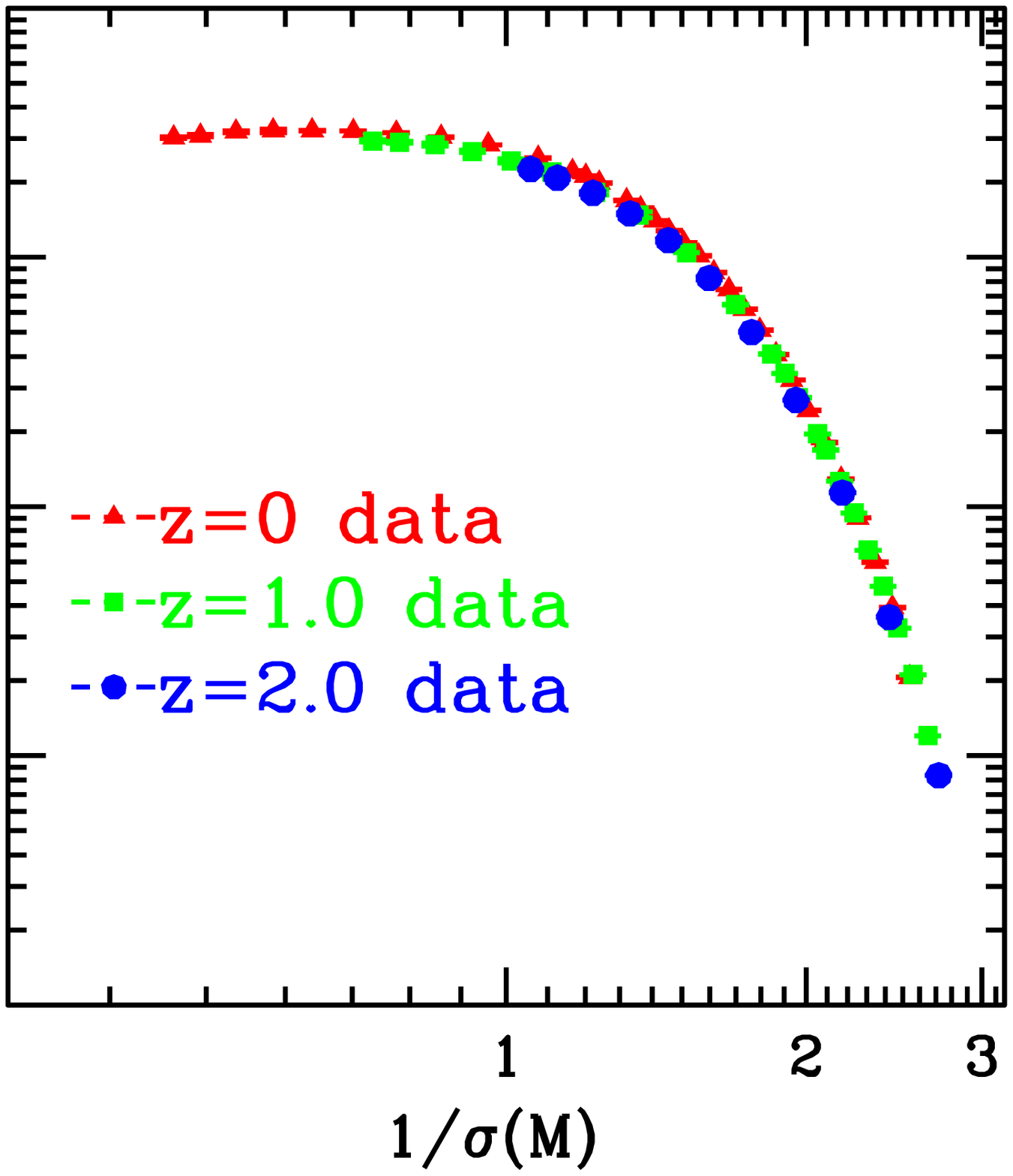}
\caption{Mass functions for the $\Lambda$CDM simulations shown 
  at redshifts $z=0$, 1, and 2, for different simulation boxes. The
  line is the mass function fit. The far right panel shows the mass
  function obtained by combining all the boxes. Note that the results
  for the different redshifts do not line up perfectly and therefore a
  redshift independent fit cannot be found at very high accuracy.}
\label{fig:fsigma}
\end{figure*}

High-statistics studies of the evolution of the FOF mass function have
been carried out previously. In an investigation focusing mainly at
high redshifts, to explain the violation of universality,
\cite{reed06}  proposed an effective spectral slope $n_{\rm eff}$ set
by the halo radius, parameterized as   
\begin{equation} 
n_{\rm eff}=6\frac{d\ln\sigma^{-1}}{d\ln M}-3.
\end{equation}
This new effective slope induces a redshift dependence in the mass
function. However, as shown in Figure~\ref{fig:ratio_mice}, the
analytic fit of \cite{reed06} is not in good agreement with our
results (see also \cite{bagla09} which studies the dependence of the
mass function universality on the initial matter power spectrum). This
discrepancy indicates that high redshift evolution of the mass
function is slower compared to that at lower redshifts.
\cite{crocce09} also use a simple power-law form to fit for redshift
evolution. Our reference model and the parameters of \cite{crocce09}
are very close, consequently, their results are significantly closer
to ours, except at very high masses, where their fit appears to
deviate. Here, discrepancies may result from the use of an approximate
transfer function \footnote{M.~Crocce, private communication} and a
small systematic offset in their fitting procedure at high masses (Cf.
their Figure 8). Taking this offset into account we find that the
actual numerical results are in agreement at the few percent level.
The expressions for the fitting functions of \cite{reed06} and
\cite{crocce09} are given in Table~\ref{tab:fits}. Figure~\ref{fig:mf}
shows the abundance $dn/d\ln M$ as measured in our simulation along
with the analytic fits. Figure~\ref{fig:fsigma} summarizes the results
from this section, showing the mass function at different redshifts
and our best fit results.

\subsection{Mass function-derived large-scale halo bias}
\label{section:bias}

The evolution of the spatial distribution of halos has been studied in
detail in \cite{cole89} and subsequently in \cite{mowhite96} and
\cite{st99}. These studies assume that dark matter halos are biased
tracers of the underlying dark matter distribution. The halo bias in
general is stochastic and a nonlinear function of the underlying
density field \citep{white05, sw04}. In addition, it also depends on
the assembly history of the halos \citep{dalal08}. As discussed in
\cite{st99}, within the halo model, the large scale deterministic bias
can be derived knowing the shape and evolution of the mass function.
It is known that the peak--background split model prediction for the
bias using the Sheth-Tormen or \cite{warren05}~mass function is in
disagreement with direct numerical calculations of this quantity
obtained from ratios of correlation functions or power spectra in
simulations (see, e.g., \citealt{lukicphd08, pw09}, and
\citealt{tinker10}). Given a more accurate mass function, it is easy
to check if the halo model approach now produces a better answer for
the halo bias. To test this -- following \cite{st99} and \cite{cole89}
-- we now obtain the expression for the large scale bias using our
expression for the mass function.

\begin{figure}[b]
\includegraphics[width=9.0cm]{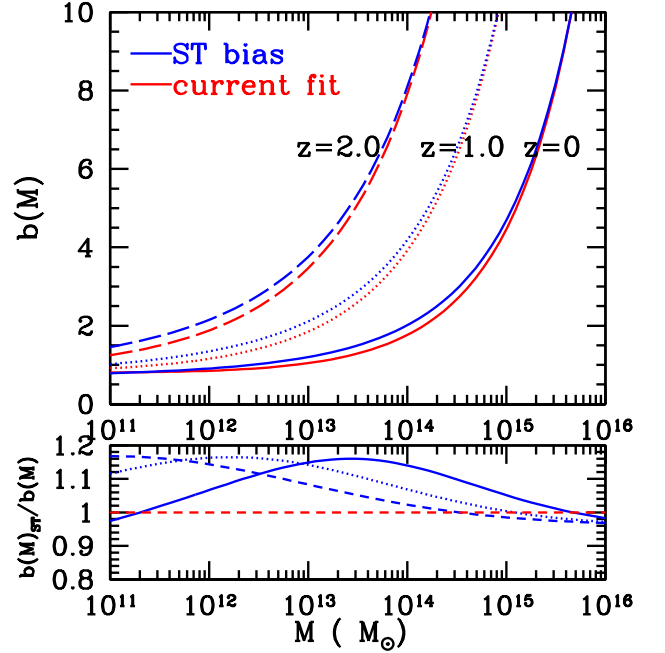}
\caption{Large scale halo bias from the peak-background
  split formalism, derived from the mass function. The upper panel shows
  the large scale bias at redshifts $z=0$, 1 and 2. The lower panel
  shows the ratio between the bias derived here and the Sheth-Tormen
  bias for $z=0$, 1 and 2. Note that we assume that the redshift of
  observation is the same as the redshift of formation of the halos
  ($z_{\rm form}= z_{\rm obs}$).} 
\label{fig:bias}
\end{figure}

\begin{figure*}[t]
\includegraphics[width=7.5cm]{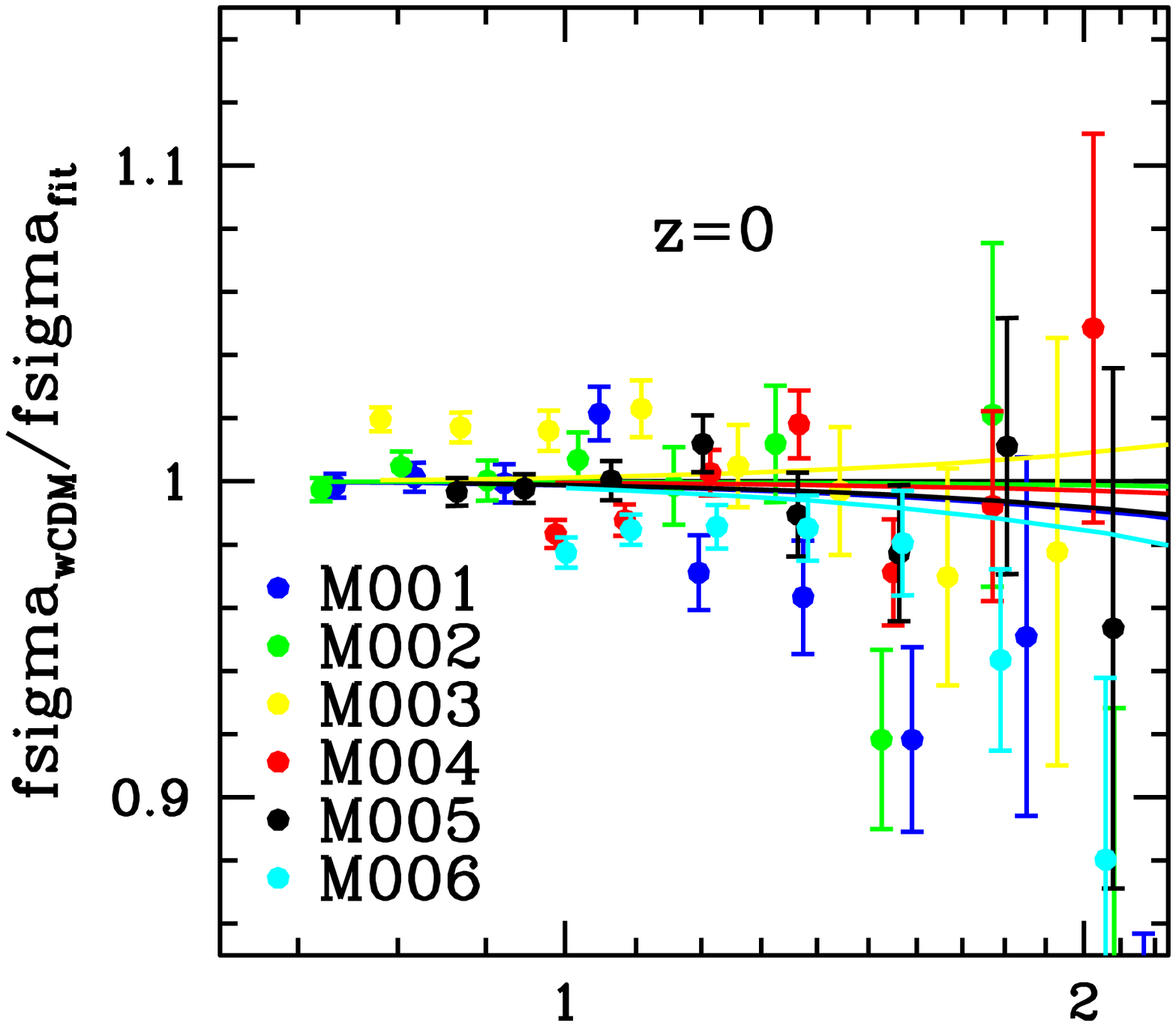}
\hspace{-2.0cm}
\includegraphics[width=7.5cm]{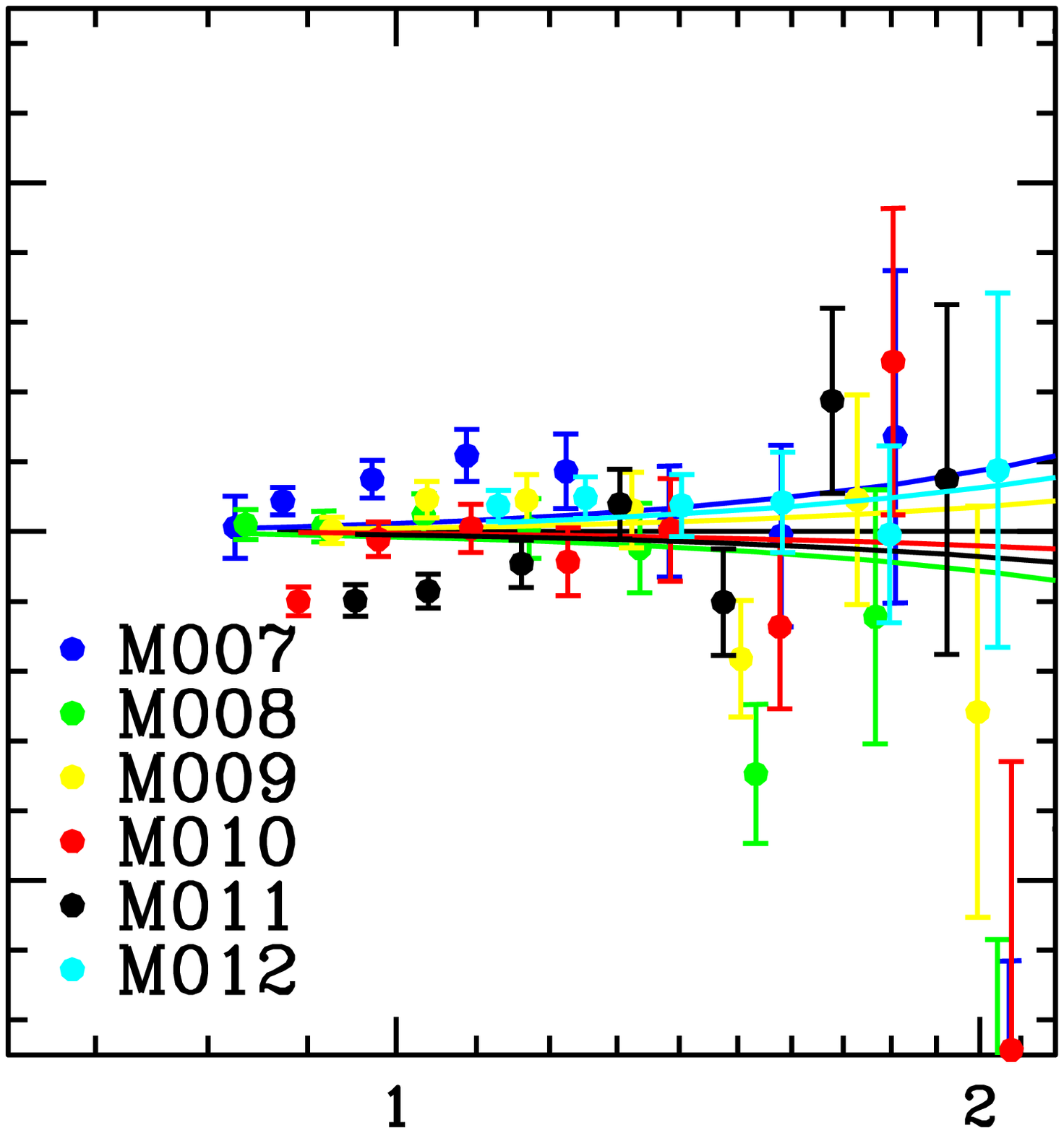}
\hspace{-2.0cm}
\includegraphics[width=7.5cm]{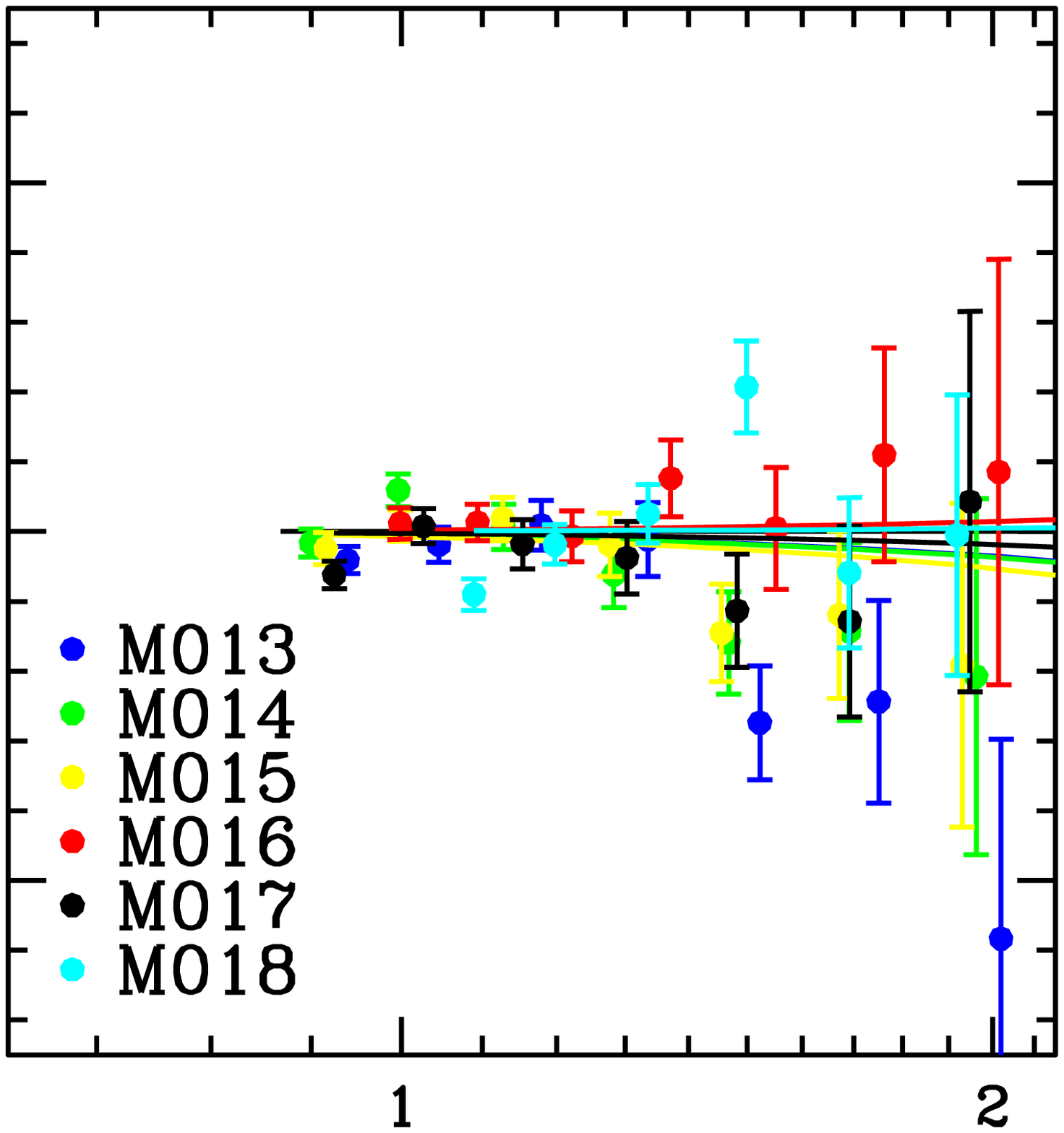}\\
\vspace{-1.85cm}

\includegraphics[width=7.5cm]{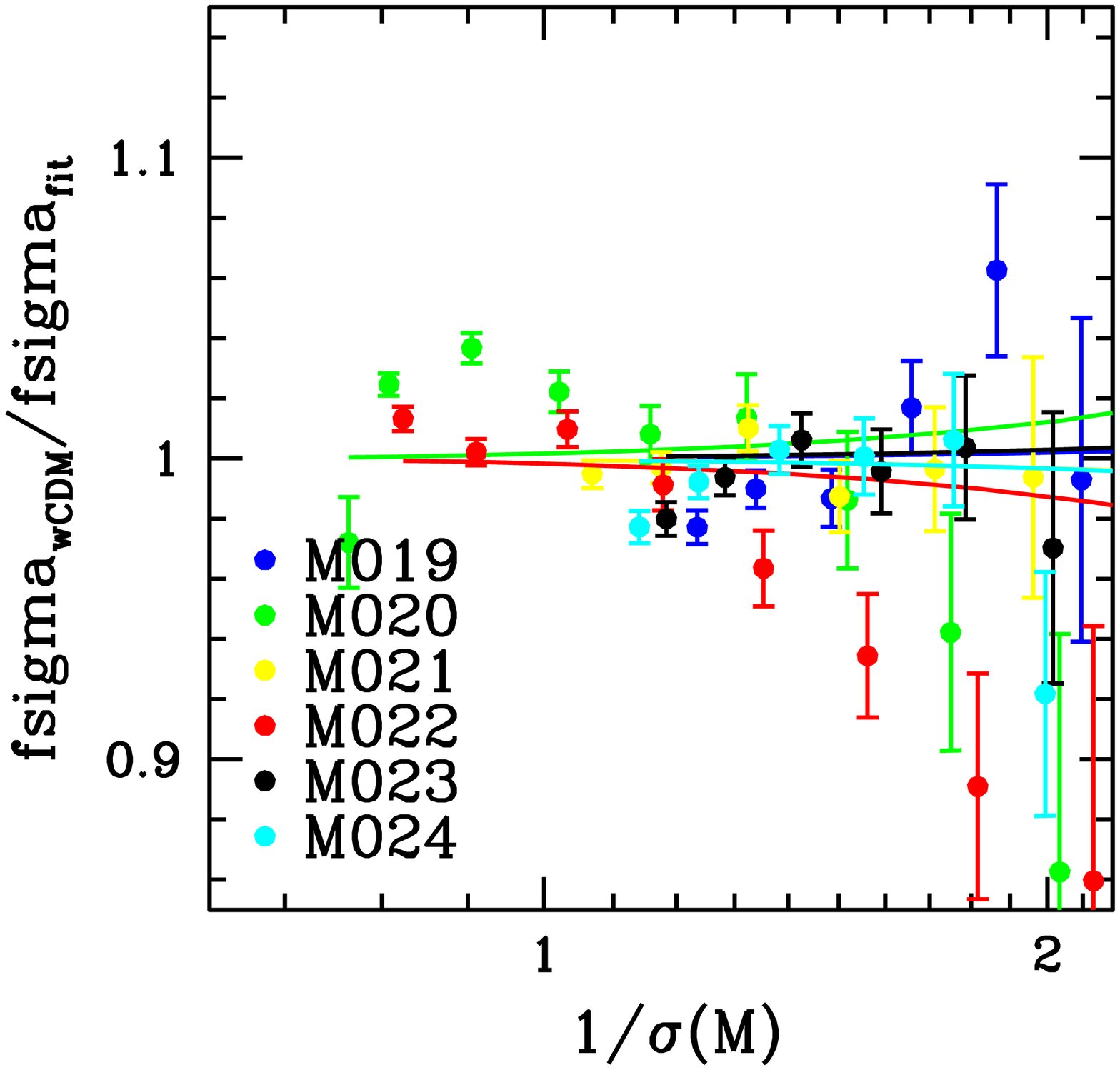}
\hspace{-2.0cm}
\includegraphics[width=7.5cm]{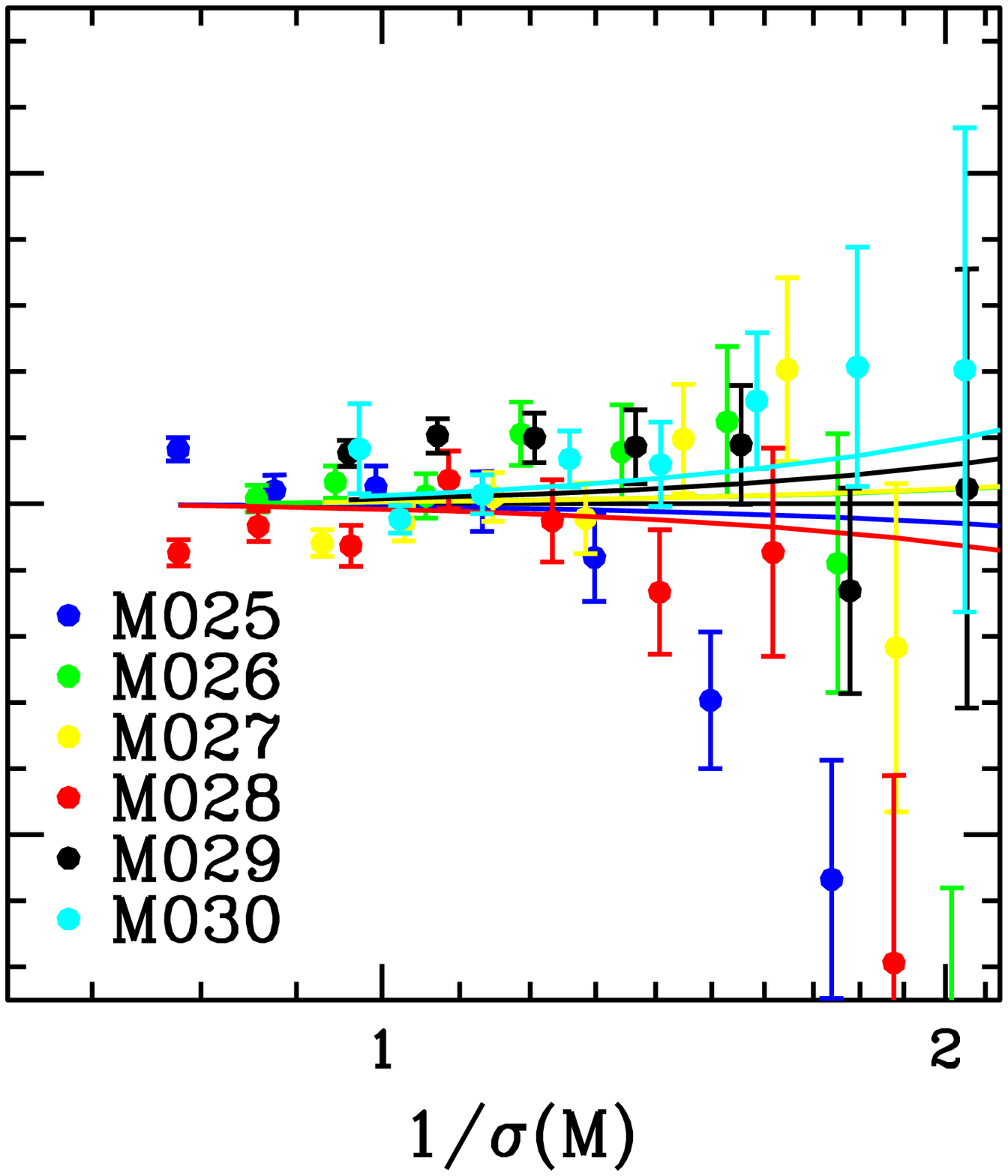}
\hspace{-2.0cm}
\includegraphics[width=7.5cm]{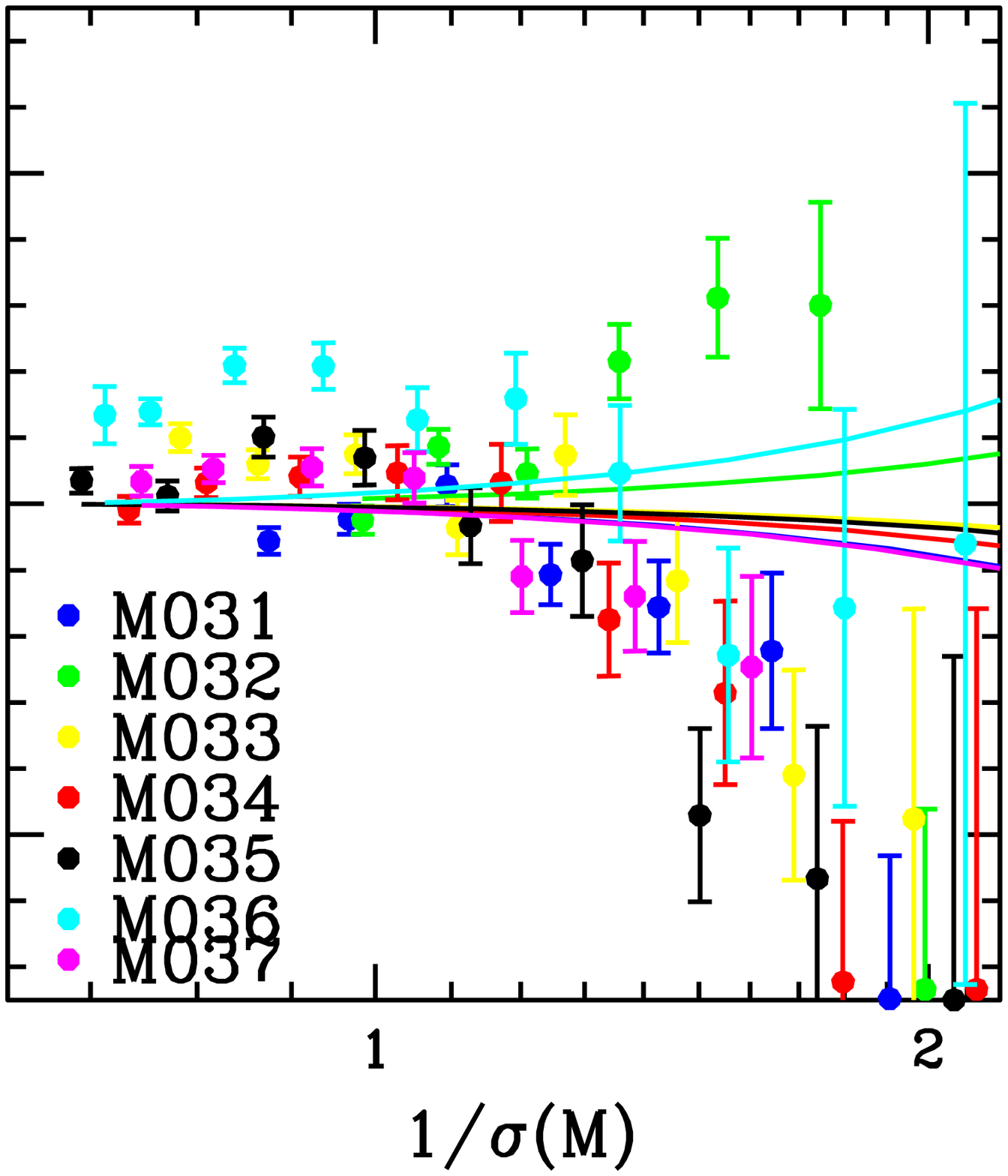}
\caption{Test of universality of the mass function for $w$CDM
  cosmologies (the G runs) at z=0. The range of the cosmological
  parameters covered by the simulations is given in
  Table~\ref{tab:basic}. The ratio is taken with respect to the best
  fit mass function for the $\Lambda$CDM case at z=0. The lines show
  the ratio of the fit if $\delta_c$  is cosmology dependent in the
  $\Lambda$CDM fit.}   
\label{fig:wcdmz}
\end{figure*}

\begin{figure*}[t]
\includegraphics[width=7.5cm]{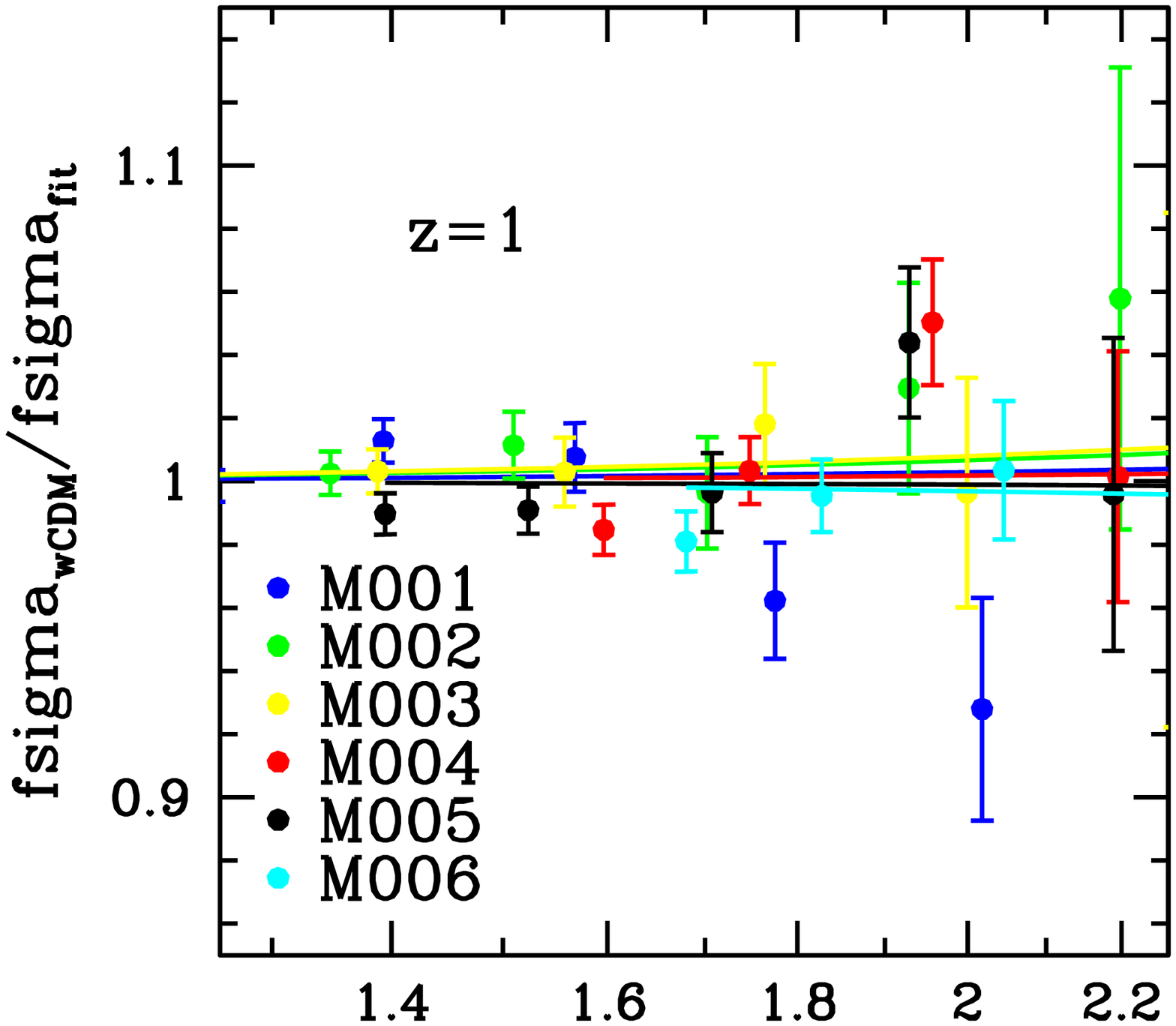}
\hspace{-2.0cm}
\includegraphics[width=7.5cm]{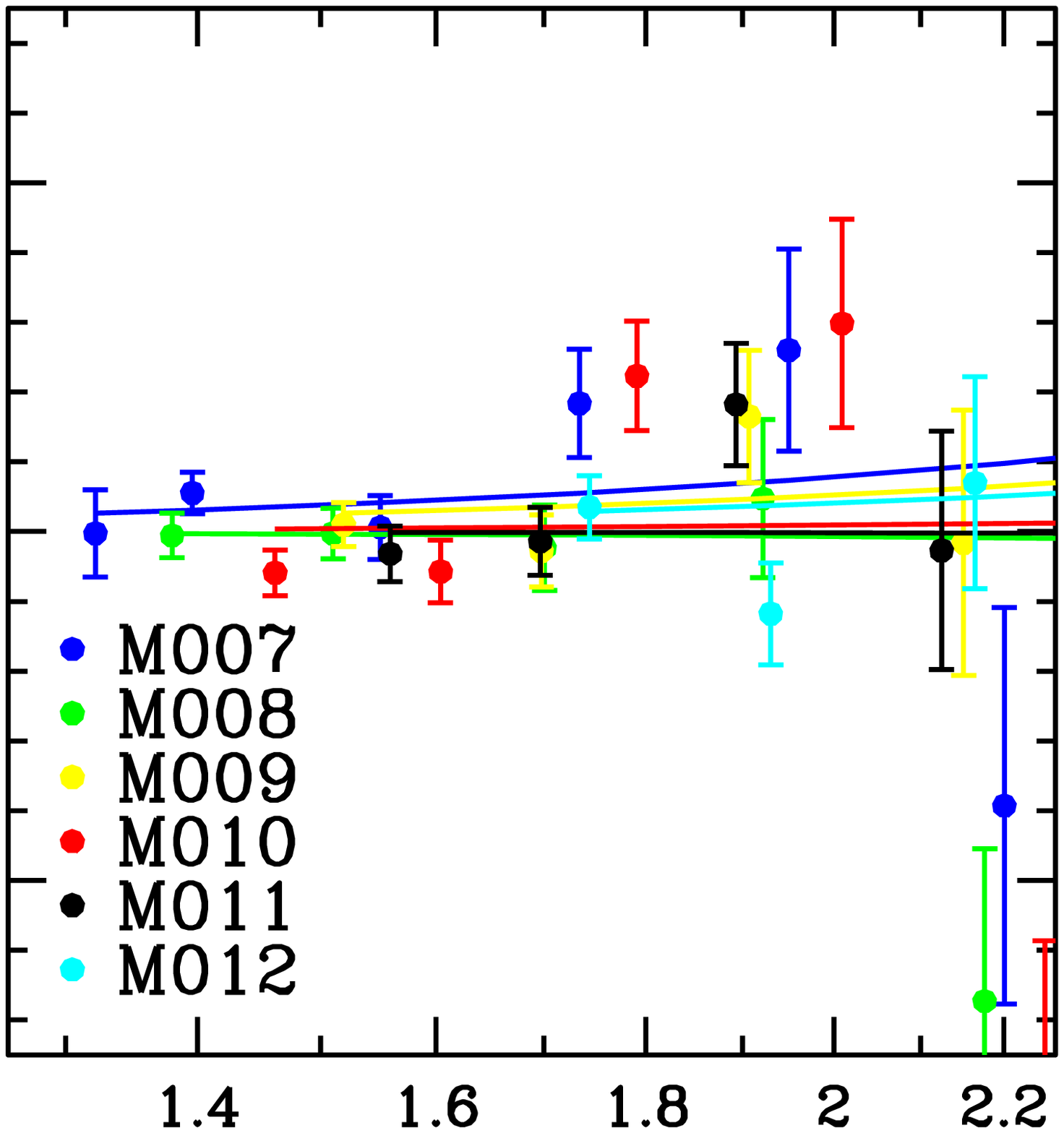}
\hspace{-2.0cm}
\includegraphics[width=7.5cm]{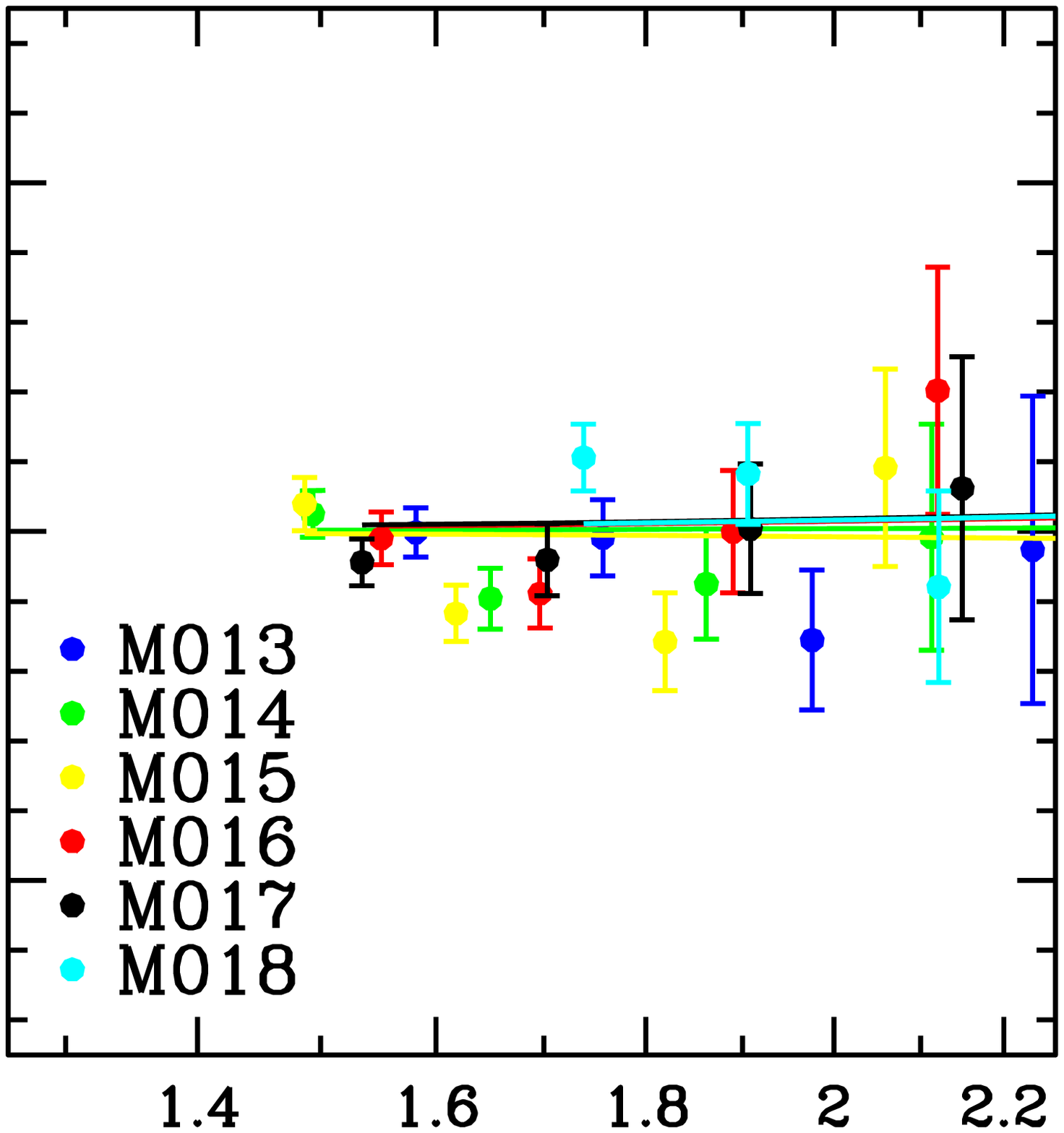}
\vspace{-1.85cm}

\includegraphics[width=7.5cm]{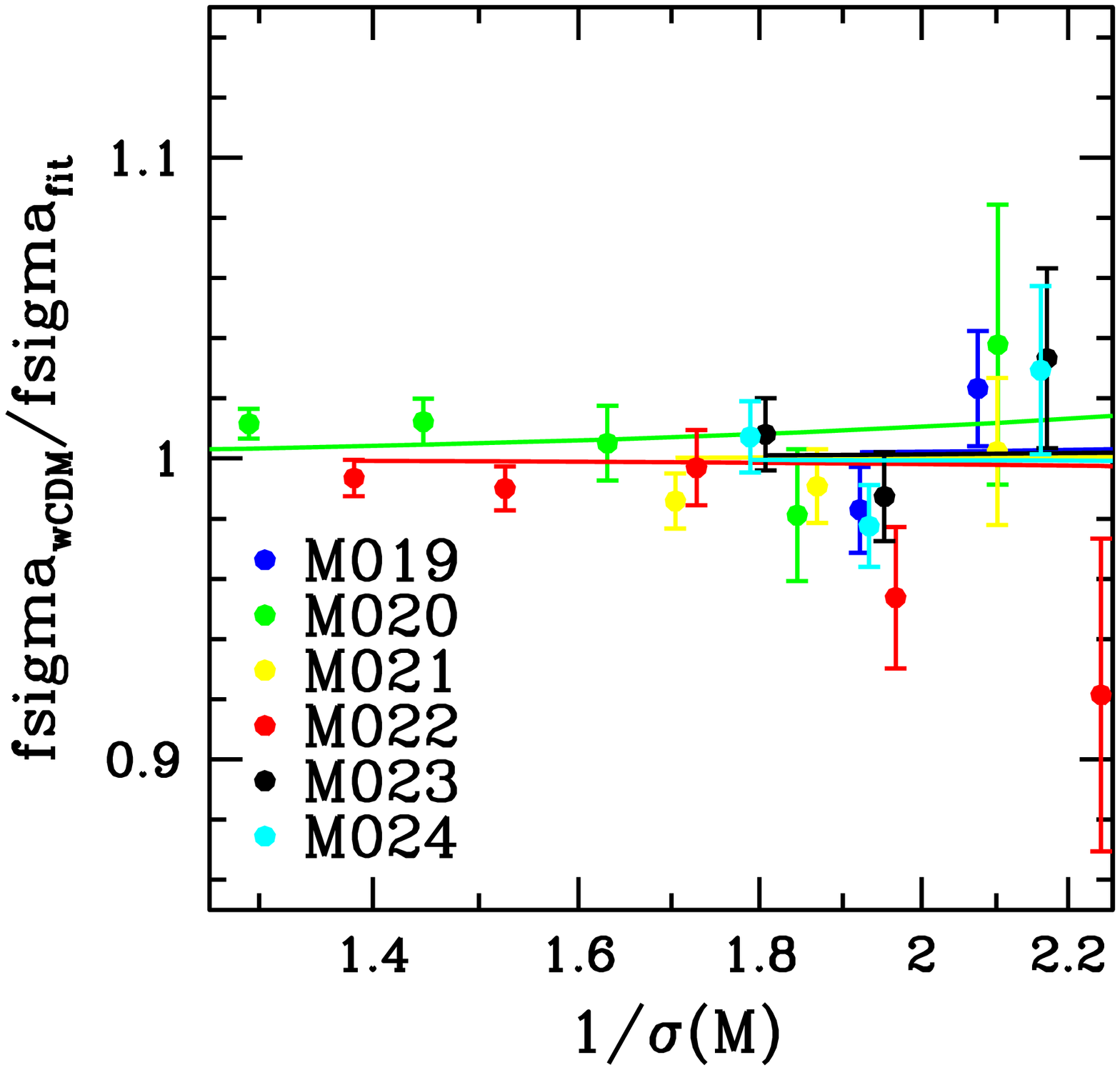}
\hspace{-2.0cm}
\includegraphics[width=7.5cm]{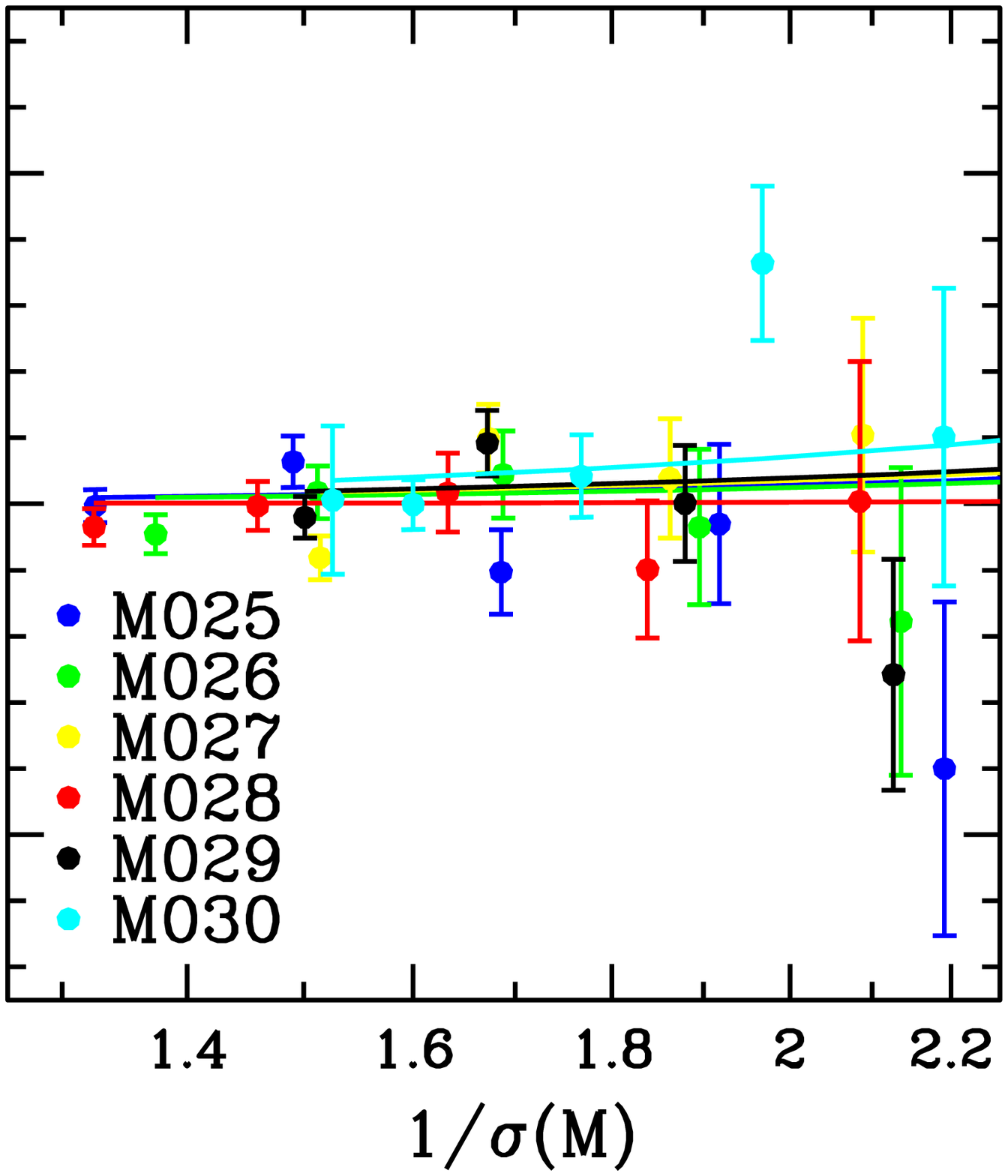}
\hspace{-2.0cm}
\includegraphics[width=7.5cm]{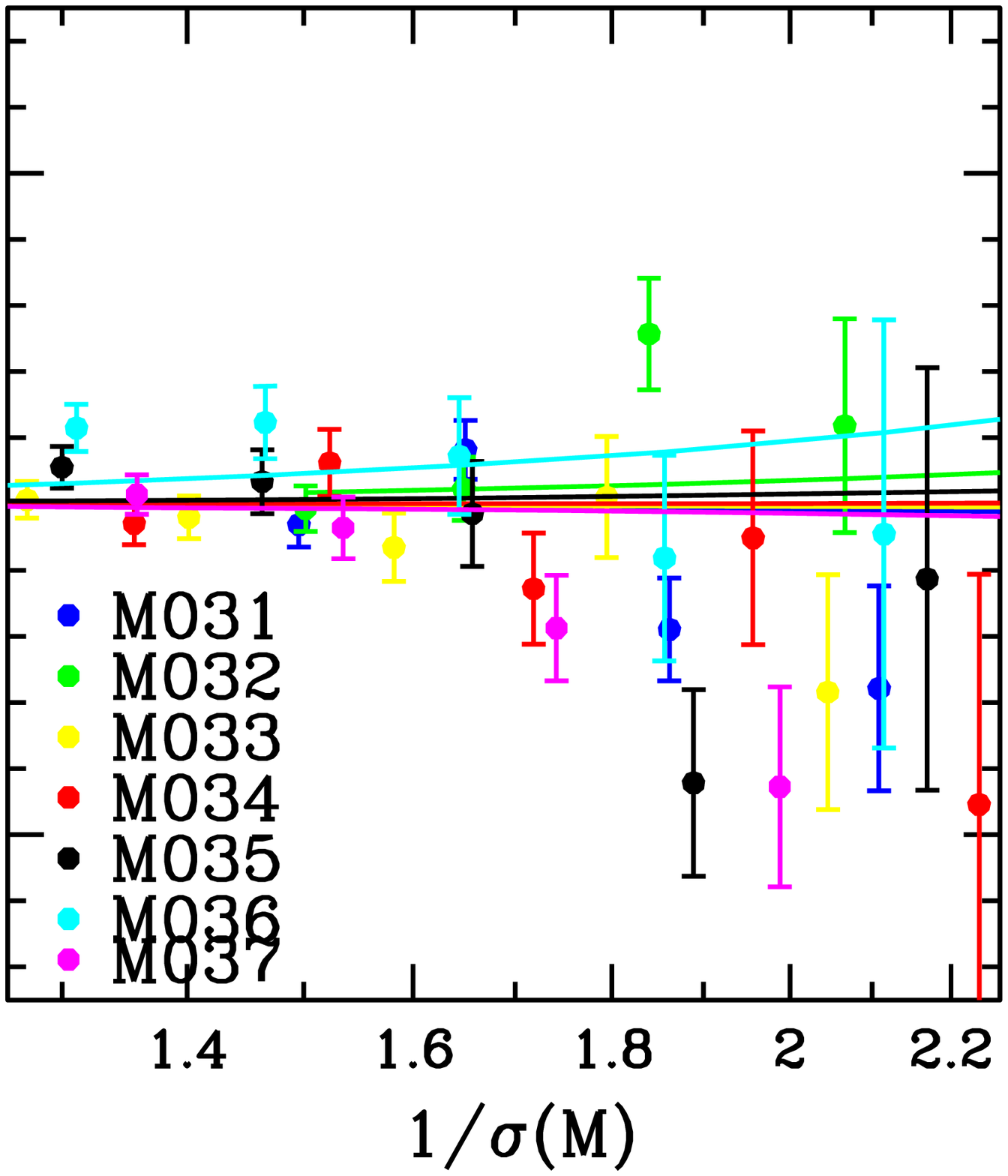}
\caption{Ratio of the mass function for  $w$CDM
  cosmologies to the best fit mass function for the
  $\Lambda$CDM case at z=1. The lines show the ratio of the fit if
  $\delta_c$ is cosmology dependent in the $\Lambda$CDM fit.}  
\label{fig:wcdmz1}
\end{figure*}

\begin{figure*}[t]
\includegraphics[width=7.5cm]{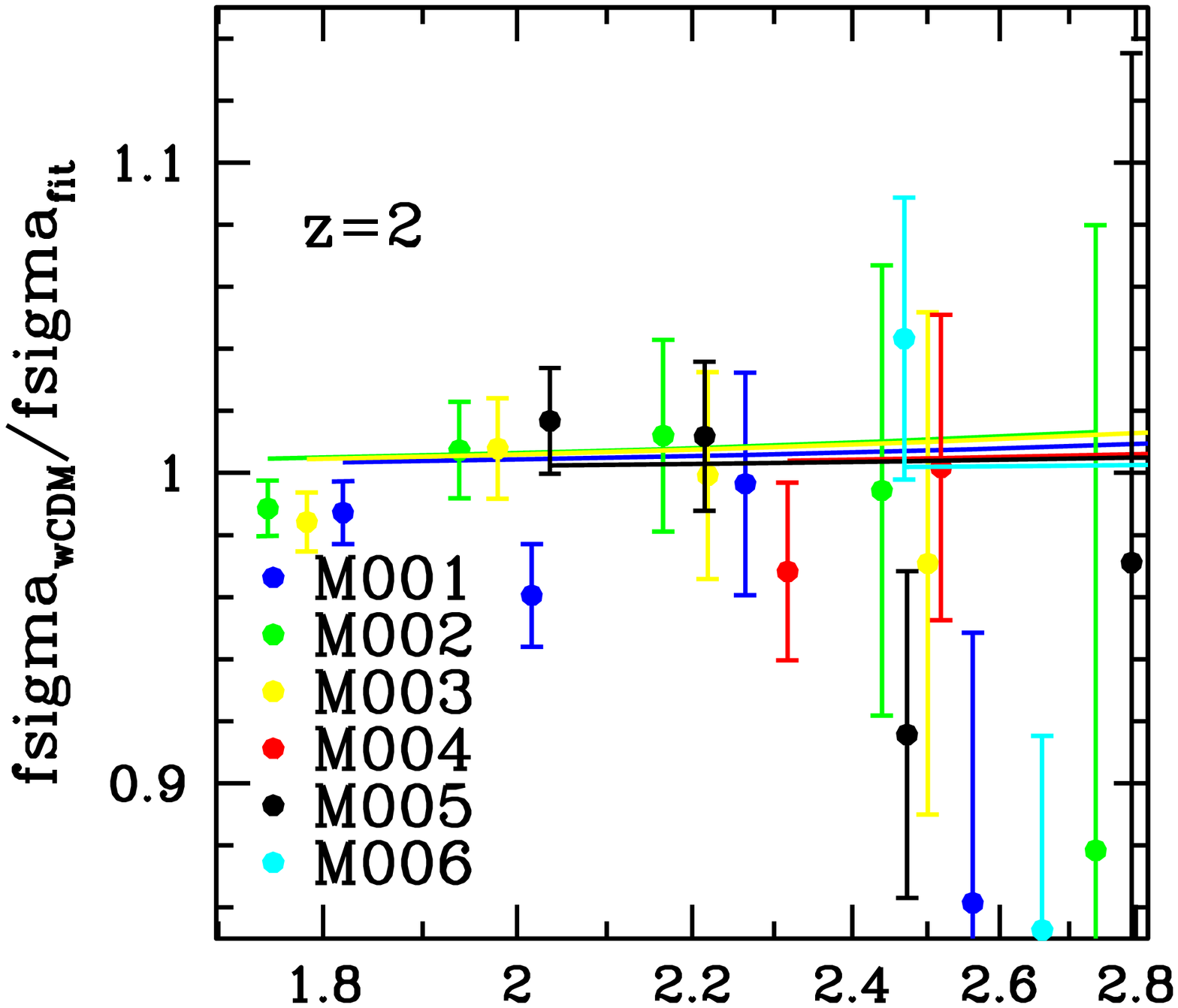}
\hspace{-2.0cm}
\includegraphics[width=7.5cm]{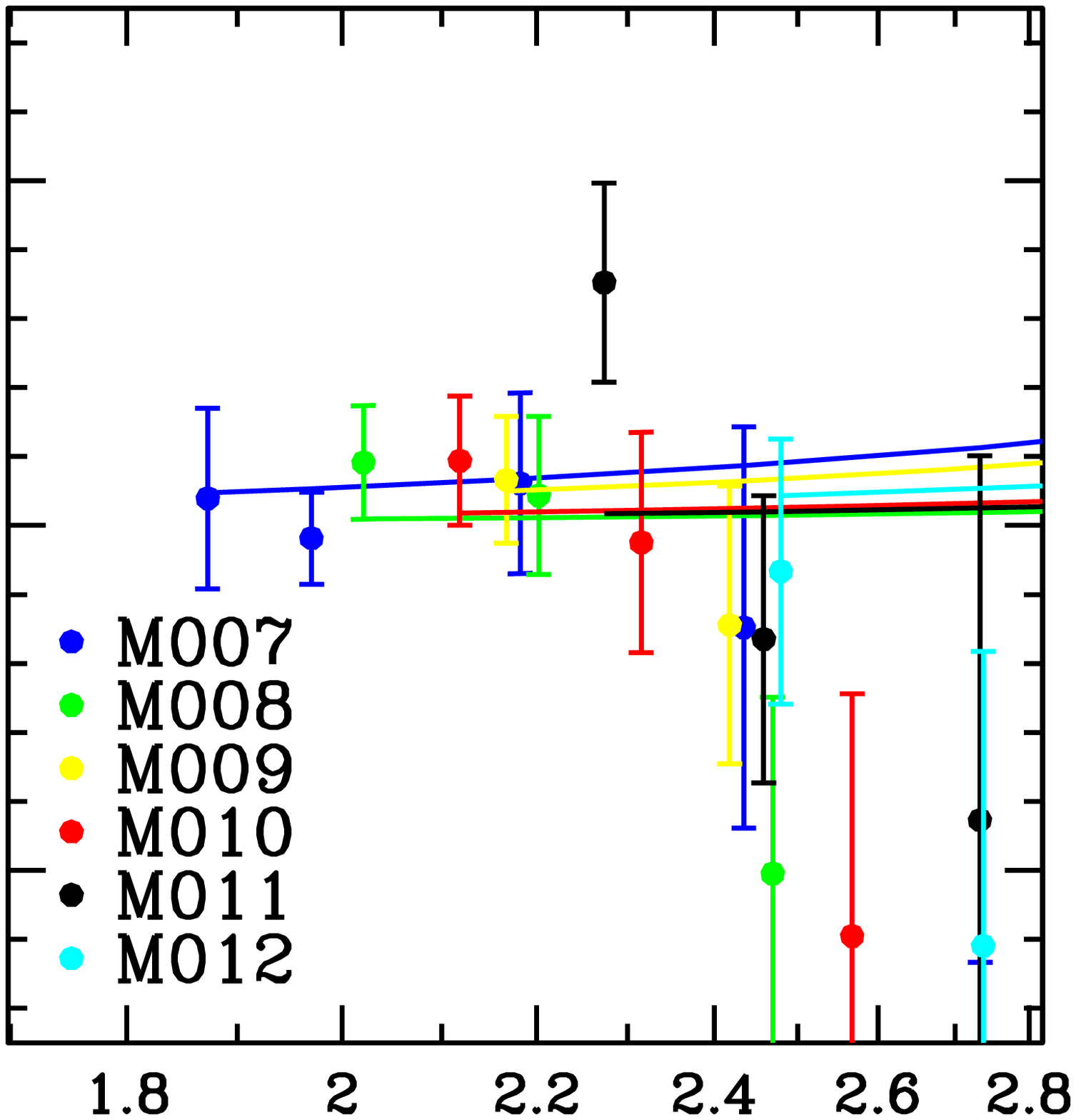}
\hspace{-2.0cm}
\includegraphics[width=7.5cm]{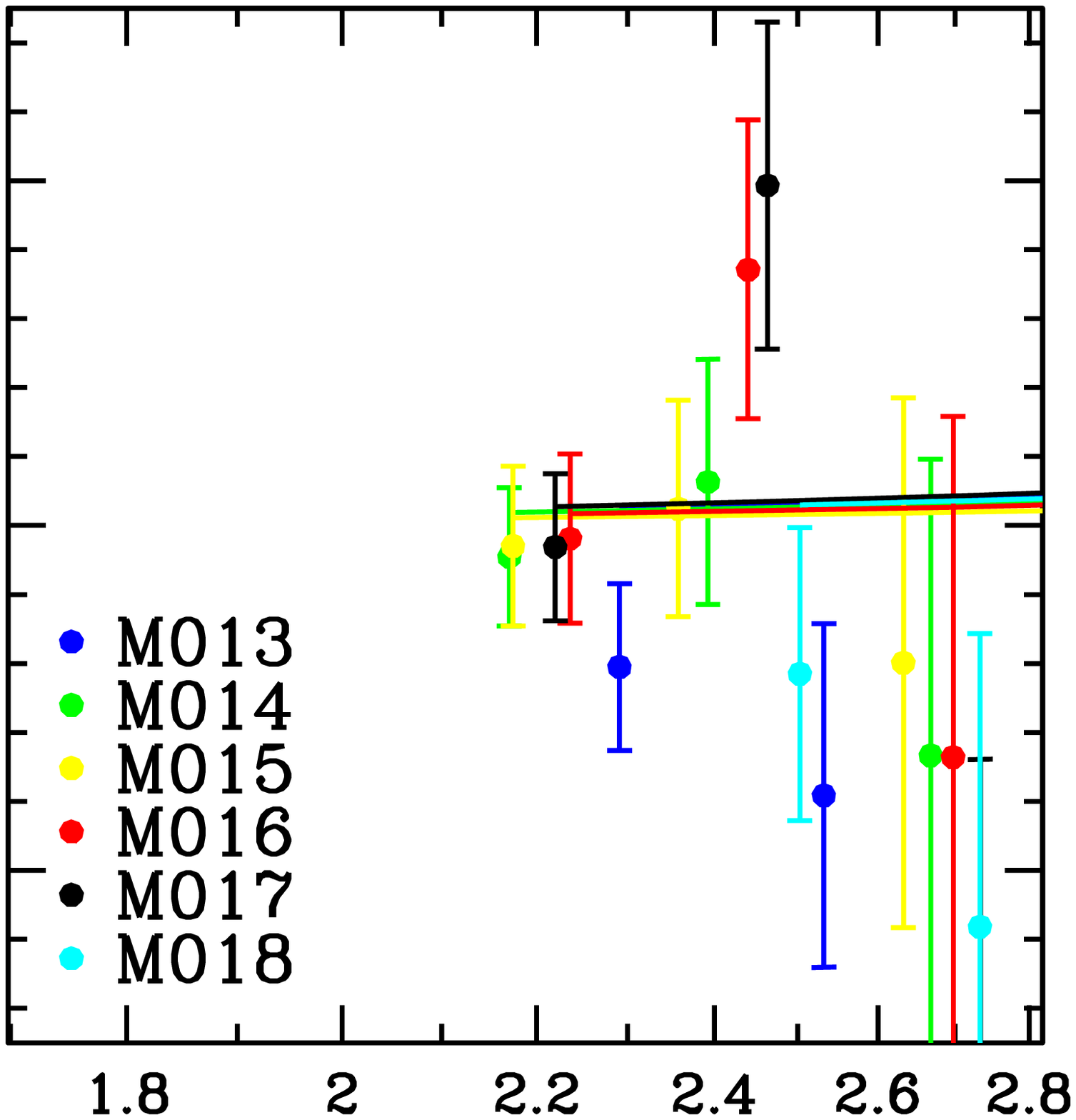}
\vspace{-1.85cm}

\includegraphics[width=7.5cm]{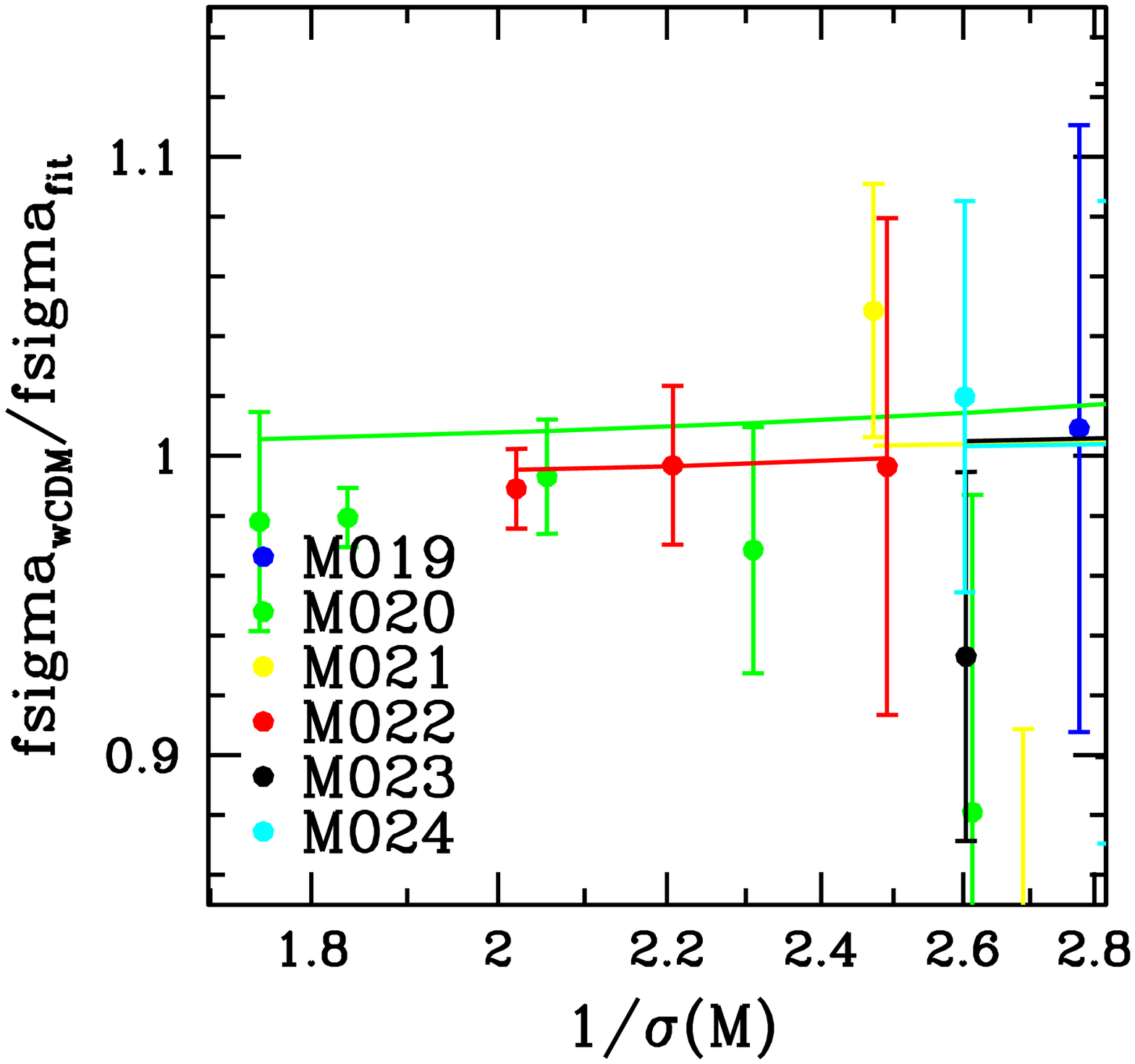}
\hspace{-2.0cm}
\includegraphics[width=7.5cm]{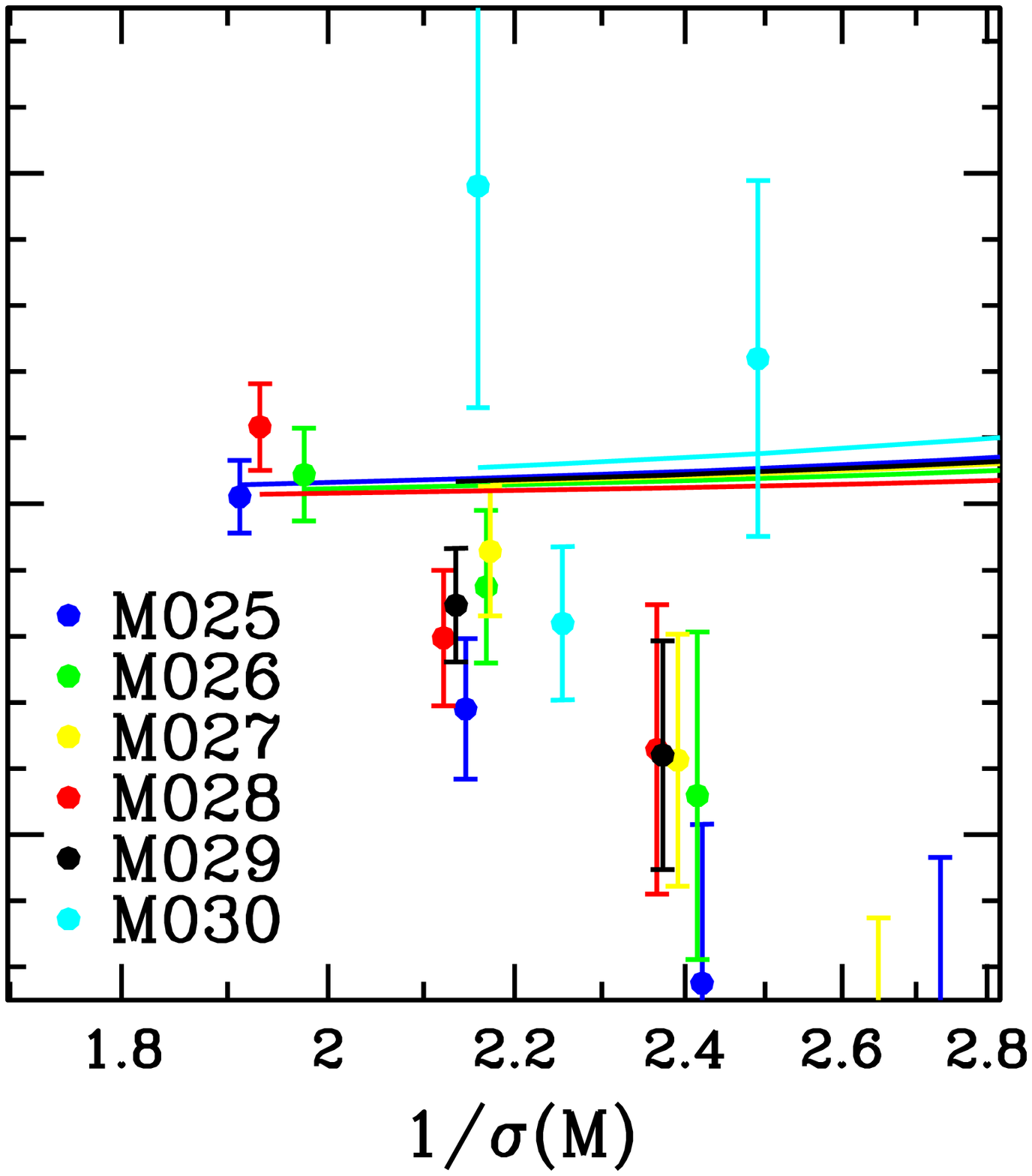}
\hspace{-2.0cm}
\includegraphics[width=7.5cm]{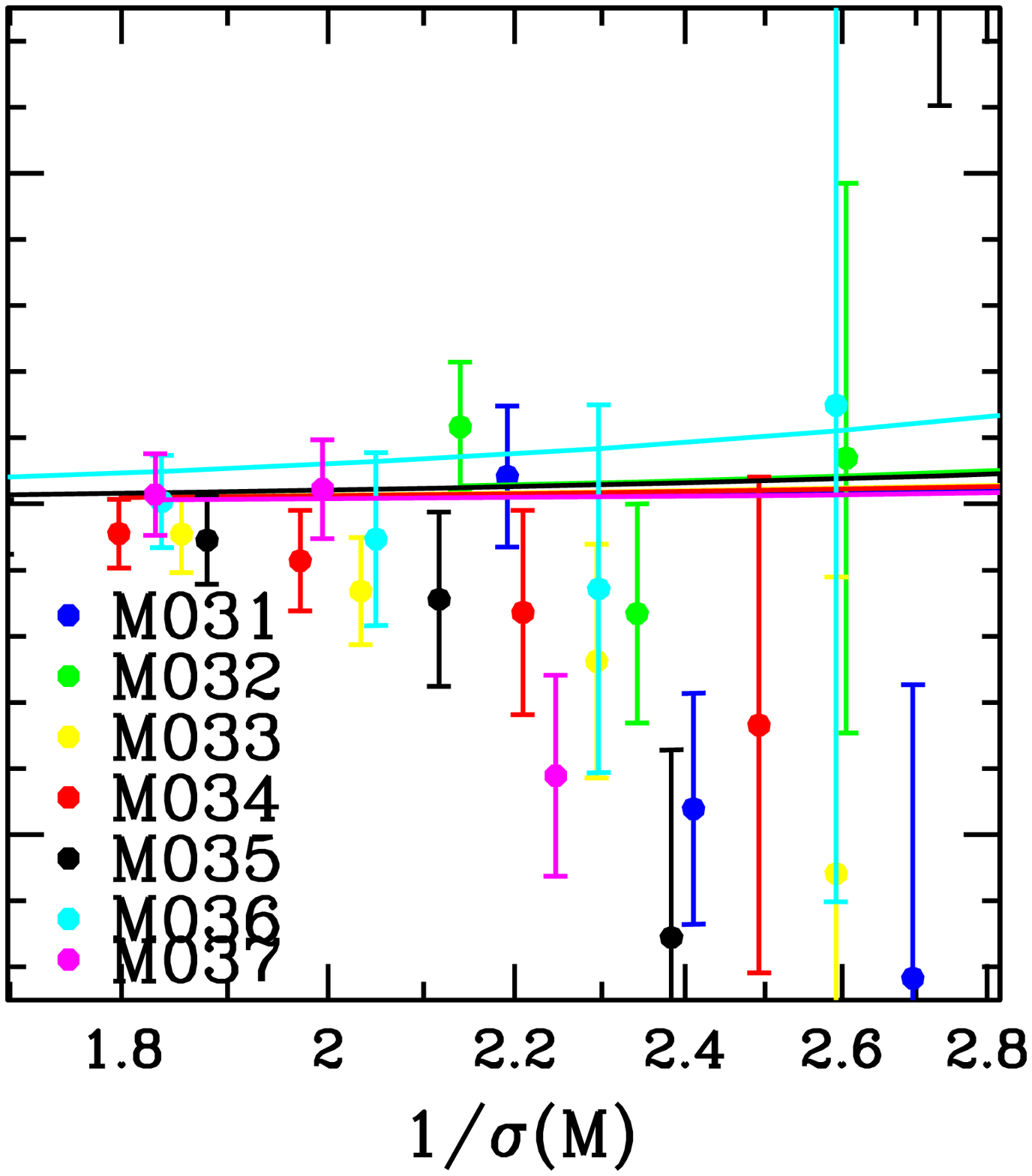}
\caption{Ratio of the mass function for  $w$CDM
  cosmologies to the best fit mass function for the
  $\Lambda$CDM case at z=2. The lines show the ratio of the fit if
  $\delta_c$ is cosmology dependent  in the $\Lambda$CDM fit.}  
\label{fig:wcdmz2}
\end{figure*}

\begin{figure*}[t]
\includegraphics[width=6cm]{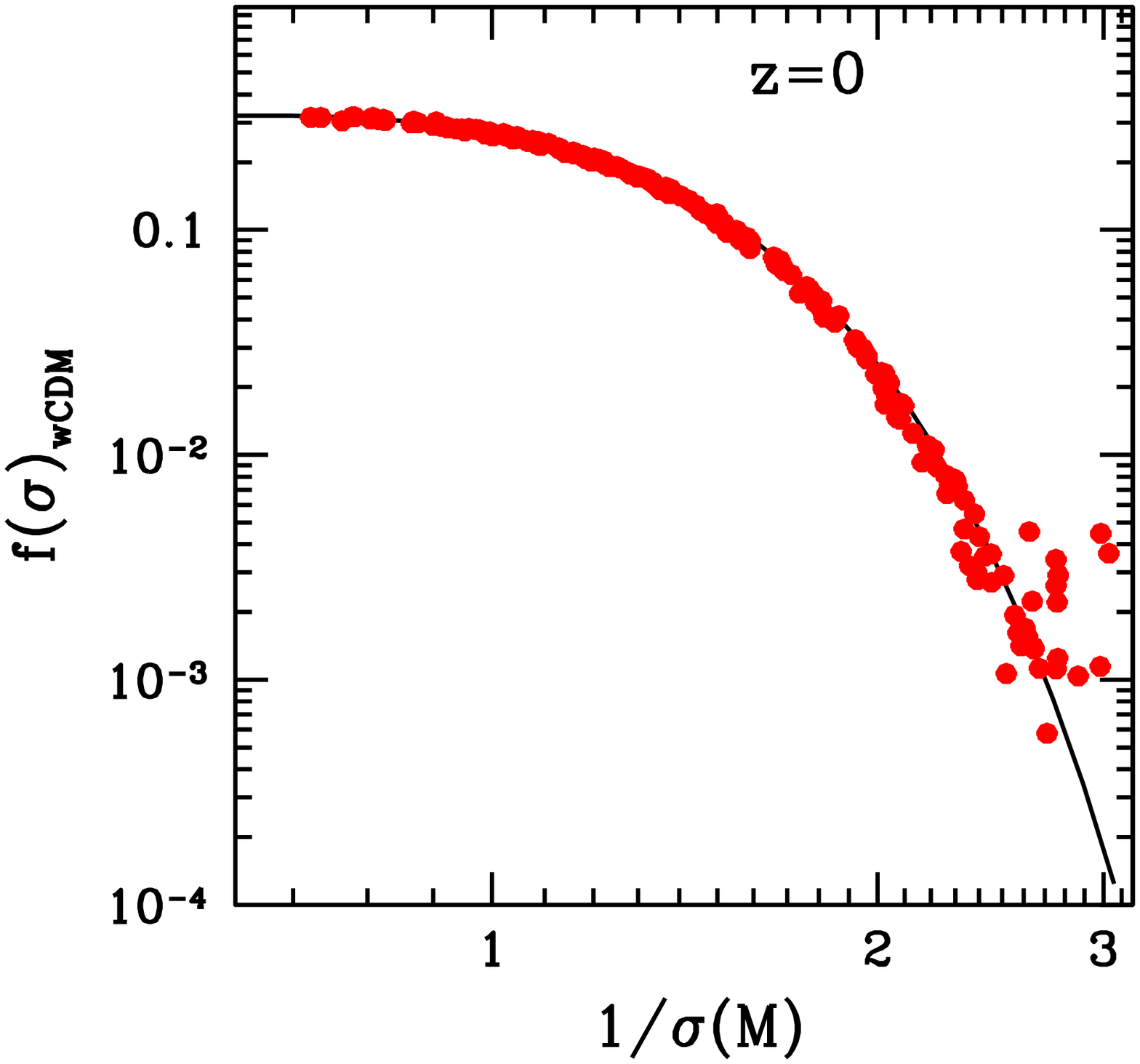}
\hspace{0cm}
\includegraphics[width=6cm]{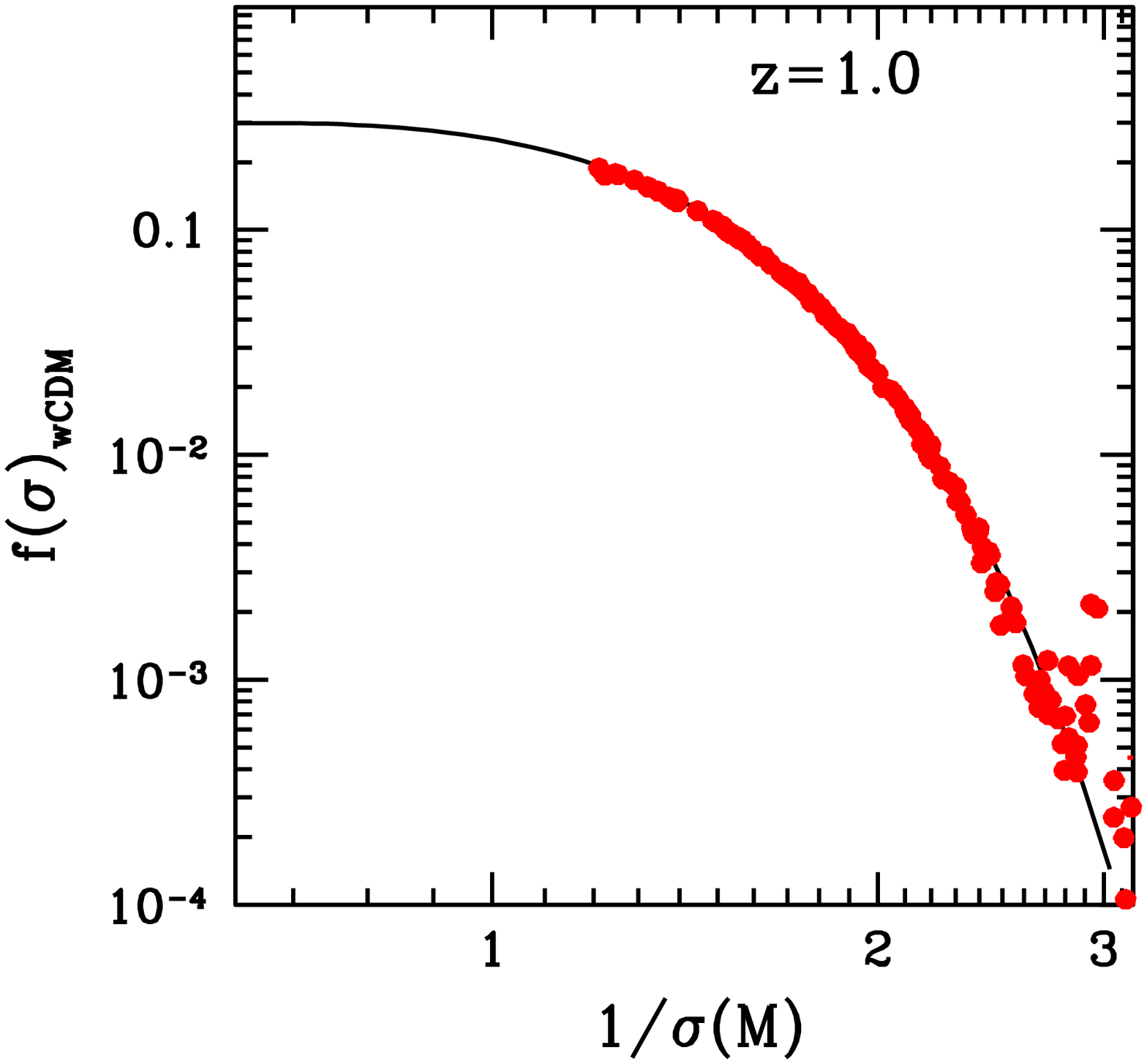}
\hspace{0cm}
\includegraphics[width=6cm]{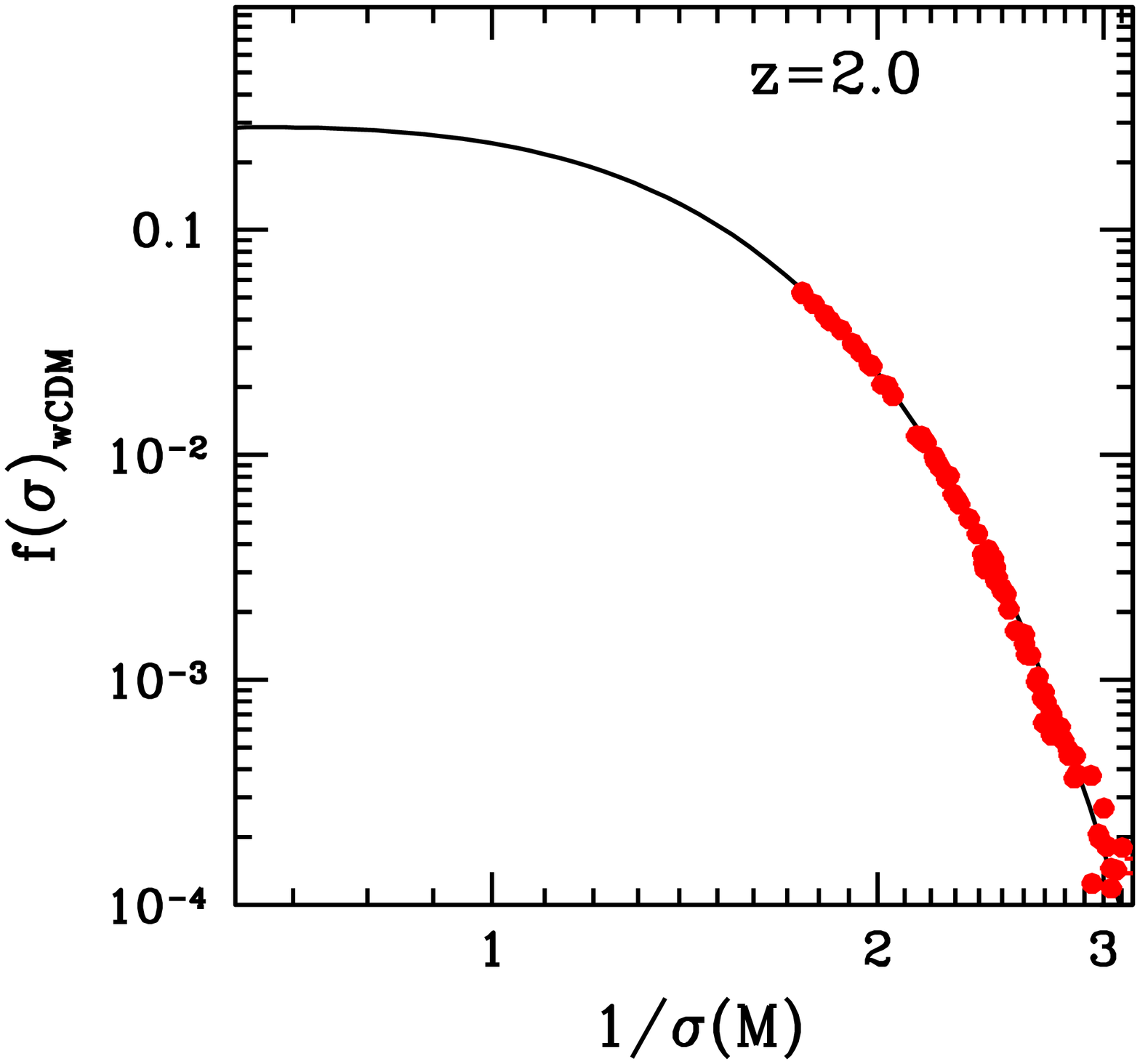}
\caption{Approximate universality of the $w$CDM mass functions, all 37 
  models shown together.}
\label{fig:wcdm_band}
\end{figure*}

The expression for bias can be written in terms of the conditional and
unconditional mass functions as
\begin{eqnarray}
b_h&=& \frac{N(m, z_1| M, V, z_0)}{n(m, z_1)V} -1 \nonumber\\
&=& \frac{\nu_{10}f(\nu_{10})}{\nu_1 f(\nu_1)}-1,
\label{eqn:bias}
\end{eqnarray}
where $N(m, z_1| M, V, z_0)$ is the average number of halos of
mass $m$ which collapsed at $z_1$ and in a cell of volume $V$ which
contains the mass $M$ at $z_0$ and $\nu= \delta_c^2/\sigma^2$.

In the large scale limit, the `peak-background split' formalism
\citep{st99, cole89} prediction for the halo bias for mass $m$ at
redshift $z$ can be derived as follows: Define the peak height $\nu_1$
relative to the background $\nu_0$ as
$\nu_{10}^2=(\delta_1-\delta_0)^2/(\sigma_1^2-\sigma_0^2)$.
Keeping the leading order terms in the expression gives $\nu_{10}^2=
\nu_1^2(1-2\delta_0/\delta_1)$.
One can then Taylor expand Equation~(\ref{eqn:bias}) and use
Equation~(\ref{eq:st}) to obtain the bias predicted by the ST mass
function in Lagrangian space. Converting to bias in Eulerian space,
the result is
\begin{equation}
  b_{\rm ST}=  1+ \frac{a\nu -1}{\delta_{c}} + \frac{2p/\delta_{c}}{1
    + (a\nu)^p}. 
  \label{eq:biasst}
\end{equation}

Similarly, using the mass function fit based on our simulation results,
Equation~(\ref{eq:st_mod}), the large scale bias can be expressed as
\begin{equation}
  b_{\rm mod}= 1 + \frac{\tilde{a}\nu - \tilde{q}}{\delta_{c}}
  +\frac{2\tilde{p}/\delta_c}{1 +(\tilde{a}\nu)^{\tilde{p}}}.
  \label{eq:biasstmod}
\end{equation} 
Figure~\ref{fig:bias} shows the predicted bias compared to the ST
result. Due to the difference in the two mass function fits, we find a
corresponding change in the large scale bias of $10-15\%$. Note that
this difference is maximal for halos in the mass range $10^{13}-
10^{14}$~M$_\odot$. Unfortunately, the fact that the (nominally)
improved prediction for the bias lies consistently below the ST
prediction, makes it deviate even further from numerical calculations
for the large-scale bias as a function of mass. This behavior is in
excellent accord with the analysis in \cite{lukicphd08} performed
using the \cite{warren05}~mass function fit and simulation data from
\cite{lukic07}. Therefore, we conclude that the simple halo model
result for the halo bias does not converge correctly as one essential
ingredient -- the mass function accuracy -- is systematically
improved. Our conclusion is consistent with other recent studies:
\cite{tinker10} have found that SO halos are systematically biased
$10\%$ higher compared to the Sheth-Tormen bias prediction.
\cite{manera09} also conclude that a prediction based on a simple
peak-background ansatz is inadequate to find agreement with their
simulation results. 

\section{Mass Function for $w$CDM Cosmologies}
\label{section:mf_wcdm}

Based on our results in Section~\ref{section:massfunction}, which
include a mass function fit at 2\% accuracy for the reference
$\Lambda$CDM cosmology, we now investigate how well this fit works as
cosmological parameters are varied. Including the dark energy equation
of state parameter, $w$, as a phenomenological constant, we consider a
suite of $w$CDM models: 37 simulations that span a wide range of
cosmological parameter values. The values of the parameters for each
simulation are given in Table~\ref{tab:basic} with the ranges being
specified in Equation~(\ref{eq:range}). Although each simulation is
individually large, the volume of each run is much smaller
(2.2~Gpc$^3$) than for the combined reference $\Lambda$CDM runs
(250~Gpc$^3$). Therefore, the statistics of the halos, especially at
high masses, is not as well-determined in this case. Consequently, we
do not attempt an as careful statistical fit for the mass function for
the $w$CDM models, although we do use all the relevant error controls.
For each run we compute the Poisson error and the sample variance
error using Equation~A5.

Interestingly, it turns out that the modified universal fit already
derived for the $\Lambda$CDM case, including the redshift evolution,
also works in a relatively unbiased way for the suite of $w$CDM
cosmologies at the $5-10\%$ level. However, clear systematic
violations of universality are observed from this suite of runs as
cosmological parameters are varied (Figures~\ref{fig:wcdmz},
\ref{fig:wcdmz1} and \ref{fig:wcdmz2}). The figures show the ratio of
the $w$CDM mass functions at $z=0,~1$, and 2 obtained from our
simulation suite with respect to the reference $\Lambda$CDM mass
function. We find that the FOF mass function shows systematic
$\sim$5-10\% variations (in both directions) with respect to the
$\Lambda$CDM reference (at higher masses, the statistics is not good
enough to make very precise statements), although universality is
preserved at the $2-3\%$ level upto $1/\sigma\sim 1.2$. Note that for
some choices of cosmological parameters the difference is as large as
10\% especially at the high mass end at $z=0$.

We also investigate the cosmology dependence of $\delta_c$ as
predicted by spherical collapse. Details of the calculation of
$\delta_c$ as a function of cosmology are given in
Appendix~\ref{sphcol}. (See also \citealt{pace10} for a detailed study.)
The lines in Figures~10, 11, and 12 represent the ratio of our fit at
a particular redshift when $\delta_c^{wCDM}$ is used in our expression
in Table~4 to the reference $\Lambda$CDM fit at that redshift. Adding
$\delta_c^{wCDM}$ agrees qualitatively with the $w$CDM runs, however the
difference predicted by the spherical collapse model is too small to
be in quantitative agreement with our results. Note that
$\delta_c^{wCDM}$ depends only on two parameters for a flat $w$CDM
cosmology, namely $\Omega_m$ and $w$. However, universality can also
break down as other parameters change, and some of these --
$\sigma_8$, $h$, and $n_s$, -- are also varied here. There is
currently no theoretical framework to understand the breaking of
universality with these parameters. At higher redshift, $w$CDM
cosmologies are expected to converge to the $\Omega_m$=1 cosmology.
This is roughly the trend seen in Fig. 11 and 12 for the case of $z=1$
and $z=2$.

Figure~\ref{fig:wcdm_band} summarizes our findings in this section
showing the best fit mass function $f(\sigma)$ at three redshifts and
the simulation results from the 37 cosmologies. Overall, all
cosmologies are described reasonably well by our new fit. Points to be
noted include: (i) An approximately universal fit describes the $w$CDM
mass function at $10\%$ accuracy over an observationally useful range
of cosmological parameters, and (ii) universality of the mass function
is systematically broken at this level of accuracy within both $w$CDM
(for example, models 22, 25, and 35) and $\Lambda$CDM cosmologies
(models 31 and 34).

Given the current observational state of the art, a $w$CDM mass
function determined at the $5-10\%$ level over the range of
cosmologies studied here appears adequate for data analysis. However,
this is not likely to be the case for the next generation of
observations. To extend our work further would require a series of
large simulations with their mass function results interpolated along
the same lines as already achieved for the power spectrum to scales of
order $k\sim 1$~$h$Mpc$^{-1}$ \citep{lawrence}.

\section{Discussion}
\label{section:disc}

In this work, we study the mass function of dark matter halos over a
wide range of masses ($6\cdot 10^{11}-3\cdot 10^{15}$~M$_{\odot}$ for
a reference $\Lambda$CDM model) and a redshift range of $z=0-2$ for a
large range of $w$CDM cosmologies. A friends-of-friends algorithm with
a linking length of $b=0.2$ is used for halo identification. The
primary aims are to control numerical errors and gain sufficient
statistical power for cosmological parameter estimation and testing of
the universality of the mass function. For a reference $\Lambda$CDM
model, we achieve a 2\% error in determining the mass function. At
this level of numerical control, deviations from universality (in both
cosmological parameters and redshift evolution) can be studied
systematically. Our level of numerical control is approaching the
$N$-body baseline level -- necessary but by no means sufficient --
required by next-generation cosmological surveys. The quest for high
accuracy in the `$N$-body' mass function has a natural stopping point
at the percent level simply because at this point many other physical
processes become important (e.g., baryonic effects,
\citealt{stanek09}). Moreover, the connections to observations need to
be directly modeled and end up adding their own significant
contribution to the overall error.

We use a large number of high resolution simulations for studying the
mass function in a single $\Lambda$CDM cosmology. These simulations
are used to establish the error control methodology in order to obtain
an accurate mass function. Error sources systematically studied here
include effects of finite force resolution, FOF particle sampling
bias, and systematics induced by finite-volume effects. 

Using the more accurate mass function fit for this reference
cosmology, we also rederive the large scale halo bias using the
`peak-backgound split' approach. Compared to previous results obtained
with the ST fit, the halo bias changes by 10-15\% between $z=0-2$, and
is systematically lower. Unfortunately, instead of improving the
agreement with direct numerical measurements of halo bias (computed by
taking ratios of correlation functions), the use of an improved mass
function only increases the discrepancy. This points to an essential
difficulty with the halo modeling approach. We will return to this
problem elsewhere~\citep{lukic10}.

After studying the mass function in detail for the reference
$\Lambda$CDM cosmology, we extend the range of cosmological parameters
by considering a suite of $w$CDM simulations. The range of parameters
is set by the current constraints on cosmological parameters. The
simulation parameters for the runs are given in Tables~\ref{tab:basic}
and \ref{tab:sim_specs}. We note that the reference mass function fit
provides a good description of the $w$CDM results at an accuracy of
$\sim 10\%$, however, with systematic deviations that point to clear
violations of the universality of the mass function, not only in
$w$CDM parameter space, but also within the set of $\Lambda$CDM models
(and models very close to $\Lambda$CDM).


The breaking of universality as a function of different $\Lambda$CDM
parameters is studied in \cite{J01, tinker08, warren05}. These studies
have shown that universality breaks down at the 20\% level when
$\sigma_8$ and $\Omega_m$ are varied. Also \cite{courtin10} have
studied how universality breaks down for quintessence cosmology and
found similar variation of universality with cosmology as reported
here. The universality of the mass function in the presence of early
dark energy has been studied in \cite{grossi_ede}. We find that
universality in the mass function holds at the 10\% level as a
function of cosmological parameters (the $w$CDM suite) and redshift
($w$CDM and the reference $\Lambda$CDM model).

Given the breaking of universality discussed here (see also
\citealt{tinker08}), it is not clear what the utility of fitting
formulae for the mass function might be for observations that actually
do require theoretical predictions at the percent level of accuracy.
An extensive future simulation campaign will be needed to properly
come to grips with this problem. Nevertheless, to provide a compact
description of our results, we derive a fitting formula that agrees
with our reference $\Lambda$CDM simulation results at 2\% accuracy
over the redshift range of $z=0-2$ and the mass range of $6\cdot
10^{11}-3\cdot 10^{15}$~M$_{\odot}$. The fitting function is a simple
modification of the Sheth-Tormen form with one extra shape parameter
to improve the fit at $z=0$ and two extra evolution parameters. This
form does not lead to divergences if the normalization condition that
all mass resides in dark matter halos is imposed. It holds at the 10\%
level of accuracy for a broad class of $w$CDM cosmologies with the
redshift evolution taken into account.

\acknowledgments

A special acknowledgment is due to supercomputing time awarded to us
under the LANL Institutional Computing initiative. Part of this
research was supported by the DOE under contract W-7405-ENG-36 and by
a DOE HEP Dark Energy R\&D award. SB, SH, KH, ZL, and CW acknowledge
support from the LDRD program at Los Alamos National Laboratory. KH
and ZL were supported in part by NASA. MW was supported in part by
NASA and the DOE. KH, SH, and MW thank the Aspen Center of Physics
where part of this work was completed. SB and ZL would like to
acknowledge useful discussions with Darren Reed. We thank the referee
for a careful reading of the manuscript and for a number of useful
insights and suggestions.

\appendix 

\section{Systematic Errors in Mass Function
  Measurements from N-body Simulations}
\label{app}

\subsection{Initial Conditions}
\label{ics}

\begin{figure}[t]
\includegraphics[width=9.0cm]{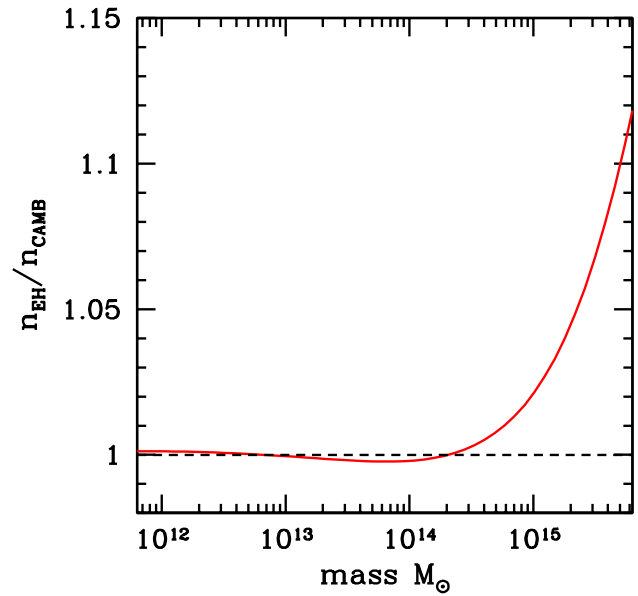}
\caption{Ratio of halo abundances using the fitting formula given in Table~3
for two different transfer functions- the analytic fitting form
of~\cite{eh} and that generated using CAMB. The ratio of the mass
function for two different transfer function choices is shown.  At the
high mass end, the mass function is overestimated by several percent
if the analytic fit is used. }
\label{fig:tk}
 \end{figure}

 One issue regarding initial conditions not considered in
 \cite{lukic07} is the accuracy of the transfer function used for
 generating the linear power spectrum. Usually computed via Boltzmann
 codes such as CMBFAST~\citep{seljak} or the related code
 CAMB\footnote{http://camb.info}, the transfer function is in
 principle known to 0.1\% accuracy~\citep{trans}, quite sufficient for
 the task at hand. Due to their simplicity, however, analytic fits
 have sometimes been used. As an example, the formula given
 in~\cite{eh} agrees with numerical results to better than 5\% (around
 a central region of parameter space). However, using such a fit (as
 for example, in the work of \citealt{crocce09}) gives two errors: (i)
 an incorrect mapping from $\sigma(M)$ to $M$ (Figure~\ref{fig:tk},
 where using the wrong transfer function yields a noticeable
 overprediction of the halo abundance at high masses), and (ii) a
 systematic error in the underlying simulation which no longer
 accurately represents the correct cosmology. It is therefore
 important to use the most accurate transfer function available to
 obtain precision results from simulations.

\subsection{Error due to Finite Force Resolution}
\label{force}

To study the effect of finite force resolution errors quantitatively,
we use two simulations with identical initial conditions and
cosmological parameters. Both the runs have a box size of length
1.3~Gpc and $1024^3$ particles. One of the simulations is run using
{\sc GADGET-2} with a force resolution of 50 kpc as specified in
Table~\ref{tab:sim}. The other was run with a particle-mesh (PM) code
with a $2048^3$ spatial grid, corresponding to a force resolution of
approximately 700 kpc. We call the two runs G and PM respectively.

\begin{figure}[t]
\includegraphics[width=9cm]{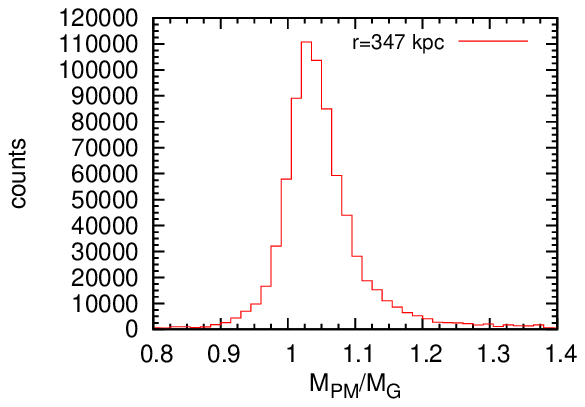}
\hspace{-1.7cm}
\includegraphics[width=9cm]{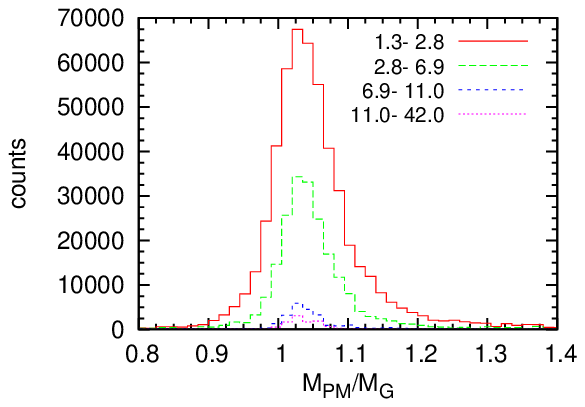}
\caption{Distribution of the ratio of halo masses in the low
  resolution (PM) and high resolution (G) runs in units of
  $10^{14}$~M$_{\odot}$. The top panel shows the distribution for all
  halos with $M \ge 1.3\times 10^{14}$~M$_{\odot}$ in run G that have
  a matching halo in the PM run; more than 95\% of the halos have
  matched pairs within 347 kpc. The bottom panel shows the
  distributions for different mass bins demonstrating that the shift
  in the halo mass between the high and low resolution runs is
  practically independent of mass.} 
\label{fig:hist}
\end{figure}

Since the runs have identical initial conditions, halos in both runs
should have approximately the same locations. For each halo in run G,
we expect to find a corresponding ``match'' in run PM. However, as
shown in~\cite{lukic07}, for a given force resolution, halos below a
certain mass (or equivalently containing a certain number of
particles) are not reliably formed. For the PM run, the associated
prediction for the minimum number of particles required for a halo to
form is $\approx 400$ (see Equation~30 in \citealt{lukic07}). In order
to make sure that the mass function is not suppressed due to this
effect, the actual number of particles in a halo should be larger than
this minimal value. We find that $\sim2000$ particles per halo is a
good choice and hunt for matched halo pairs only above a corresponding
mass of $2\cdot 10^{14}$~M$_\odot$ (or halos containing more than 2500
particles). This being the case, we find that nearly 95\% of the halos
have their centers within $\sim400$~kpc (consistent with the force
resolution of the PM run). The ratio between the matched halo masses
averaged over all the matched pairs is shown as a histogram in
Figure~\ref{fig:hist}. As a function of mass bin (lower panel), the
peak of the ratio distribution remains essentially unchanged, with
median values ranging from 1.041 to 1.034. Roughly speaking, the data
is therefore consistent with an overestimate by approximately 4\% in
individual halo masses in the PM run, independent of the mass. (Note
that the mean of the ratio distribution has a slightly stronger
dependence on halo mass, ranging from 1.06 to 1.036, resulting from
tail effects in the distribution.)

\begin{figure*}
\includegraphics[width=7.5cm]{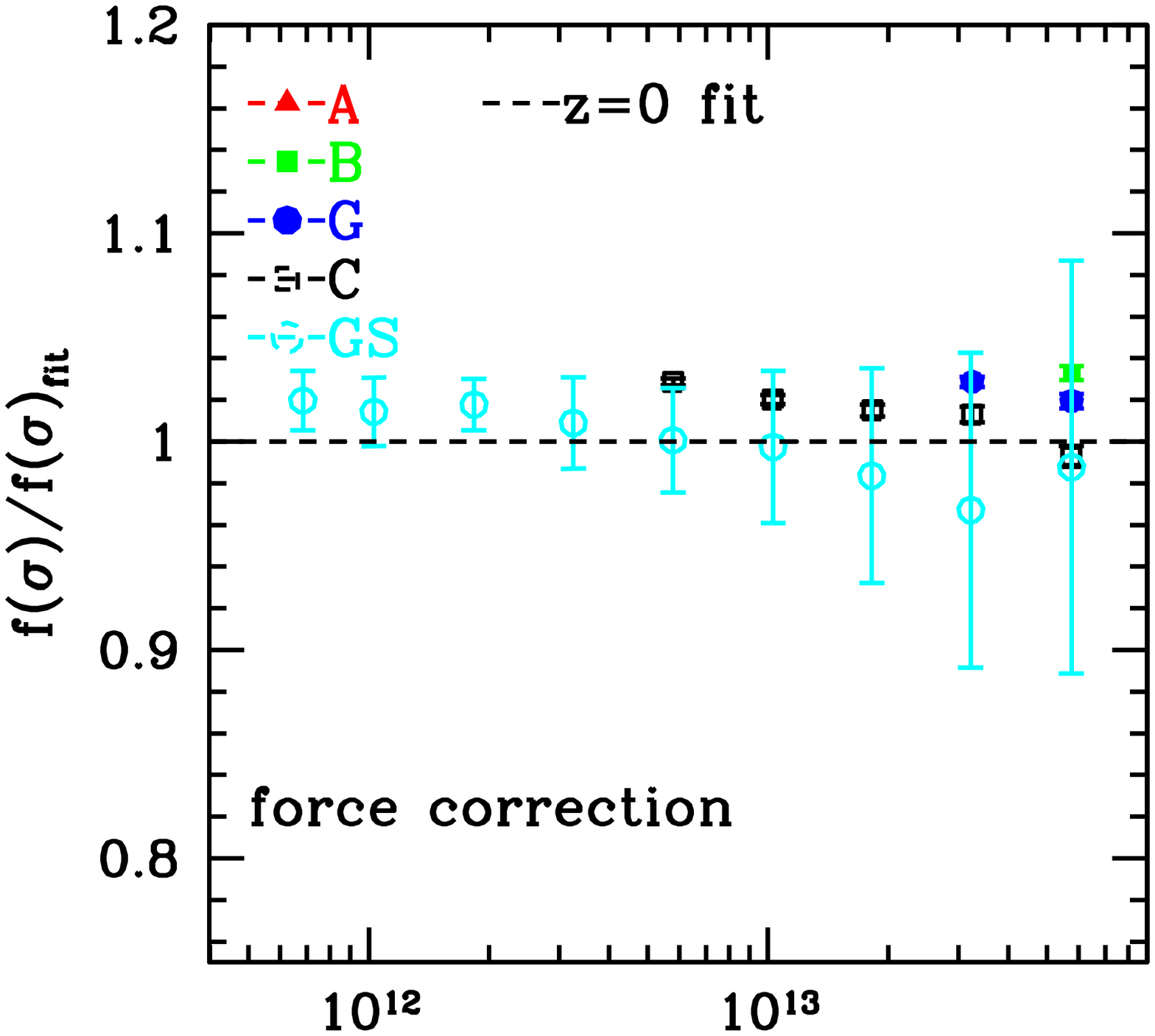}
\hspace{-2.2cm}
\includegraphics[width=7.5cm]{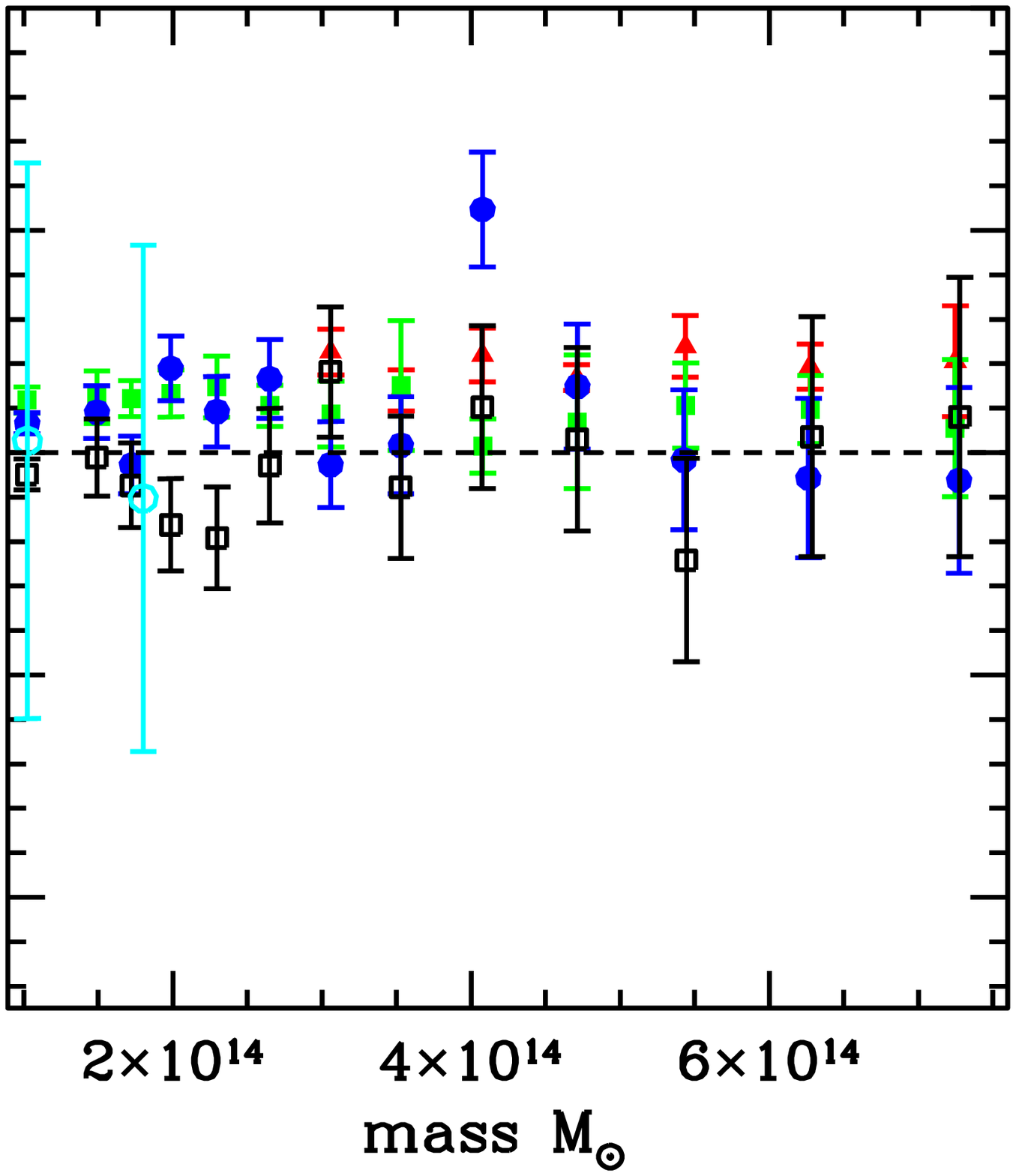}
\hspace{-2.2cm}
\includegraphics[width=7.5cm]{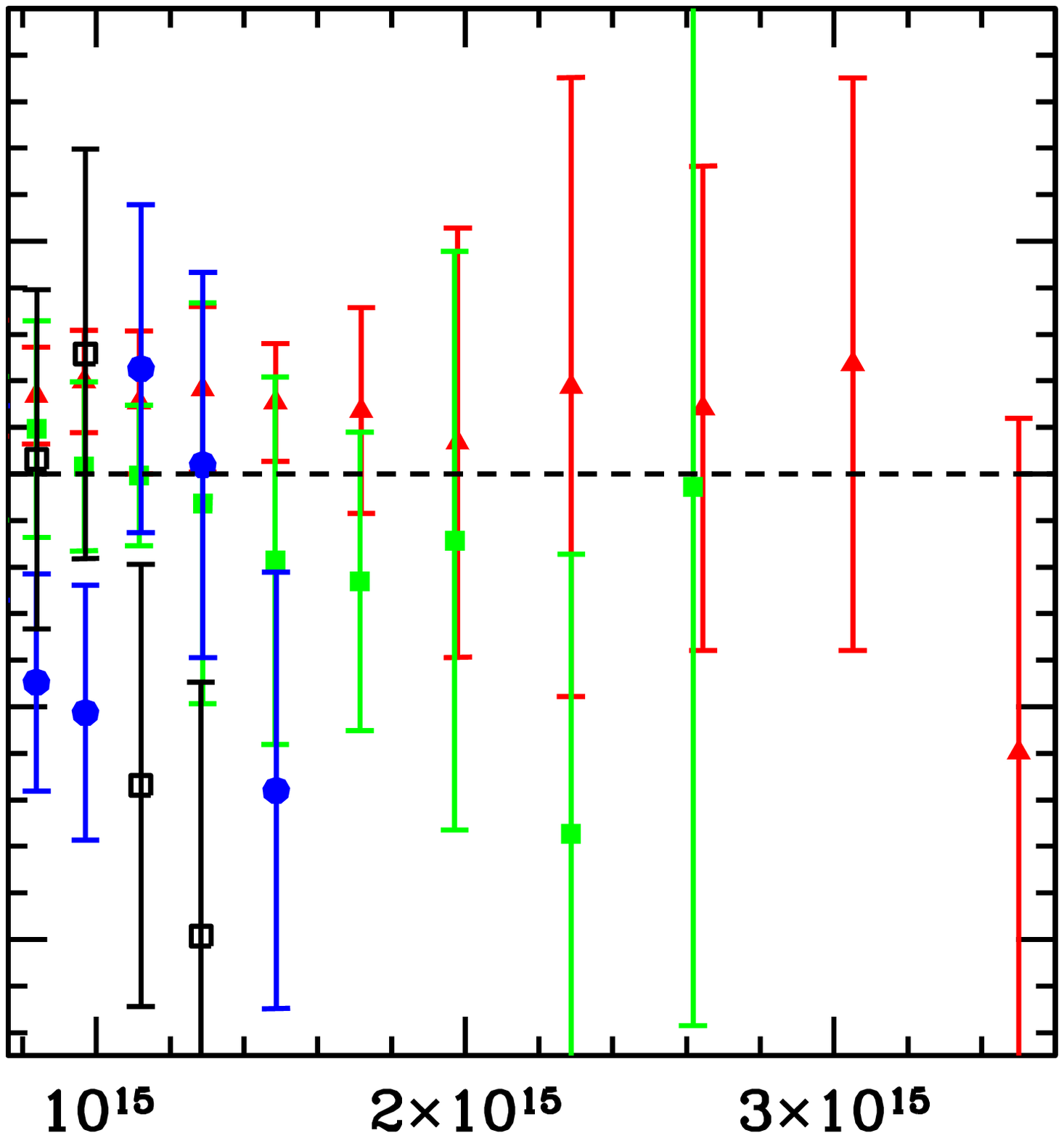}
\caption{ Impact of the correction for finite force resolution errors
  on the mass function. The ratio of the simulation results with
  respect to the best-fit $z=0$ mass function given in
  Equation~(\ref{eq:st_mod}) is shown. We display results for different box
  sizes separately to study possible numerical artifacts in the
  overlapping regions of different boxes. The effect only alters high
  mass halos (see for a comparison the uncorrected mass function in
  Figure~\ref{fig:mass_corr}) and leads to a systematic lowering of
  the mass function by 2\% at the high mass end.} 
\label{fig:force_corr}
\end{figure*}

Since the effect is small, a simple linear extrapolation is sufficient
to correct each of the runs in Table~\ref{tab:sim} in order to predict
the halo mass in the limiting case.  The corrected mass $M_c$ is given by
\begin{equation}
M_c/M= 1.0- 0.04(\epsilon/650 ~{\rm kpc}).
\end{equation}
Here $M$ is the uncorrected halo mass and $\epsilon$ is the force
resolution measured in kpc of the different runs as specified in
Table~\ref{tab:sim}. (We have checked that this formula is consistent
with another smaller set of PM simulations -- $256^3$ particles,
$1024^3$ mesh, with a force resolution of 334~kpc.) The biggest
correction needed is $\approx 0.6\%$ for individual halo masses for
the $A$ runs (with a force resolution of 97~kpc). This results in a
systematic lowering of approximately 2\% in the high-mass tail of the
mass function (primarily run A) as shown in
Figure~\ref{fig:force_corr}.

\subsection{Error due to Finite Number of Particles in a Halo}
\label{fnum}

The reason for the scatter and bias in FOF masses due to the finite
number of particles in a halo is that the FOF algorithm aims to
capture the mass of a halo within a certain iso-density contour,
rendering the halo mass sensitive to the accurate determination of
this boundary. Undersampling of the halo will lead to particles on the
halo boundary tending to link more to particles close by than in the
case of a well-sampled halo with the end result of overestimating the
halo mass (see \citealt{lukic08} for details).

In their work on the mass function, \cite{warren05} suggested a
correction for the FOF halo mass of the form, $ n_h^{\rm
  corr}=n_h(1-n_h^{-0.6}),$ where $n_h$ indicates the number of
particles in a halo. This adjustment is an empirical finding using a
set of nested volume simulations. This correction factor has been used
in recent studies, for example \cite{tinker08, crocce09}. In broad
outline this result is consistent with the findings of \cite{lukic08}.
The adjustment formula lowers masses for halos with smaller numbers of
particles and depends only on the number of particles within a halo.
Unfortunately, this adjustment cannot be applied in all circumstances,
as the details of the correction can depend on the details of the
individual simulations. Additionally, this adjustment should not be
applied at small particle numbers, where it is known to
overcompensate.

The problem with FOF halo masses described above is best seen in our
simulations by comparing results for the mass function in overlapping
mass bins across the different-sized simulation boxes (with differing
mass resolutions, see also \citealt{warren05}). This is shown in the
upper panel in Figure~\ref{fig:mass_corr}. In our work we combine five
different box sizes (multiple runs for each box size) with different
mass and force resolution to cover a large range of halo masses. The
figure shows the ratio of the raw simulation results from the five
different box sizes with respect to our best-fit mass function. In the
absence of a systematic bias across the boxes, the results should
match up within Poisson errors, but this is clearly not the case.

In order to correct for the finite particle sampling of the halos we
investigate different strategies. If all halos are NFW, then following
the analysis in \cite{lukic08} we can implement a correction which
takes into account the concentration and the number of particles in a
halo. The general trend is that higher concentration halos should have
smaller corrections while, overall, the corrected halo masses would be
lower (as in the simulations). Since our force resolution is not
sufficient to reliably determine the concentration of the halos, we
can use the known mass-concentration relation (including scatter) to
estimate a concentration for each halo. It turns out that such a
correction lowers the mass function overall by almost 10\% but does
not provide an accurate match between different boxes. Therefore, this
simple modeling fails to explain the quantitative results from the
simulations, presumably because the concentration estimates are too
crude and because of the irregularities of roughly 20\% of the halos.

\begin{figure*}
\includegraphics[width=7.5cm]{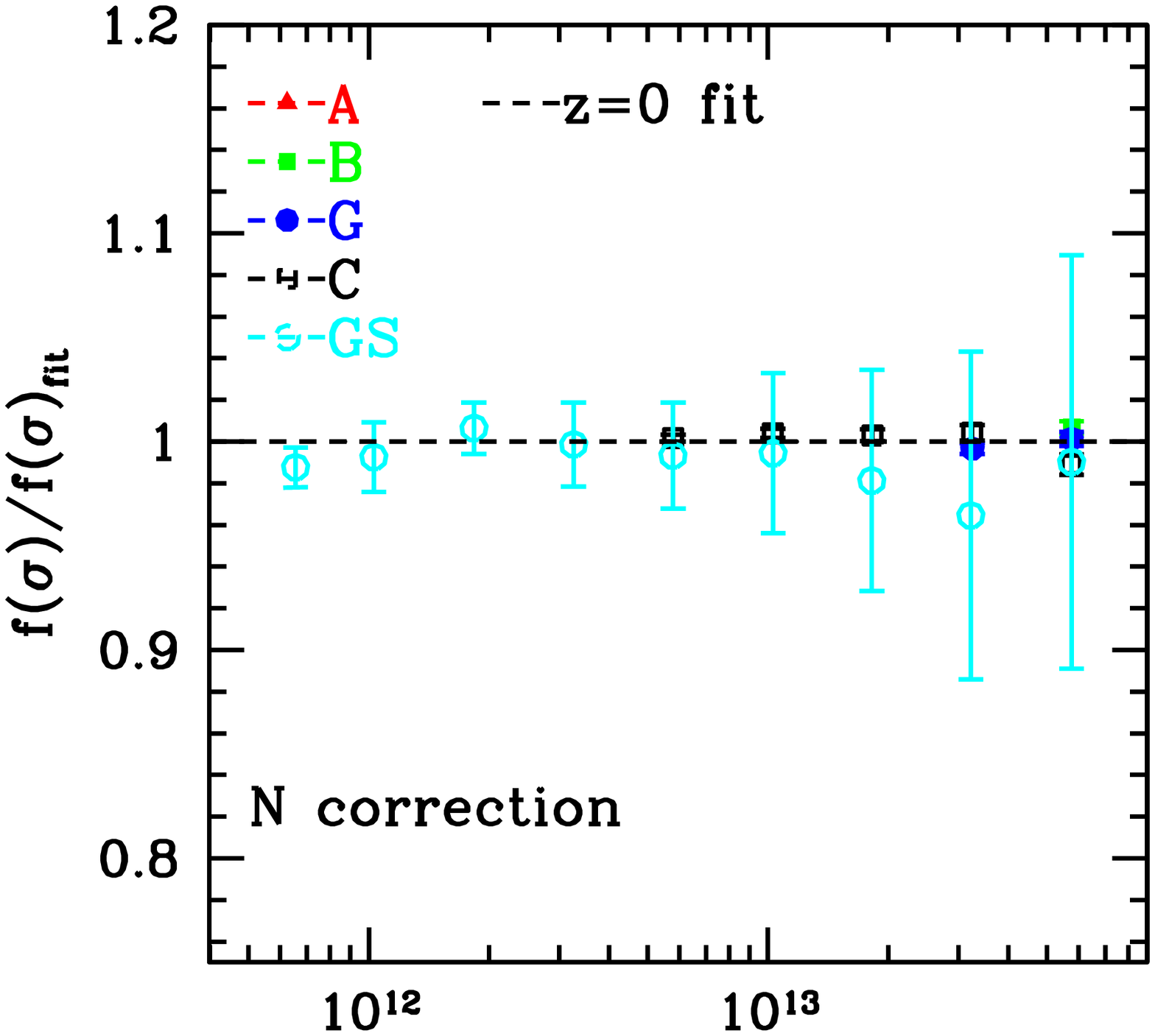}
\hspace{-2.2cm}
\includegraphics[width=7.5cm]{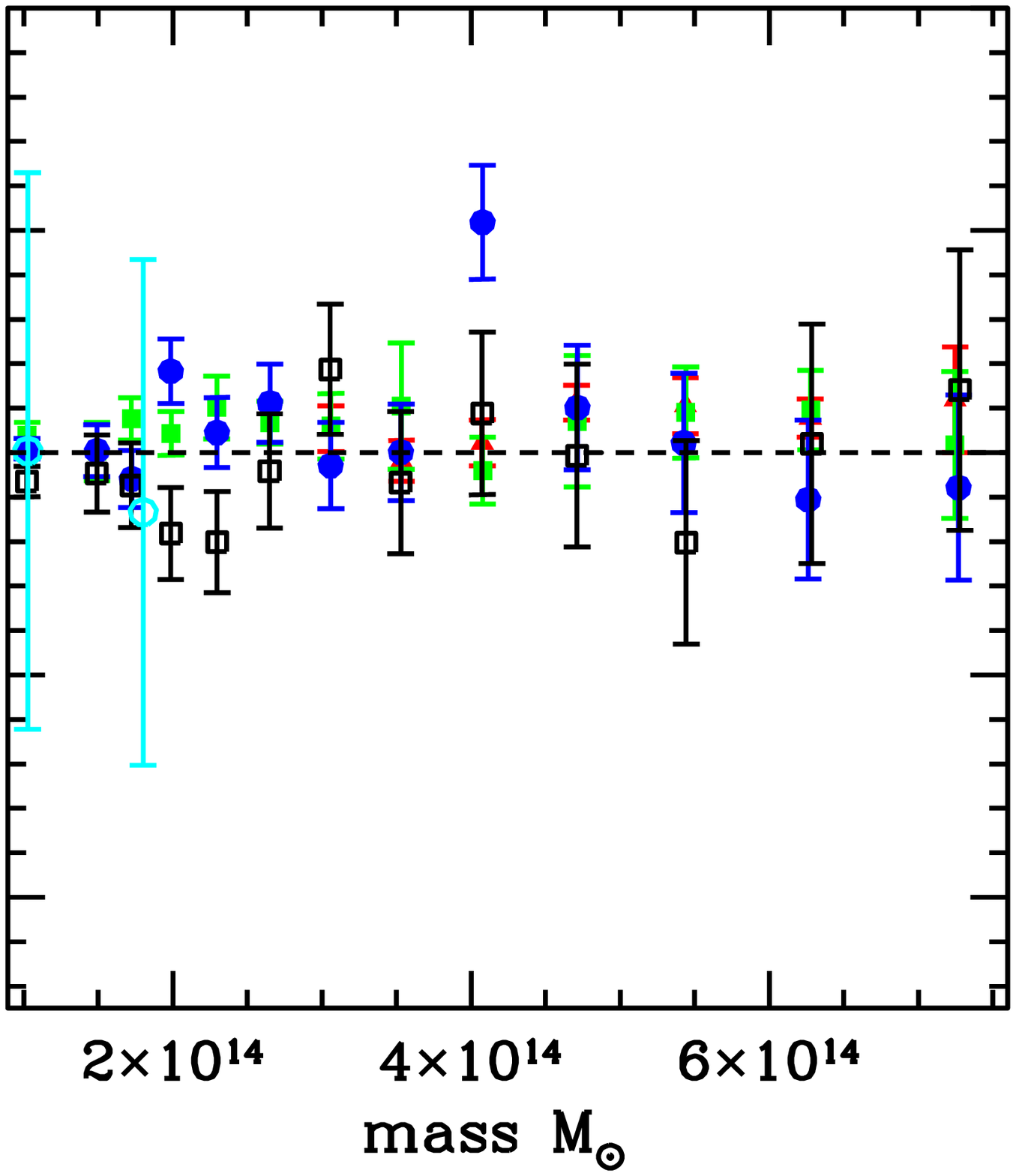}
\hspace{-2.2cm}
\includegraphics[width=7.5cm]{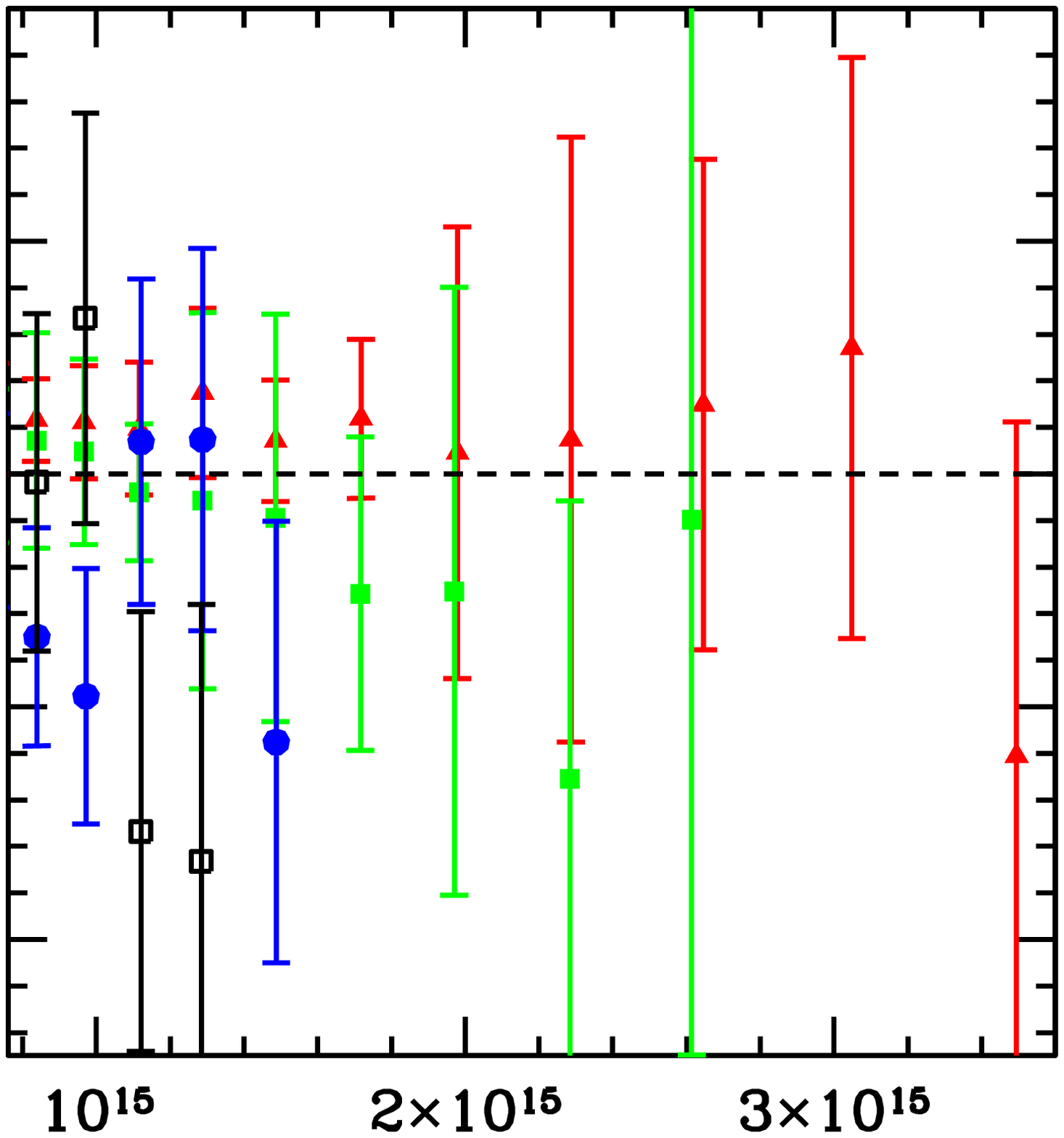}
\caption{ Impact of finite sampling errors on the mass function. The
  ratio of the simulation results with respect to the best-fit $z=0$
  mass function given in Equation~(\ref{eq:st_mod}) is shown. The
  effect from finite sampling is much larger than the effect due to
  finite force resolution.} 
\label{fig:n_corr}
\end{figure*}

We therefore use a conservative strategy of a large minimal number of
particles per halo, $n_h>400$, and apply the empirical $n_h$ dependent
correction. We find that the suggested form in \cite{warren05} can be
slightly improved by changing it to: 
\begin{equation}
n_h^{\rm corr}=n_h(1-n_h^{-0.65}).
\end{equation}
The result of applying the correction according to the above formula
is shown in Figure~\ref{fig:n_corr}.

\subsection{Finite Volume Effects}
\label{fvol}

In this appendix we analyze the third finite volume effect mentioned
in Section~\ref{section:halos} in detail, related to the (effective)
number of independent realizations, i.e., the sample variance.
Because we are using a simple, independent simulation box technique,
we compute the variance for each of the runs -- separately for each
fixed value of the simulation volume -- over the number of
realizations of each run. Thus for each run (A, B, C, etc.) we
calculate 
\begin{equation}
  \frac{\Delta n}{n}= \frac{1}{\bar n}\sqrt {\sum_{N_{\rm runs}} \frac{[(n(M,z) - {\bar
      n(M,z)})^2]}{N_{\rm runs}}} 
\end{equation}
where $N_{\rm runs}$ is the number of realizations for a particular
fixed-volume run (e.g., $N_{\rm runs}=10$ for the A runs). ${\bar n}$
is the average number density of halos in a bin averaged over all the
realizations and $n$ is the number density for a single realization.
Note that this error includes both Poisson fluctuations of halos and
the separate sample variance contribution. The fractional error on
$f(\sigma)$ is then calculated as 
\begin{equation}
\frac{\Delta f_{\rm data}}{f_{\rm data}}= \frac{\Delta n_{\rm
    bin}}{n_{\rm bin}} ,
\label{eq:err}
\end{equation}
where we include only Poisson errors and ignore the error in
determining $M_{\rm bin}$ due to Poisson fluctuations. 

\begin{figure}[t]
\includegraphics[width=9.0cm]{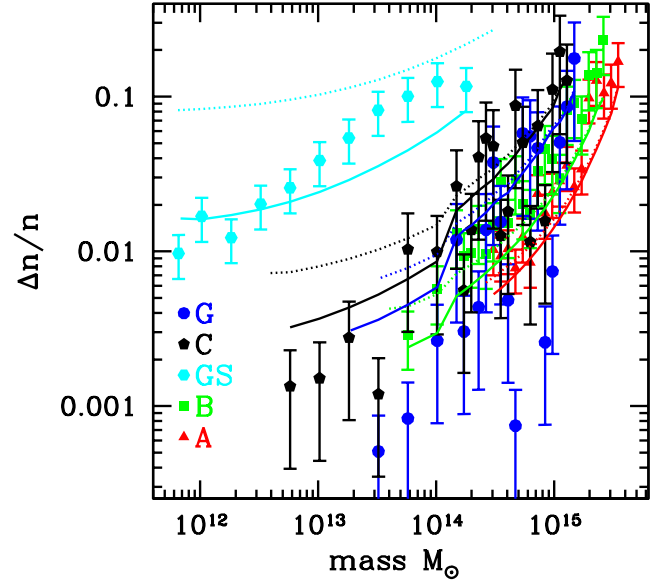}

\caption{Total fractional error (Poisson plus sample variance) for
  each simulation run. Here, the error bars denote the ``error on the
  errors'' which is simply the Poisson error. As an example, for the A runs
  with 10 realizations, the ``error on the errors'' is $\sim$ 30\%.
  The dotted lines represent the corresponding halo model predictions
  from Equation~\ref{hmerr} for an infinite volume box, while the solid
  curves represent the same predictions for the actual simulation
  sizes used. The two remain close except for the GS runs; see
  text for a discussion. For all the runs, there is satisfactory
  agreement between the theoretical estimates (solid curves) and the
  actual results from the simulations.}
\label{fig:frac_err}
\end{figure}

The fractional errors are shown in Figure~\ref{fig:frac_err}. The
results can be directly compared to simple estimates of the variance
using the halo model following \cite{hukravtsov03}. With this
approach, one finds 
\begin{equation}
\frac{\Delta n}{n}= \sqrt{\frac{1}{nV} +
  \frac{ b^2(M,z)}{(2\pi)^3}{\int d^3k\, P(k) W^2(kR_{box})}} , 
\label{hmerr}
\end{equation}
where $V$ is the volume of the simulation box, $R_{box}=
(3V/(4\pi))^{1/3}$, $W(x)$ is the spherical top-hat window function
and $b(M,z)$ is the large scale bias from Equation~\ref{eq:biasstmod}.
The first term above is the shot noise contribution and the second
term gives the sample variance contribution to the total variance.
Note that the lower limit of the integral is specified by the
simulation box size. For the larger boxes, this is irrelevant, but for
the GS runs, the effect is significant, as also shown in
Figure~\ref{fig:frac_err}, which demonstrates good agreement between
our results and the theoretical estimate. Note that by using
independent smaller volume runs we can reduce the error due to sample
variance by a significant amount (of course one should keep in mind
that too-small boxes will systematically bias the mass function low, a
small and correctable bias in our case, as shown in
Fig.~\ref{fig:mass_corr}).

\section{Spherical Collapse Calculation of $\delta_c^{wCDM}$}
\label{sphcol}

In this section we summarize the calculation of $\delta_c$ for any
cosmology as predicted by the spherical collapse model. We are
interested in perturbations in a homogeneous sphere. The full
nonlinear equation for perturbations \citep{pace10} is

\begin{equation}
\delta'' + \left ( \frac{3}{a} +
  \frac{E'(a)}{E(a)}\right)\delta'-\frac{4}{3}\delta'^2/(1+\delta)-
\frac{3}{2}\frac{\Omega_m}{a^5 E^2(a)}\delta(1+\delta)=0
\label{eq:nl}
\end{equation} 

Keeping only the linear terms in Equation~\ref{eq:nl} gives

\begin{equation}
\delta'' + \left(\frac{3}{a}+\frac{E'(a)}{E(a)}\right)\delta'-
\frac{3}{2} \frac{\Omega_m}{a^5E(a)^2}\delta=0 
\label{eq:lin}
\end{equation}

where $E(a)=\sqrt{\Omega_m/a^3+ (1-\Omega_m)}$ is the Hubble factor at
a redshift $z=1/a-1$. Assuming we want to calculate the value of
$\delta_c$ at $z=0$, we require the sphere to collapse at $z=0$. We
set the initial conditions $\delta_i'=0$. We varied $\delta_i'$ and
checked that the final value of $\delta_c$ is quite insensitive to the
value of $\delta_i'$. We solve Equation~\ref{eq:nl} such that $\delta$
becomes a large number at $z=0$ (numerically we attain convergence for
$\delta=10^8$) for a chosen initial perturbation, $\delta_i$. We
iteratively converge to the value of $\delta_i$ such that collapse
happens at $z=0$, i.e., $\delta=10^8$ at $z=0$. Once $\delta_i$ is
computed by solving Equation~\ref{eq:nl}, we use that value to solve
Equation~\ref{eq:lin}. The solution of Equation~\ref{eq:lin} gives the value of
$\delta_c$ at $z=0$. Thus, for a given cosmology, i.e., for given
values of $\Omega_m$ and $w$, solving Equations~\ref{eq:nl} and
\ref{eq:lin} gives $\delta_c$ at a given redshift, $z$.

\end{document}